\documentclass{aa}

\usepackage{txfonts}
\usepackage{natbib}
\usepackage{amsmath,amsxtra}	
\usepackage{booktabs}
\usepackage{graphicx}
\usepackage{multirow}
\usepackage[T1]{fontenc}	

\usepackage{longtable}
\usepackage{pdflscape}

\usepackage[colorlinks, citecolor = blue, linkcolor = blue, urlcolor = blue]{hyperref}

\bibpunct{(}{)}{;}{a}{}{,} 

\newenvironment{Contfigure*}{%
\addtocounter{figure}{-1}%
\begin{figure*}}{%
\end{figure*}}

\newcommand{\tfm}[1]{\tablefootmark{#1}}

\begin{document}


\title{Approaching hell's kitchen: Molecular daredevil clouds in the vicinity of Sgr~A*\thanks{based on ALMA observations under the project number 2011.0.00887.S, which were executed on 18 May 2012.} 
\thanks{Supplementary data (reduced FITS images and datacubes) of the continuum and line emission listed in Tables \ref{Obs-para} and \ref{ALMA-lines} are available at the CDS via anonymous ftp to \texttt{cdsarc.u-strasbg.fr (130.79.128.5)} or via \texttt{http://cdsweb.u-strasbg.fr/cgi-bin/qcat?J/A+A/}
}
}


\author{Lydia Moser \inst{1,2,3}
        \and \'Alvaro S\'anchez-Monge \inst{2} 
        \and Andreas Eckart \inst{2,3}
        \and Miguel A. Requena-Torres \inst{4,5}
        \and Macarena Garc\'ia-Marin \inst{6}   
        \and Devaky Kunneriath \inst{7,8}
        \and Anton Zensus \inst{2,3}
        \and Silke Britzen \inst{3}
        \and Nadeen Sabha \inst{2}
        \and Banafsheh Shahzamanian \inst{2,3}  
        \and Abhijeet Borkar \inst{2,8} 
        \and Sebastian Fischer \inst{9}
}               
          

\institute{Argelander-Institut f\"ur Astronomie, University of Bonn, Auf dem H\"ugel 71, 53121 Bonn, Germany;\\ \email{moser@astro.uni-bonn.de}
        \and I. Physikalisches Institut, Universit\"at zu K\"oln, Z\"ulpicher Str. 77, 50937 K\"oln, Germany
        \and Max-Planck-Institut f\"ur Radioastronomie, Auf dem H\"ugel 69, 53121 Bonn, Germany
        \and Space Telescope Science Institute, 3700 San Martin Dr., Baltimore, 21218 MD, USA
        \and Department of Astronomy, University of Maryland, College Park, MD 20742, USA
        \and European Space Agency, 3700 San Martin Drive, Baltimore, 21218 MD, USA
        \and National Radio Astronomy Observatory, 520 Edgemont Road, Charlottesville 22903, USA
        \and Astronomical Institute, Academy of Sciences, Bo{\v c}n\'i II 1401, CZ-14131 Prague, Czech Republic
        \and German Aerospace Center (DLR), K\"onigswinterer Str. 522-524, 53227 Bonn, Germany
}

\date{Received 26 February 2016 / Accepted 19 December 2016}

\abstract {We report serendipitous detections of line emission with the Atacama Large Millimeter/submillimeter Array (ALMA)
in bands 3, 6, and 7 in the central parsec down to within 1$''$ around Sgr~A* at an up to now highest resolution (<0.5$''$) view of the Galactic center (GC) in the submillimeter (sub-mm) domain.

From the 100 GHz continuum and the H39$\alpha$ emission we obtain a uniform electron temperature around $T_e \sim 6000$ K for the minispiral. The spectral index \textbf{($S \propto \nu^\alpha$)} of Sagittarius A* (Sgr~A*) is $\sim$ 0.5 at 100 - 250 GHz and $\sim$ 0.0 at 230 - 340 GHz. The bright sources in the center show spectral indices around -0.1 implying Bremsstrahlung emission, while dust emission is emerging in the minispiral exterior.  

Apart from CS, which is most widespread in the center,  H$^{13}$CO$^+$, HC$_3$N, SiO, SO, C$_2$H, CH$_3$OH, $^{13}$CS and N$_2$H$^+$ are also detected. 
The bulk of the clumpy emission regions is at positive velocities and in a region confined by the minispiral northern arm (NA), bar, and the sources IRS 3 and 7. Although partly spatially overlapping with the radio recombination line (RRL) emission at same negative velocities, the relation to the minispiral remains unclear. A likely explanation is an infalling clump consisting of denser cloud cores embedded in diffuse gas. 
This central association (CA) of clouds shows three times higher CS/X (X: any other observed molecule) ratios than the circumnuclear disk (CND) suggesting a combination of higher excitation, by a temperature gradient and/or infrared (IR) pumping, and abundance enhancement due to UV and/or X-ray emission. Hence, we conclude that this CA is closer to the center than the CND is to the center. Moreover, we find molecular line emission at velocities up to 200 km~s$^{-1}$. 

Apart from the CA, we identified two intriguing regions in the CND. One region shows emission in all molecular species and higher energy levels
tested in this and previous observations and contains a methanol class I maser. The other region shows similar behavior of the line ratios such as the CA. 
Outside the CND, we find the traditionally quiescent gas tracer N$_2$H$^+$ coinciding with the largest IR dark clouds (IRDC) in the field. Methanol emission is found at and around previously detected methanol class I masers in the same region.
We propose to make these particular regions subject to further studies in the scope of hot core, cold core, and extreme photon and/or X-ray dominated region (PDR/XDR) chemistry and consequent star formation in the central few parsecs.
} 


\titlerunning{Molecular gas in the vicinity of Sgr~A* seen with ALMA}

\maketitle


\section{Introduction}
\label{sec:Intro}

The inner few parsecs of our Milky Way are a very fascinating and complicated region where we have the opportunity to study the interaction of a supermassive black hole (SMBH) with its environment on the smallest possible spatial scales. 
Our SMBH, \object{Sagittarius A*} (\object{Sgr~A*}), is located at the heart of a nuclear stellar cluster of massive stars \citep[e.g.,][]{Serabyn1985,Krabbe1991,Krabbe1995}, and these are both situated at the focal point of three infalling ionized gas streamers, called the \object{minispiral} or \object{Sgr A West} \citep[e.g.,][]{Lo1983,Roberts1993,Zhao2009}. 
The minisprial is not only visible in the thermal emission of ionized gas but also in the thermal emission of hot dust \citep[$T \sim 200$ K;][]{Gezari1985,Cotera1999,Viehmann2006,Lau2013}. 
Several stars from the nuclear stellar cluster are interacting with the streamers by either just passing through them and forming bowshocks or by being embedded into them. Furthermore, the extreme youth \citep[$\sim 6 \times 10^6$~yr; ][]{Paumard2006,Bartko2009} of many of these stars suggests they formed in situ and challenges our current understanding of star formation in such environments \citep[e.g.,][]{Nayakshin2007}. 
Two of the minispiral arms seem to be part of the inner ionized edge of the \object{circumnuclear disk} (\object{CND}) of molecular gas. The CND structure extends from 1.5 pc to about 2.5 pc \citep[e.g.,][]{Guesten1987,Jackson1993,Marr1993,YZ2001,Wright2001,Christopher2005} 
and is connected via molecular gas streamers, especially in the south and west, to larger cloud associations ultimately related to the giant molecular clouds (GMCs) - \object{M-0.02-0.07} ($50$ km~s$^{-1}$-cloud) and \object{M-0.13-0.08} ($20$ km~s$^{-1}$-cloud) in the east and south of the CND \citep{Coil1999,Coil2000,Liu2012}.

Earlier large-scale molecular studies of the Galactic center (GC), such as those mentioned above, suggested that the central cavity inside the CND is devoid of molecular gas and contains only neutral and ionized material.
However, indications for the opposite have been accumulating in the past in the form of CO, H$_2$CO, H$_3^+$, and OH absorption features \citep{Geballe1989,Goto2014,Karlsson2003,Karlsson2015} as well as high resolution maps of CN, high energy transitions of HCN and CS, and near-infrared (NIR) transitions of H$_2$ \citep[][]{Montero2009,Martin2012,Ciurlo2016}. In addition, \citet{Moultaka2004,Moultaka2005,Moultaka2015H2O,Moultaka2015CO} confirmed the ubiquitous presence of water and CO ice in the minispiral.
With the advent of the Atacama Large Millimeter/ submillimeter Array (ALMA) we have access to the spatial resolution and sensitivity required to study the molecular gas content in this region.
We report the serendipitous detection of line emission in the central parsec up to within 1$''$ ($\sim$ 0.04 pc) around Sgr~A* in projection.
The ALMA data used was obtained from the archive at an angular resolution of down to $\lesssim$0.5$''$.
At the distance to the GC of - here we adopt 8$\pm$0.3kpc \citep{schoedel2002,eisenhauer2003,horrobin2004,Ghez2008,gillessen2009a,gillessen2009b} - this corresponds to about 28 mpc or 5775 AU. 
This allows the so far highest angular and spatial view on the GC in the sub-mm domain. Among the highlights are the very first 340 GHz map of the minispiral, the very first and highly resolved detection of molecular emission in the immediate vicinity of the SMBH, and the highly resolved structures of CND features, especially of a region comprising a methanol class I maser closest to the SMBH.

The paper is organized as follows: In Section \ref{sec:archivedata} we give a detailed description of the observations including the calibration and line and continuum imaging. Results and data analysis are given in Section \ref{sec:results}. The global results on radio continuum and radio recombination lines (RRL) are given in Sections \ref{sec:cont} and \ref{sec:RRL}. The molecular gas emission in the outer and inner region (i.e., beyond and within the central 40'' of the GC), including results from a few special regions, are described in Section \ref{sec:outer} and \ref{sec:inner40}. The results and analysis of the molecular gas kinematics are presented in Section \ref{sec:kin}. In the discussion in Section \ref{sec:Diss} we concentrate on the continuum spectral index (Sect. \ref{sec:spex}), the electron temperature (Sect. \ref{sec:Te}), and the emission toward some of the most prominent stellar sources in the central stellar cluster (Sect. \ref{sec:IRS}). Molecular line ratios including molecular excitation and abundances are discussed in Section \ref{sec:ratio}.
In Section \ref{sec:molgas} we describe the general nature of the molecular gas toward the central region of the Milky Way. Here we concentrate on the IR dark clouds (IRDC) and methanol masers (Sect. \ref{sec:dark}), the high velocity clouds (Sect. \ref{sec:HVC}), and the origin of the molecular gas in the central region (Sect. \ref{sec:Loc}).
A summary is given in Section \ref{sec:sum}. We present additional images and tables in online appendices \ref{app:img} - \ref{sec:app-ratio}.

\begin{table*}[tb]
        \centering
        \caption{Observational parameters}
        \tabcolsep=0.09cm
        
        \begin{tabular*}{0.95\textwidth}{@{\extracolsep{\fill}} cccccccccc}
                \toprule
                band    & $\nu_\textrm{c}$ & $t_\textrm{total}$ &  $FOV$ &  $FOV10$ & $\theta_\textrm{beam-c}$ & $v_\textrm{ch}$ & $peak_\textrm{c}$     & $rms_\textrm{c}$        & $rms_{\textrm{ch}}$ \\ 
                & [GHz] & [min]  &  [$''$] &  [$''$] & [$''\times''$] & [km~s$^{-1}$] & \multicolumn{3}{c}{[mJy~beam$^{-1}$]} \\                                                                        
                \midrule                                                                              
                3       & 100   & 18 & 59.2 & 98.4 & 1.83 $\times$ 1.51 & 46.88 & 2.42    &       0.50 &  0.53 \\           
                6       & 250   & 18 & 22.8 & 37.9 & 0.72 $\times$ 0.57 & 18.75 & 4.13    &       0.21 &  1.00 \\            
                7       & 340   & 27 & 16.7 & 27.9 & 0.49  $\times$ 0.41  & 13.79 & 4.26 &       0.27 &  1.60 \\  
                \bottomrule
        \end{tabular*}
        \tablefoot{$\nu_\textrm{c}$ is the central frequency, $t_\textrm{total}$ the target integration time, $FOV$ the average field of view, $FOV10$ the average field of view at the 10\% primary beam power level ($\Delta FOV = \Delta FOV10 = 10\%$), $\theta_\textrm{beam-c}$ the continuum beam size, $v_\textrm{ch}$ the average channel resolution, $peak_\textrm{c}$ the peak (Sgr~A*) flux density of the continuum,  $rms_\textrm{c}$  the noise $\sigma$ of the continuum, and $rms_{\textrm{ch}}$ the average noise $\sigma$ of the channel.
        }
        \label{Obs-para}
\end{table*}

\begin{table*}[tb]
        \centering
        \caption{Overview of the properties of the lines detected with ALMA}
        \tabcolsep=0.10cm
        \begin{tabular*}{0.95\textwidth}{@{\extracolsep{\fill}} lcccccccc}
                \toprule
                &                                               &       \multicolumn{3}{c}{physical}                            &       \multicolumn{4}{c}{technical}                                           \\
                \cmidrule(l{1pt}r{1pt}){3-5}    \cmidrule(l{1pt}r{1pt}){6-9}
                molecule                & transition    &$\nu$  & $E_\textrm{u}$          & $n_\textrm{crit}(100~$K$)$           & $\theta_\textrm{beam}$& $\sigma_\textrm{ch-rms}$        & $n_\textrm{clip}$     & v$_\textrm{int}$                                              \\      
                &                                       & [GHz] &       [K]              &       [cm$^{-3}$]                     &  [$''\times''$]       & [mJy~beam$^{-1}$]             &                         & [km~s$^{-1}$]                                                 \\      
                \midrule                                                                                                                                                                                                                         
                $^{13}$CS               & $ 2                   $--$ 1                  $ & 92.494        & 6.66   & 3.3$\; \times \; 10^{5}$             & 1.93 $\times$ 1.57    & 0.52  & 3     & -130 - 130                                            \\      
                N$_2$H$^+$              & $ 1                   $--$ 0                  $ & 93.174        & 4.47   & 2.0$\; \times \; 10^{5}$             & 1.93 $\times$ 1.57    & 0.53  & 4     & -150 - 100                                            \\      
                H51$\beta$              &                                                 & 93.607       &        &                                      & 1.93 $\times$ 1.57    & 0.54  & 3     & -370 - 340                                            \\      
                CH$_3$OH-A              & $ 8_0                 $--$ 7_1                $ & 95.169        & 83.50  & 7.9$\; \times \; 10^{5}$             & 1.89 $\times$ 1.55    & 0.52  & 3     & -120 - 130                                            \\      
                H49$\beta$              &                                                 & 105.302      &        &                                      & 1.72 $\times$ 1.44    & 0.54  & 3     & -350 - 320                                            \\      
                H39$\alpha$             &                                                 & 106.737      &        &                                      & 1.75 $\times$ 1.48    & 0.62  & 3     & -440 - 400                                            \\      
                CS                      & $ 5                   $--$ 4                  $ & 244.936       & 35.30  & 5.4$\; \times \; 10^{6}$             & 0.76 $\times$ 0.59    & 0.91  & 5     & -150 - 200                                            \\      
                HC$_3$N         & $ 27                  $--$ 26                 $ & 245.606       & 165.03 &                                      & 0.76 $\times$ 0.59    & 0.87  & 3     & -70 - 100                                             \\      
                H36$\beta$              &                                                 & 260.033      &        &                                      & 0.70 $\times$ 0.55    & 1.10  & 3     & -330 - 340                                            \\      
                H$^{13}$CO$^+$          & $ 3                   $--$ 2                  $ & 260.255       & 24.98  & 3.3$\; \times \; 10^{6}$             & 0.70 $\times$ 0.55    & 1.10  & 3     & -80 - 110                                             \\      
                SiO                     & $ 6                   $--$ 5                  $ & 260.518       & 43.76  & 8.3$\; \times \; 10^{6}$             & 0.70 $\times$ 0.55    & 0.90  & 4     & -150 - 210                                            \\      
                SO $^3\Sigma$           & $ 6_{7}               $--$ 5_{6}               $ & 261.844     & 47.60  & 3.7$\; \times \; 10^{6}$             & 0.70 $\times$ 0.56      & 1.00  & 4     & -80 - 100                                             \\                      C$_2$H                  & $ 3_{3.5, \, 4} $--$ 2_{2.5, \, 3}      $ & 262.004     & 25.15  & 6.3$\; \times \; 10^{6}$               &                       &       &       &                                                         \\      
                C$_2$H                  & $ 3_{3.5, \, 3}       $--$ 2_{2.5, \, 2}   $ & 262.006     & 25.15  & 6.5$\; \times \; 10^{6}$             &                         &       &       &                                                       \\      
                C$_2$H                  & $ 3_{2.5, \, 3}       $--$ 2_{1.5, \, 2}   $ & 262.065     & 25.16  & 6.7$\; \times \; 10^{6}$             & 0.70 $\times$ 0.56      & 1.00  & 4     & -190 - 80                                             \\      
                C$_2$H                  & $ 3_{2.5, \, 2}       $--$ 2_{1.5, \, 1}   $ & 262.067     & 25.16  & 6.9$\; \times \; 10^{6}$             &                         &       &       &                                                       \\                  
                H33$\beta$              &                                                 & 335.207      &        &                                      & 0.50 $\times$ 0.42    & 1.28  & 5     & -390 - 100                                            \\      
                CH$_3$OH-E              & $ 7_{-1}              $--$ 6_{-1}             $ & 338.345       & 70.60  & 1.3$\; \times \; 10^{6}$             & 0.50 $\times$ 0.42    & 1.28  & 3     & -80 - 100                                             \\      
                CH$_3$OH-A              & $ 7_{0}               $--$ 6_{0}              $ & 338.409       & 65.00  & 1.4$\; \times \; 10^{6}$             &                        &       &       &                                                       \\      
                SiO                     & $ 8                   $--$ 7                  $ & 347.331       & 75.02  & 2.0$\; \times \; 10^{7}$             & 0.48 $\times$ 0.41    & 1.90  & 4     & -150 - 200                                            \\      
                \bottomrule                                                                                                                               
                
        \end{tabular*}
        \tablefoot{Listed are the frequencies $\nu$; upper state energies $E_\textrm{u}$; critical densities $n_\textrm{crit}$ in the two-level approximation for given kinetic temperature based on data from the Leiden Atomic and Molecular Database 
                \citep[LAMDA;][http://home.strw.leidenuniv.nl/$\sim$moldata/]{Schoier2005}; beam sizes $\theta_\textrm{beam}$, channel noise $\sigma_\textrm{ch-rms}$, and noise clipping level from $n_\textrm{clip}\sigma_\textrm{ch-rms}$ per line cube; 
                velocity range v$_\textrm{int}$ over which the integrated flux maps were obained.
        }
        \label{ALMA-lines}
\end{table*}

\section{ALMA observations and data reduction}
\label{sec:archivedata}

The GC is one of the most complex regions in the Milky Way. A detailed analysis requires high sensitivity and angular resolution. Both can now be obtained using ALMA. For the presented description and analysis we used GC data obtained in a monitoring campaign,
under the project 2011.0.00887.S (PI: Heino Falcke), on 18 May 2012 with ALMA in Band 3, 6, and 7 alternatingly. A summary of the observational properties per band is given in Table \ref{Obs-para}.

\subsection{Observation}
\label{sec:obs}

For the observations in the three bands the corresponding central frequencies were tuned to $\sim$ 100, 250, and 340 GHz, respectively.
Using the Time Division Mode, the correlator provides 128 channels with a width of 15.625 MHz resulting in a total bandwidth of the spectral window of 2 GHz. This translates to velocity channel widths of $\sim$ 50, 20, and 15 km~s$^{-1}$, respectively.
Each band contains four spectral windows yielding a total effective bandwidth of 8 GHz.
The single pointing on Sgr~A* (17\fh45\fm40\fs040, -29\fdg00\farcm28\farcs118, J2000) covers a field of view (FOV) of $\sim$ 60, 23, and 17$''$, respectively. 
During the observations, the array was in its most extended configuration (in cycle 0) with baselines ranging from 36 to 400 m. This results in angular resolutions of about 1.5$''$, 0.7$''$, and 0.5$''$, respectively.

The total observation time is 7.5 h and the integration time per target and band is on the order of 20 min. Using 19 antennas and distributing the scans per band equally over the whole observation time yields a superb $uv$ coverage (see Fig. \ref{psf-uv}). 
With this configuration, the instrument is sensitive in the different FOVs to angular scales $< 10''$, $< 4''$ and $< 3''$, respectively.

\subsection{Data calibration}
\label{sec:cali}

The basic reduction and calibration was performed with the Common Astronomy Software Application \citep[CASAv3.4;][]{McMullin2007} using the reduction scripts provided with the archive data. The uncertainty in the flux calibration is less than 17\%, 9\%, and 10\%  in band 3, 6, and 7, respectively.
All further reduction, imaging, and analysis were carried out in CASAv.4.3.
We inspected the visibilities and flagged residual noisy data.
The flux variability of Sgr~A* produces strong side lobes in the time-integrated image such that the underlying fainter extended emission, which is a main focus of our investigation, is not visible.
Hence we performed an elaborate and careful self-calibration not only on phase but also on amplitude, despite the variable nature of the source.
Phase-only self-calibration enhances the dynamic range (DR) only by a factor of 1.7, 6.4, and 6.2 at 100, 250, and 340 GHz, respectively. Combining it with a subsequent amplitude and eventually an amplitude-phase self-calibration yields a total improvement of the DR by a factor of 11.4, 31.9, and 57.7, respectively. 
Therefore, by trading astrometric and time resolution information of the data, we were able to increase the DR dramatically, i.e., to 4500, 7600, and 11200, respectively, resulting in an overall data quality sufficient for our purpose.

The subtraction of the continuum from the $uv$ data gave access to the emission line information. A description of the procedures for the extraction of the line and continuum maps is given in the following section.

\subsection{Line and continuum imaging}
\label{sec:imaging}

Apart from the continuum emission maps created using the line-free channels, we indeed identified a large number of lines above the 3 - 5$\sigma$ noise level in the different bands (see Table \ref{ALMA-lines}).
Here we describe the general procedure we followed for the creation of line and continuum images. We also outline some general properties of the different spacial and spectral features in the data cubes as far as they are relevant for the calibration of image formation.
We weighted the line and continuum data naturally and restored the images with 768 $\times$ 768 pixels and a pixel size of 0.15, 0.075, and 0.05 $''$, respectively. In this way the images cover twice the FOV. 
The channel resolution corresponds to the natural channel size of the corresponding spectral window. 
The noise rms is 0.5, 0.2, and 0.3 mJy~beam$^{-1}$ for the continuum, and 0.5, 1.0, and 1.6 mJy~beam$^{-1}$ for a channel.

All line (cube channels) and continuum data were clipped at $\geq 3 \sigma$ to avoid the inclusion of noise and dilution effects.
From the clipped line cubes, we created the moment maps, i.e., integrated flux (0th order) and velocity field (1st order), for the velocity ranges given in Table \ref{ALMA-lines} using the CASA task \texttt{immoments}.

Subsequently, all moment 0 maps, i.e., integrated flux maps and the continuum maps are corrected for the primary beam attenuation out to a primary beam level of 10\%. In the following, we refer to the field given by the 10$\%$ power full width as FOV10 (see Table \ref{Obs-para}). The accuracy of the primary beam corrected fluxes suffer with distance from the pointing center owing to the decreasing sensitivity of the primary beam. Nevertheless, many bright features are detected beyond the FOV making an extension of the map to the FOV10 preferable.

The minispiral, as the outstanding extended emission component in the region, is well detected in the 100, 250, and 340 GHz continuum (Fig. \ref{cont-full}) and in the H$\alpha$ and $\beta$ RRL emission in band 3 (Fig. \ref{recomb}). 

The H49$\beta$ and H51$\beta$ lines are partly blended by faint H61$\delta$ and H58$\gamma$ line emission, respectively, which are offset by about -300 and -500 km~s$^{-1}$ compared to H49$\beta$ and H51$\beta$, respectively.

In band 6, the H$\beta$ line is fainter and rather tracing compact features. Furthermore, it overlaps with the H$^{13}$CO$^+$ line emission in frequency space such that emission from both, H$\beta$ and H$^{13}$CO$^+$ lines, appear in the integrated flux maps. The H$\beta$ line in band 7 is very weak and only reliably detected in the IRS 2 and 13.\\
The hydrogen RRL emission is visible from $-400$ to 400 km~s$^{-1}$.
The molecular emission lines in band 3 cover the whole CND velocity range of -150 to 150 km~s$^{-1}$.
Because of the smaller FOV10, molecular emission lines in band 6 and 7 typically appear in channel ranges from -80 (-100) km~s$^{-1}$ to 80 (100) km~s$^{-1}$. Only CS extends to higher velocities up to 200 km~s$^{-1}$ and SiO covers the largest range from -160 to 200 km~s$^{-1}$. 
The resulting images are shown in Figure \ref{mols-full1}. The image cubes of N$_2$H$^+$ and C$_2$H suffer from strong side lobe artefacts in some channels, which are visible as ring structure around the position of Sgr~A* (compare Fig. \ref{psf-uv}). Therefore, for N$_2$H$^+$ we doubt the ring-shaped structures in the inner 12$''$ and the emission of C$_2$H needs to be treated with caution since obvious real emission is blended with the side lobes. These artefacts and the emission on Sgr~A* in several band 6 and 7 lines are likely to be residuals of an imperfect and insufficient continuum subtraction.

\subsection{Obtaining spectral and spatial properties of the sources}
\label{sec:specrat-how}

We catalogued all clumps with prominent molecular emission with their positions, source sizes, fluxes, and spectral features in all the molecular emission lines (see Tables \ref{mol-src} and \ref{mol-src-vel-1} - \ref{mol-src-vel-3}). In the continuum images, we looked for counterparts of the sources detected in the infrared (IR) regime \citep[][finding charts therein]{Muzic2007,Viehmann2006} and cm emission \citep[][finding charts therein]{Zhao2009} to obtain their positions, source sizes, and fluxes (see Table \ref{cont-src}). For a selection of IR sources (IRS) in the inner 10$''$, we extracted the spectral properties (see Fig. \ref{mol-IRS-vel}). Positions, sizes, and contained fluxes were derived with the CASAv4.3 task \texttt{imfit} fitting a 2D Gaussian to the naturally weighted, untapered (integrated) image. Spectra were obtained from a beamsized aperture as given in Table \ref{ALMA-lines} centered on the average position of each clump based on all emission lines and were fitted with the MPFIT module in Python\footnote{https://code.google.com/archive/p/astrolibpy/downloads}. In this context, one has to bear in mind that the emission lines often only cover 2 - 3 channels so that the spectral line fitting can be challenging and the actual uncertainty in the fit is larger than calculated. In such cases and in cases, where the fit fails despite a 3$\sigma$ detection of a line, we estimate the uncertainties and/or the parameters by visual inspection. \\
We calculated the line luminosity ratios for selected molecules based on the channel fluxes (Table \ref{mol-rat}) as well as continuum spectral index maps and integrated line flux to continuum ratio maps. The results are discussed in section \ref{sec:Diss}.
For ratios between two maps, we $uv$ tapered and restored the images and image cubes with the same beam size per ratio and clipped them at the flux levels given in Table \ref{ALMA-lines} and at 3$\sigma$ for the continuum maps.
Ideally, the $uv$ planes should be clipped to the same baseline ranges for ratios between different ALMA bands. However, this reduced the image quality so significantly, i.e., distortions and flux losses of up to 50 \% even when clipping only large baselines, that we refrained from applying this procedure.
Therefore, ratios between a higher and a lower band can only be considered as lower limit owing to the missing flux resolved out in the higher band data.
In contrast to this, ratios within the same band is very accurate in terms of the similarity of the $uv$-plane coverage and flux calibration. The only bias is introduced by the channel binning on the same grid because of unreliable interpolation in $uv$-plane regions with scarce sampling and/or low signal-to-noise ratio (S/N), and the cleaning procedure. These effects are difficult to constrain. Since the S/N per channel is less than for the total band, we treat the molecular line ratios per channel as rough values and omit the uncertainties. For tracking trends in the molecular line ratios by region this procedure is sufficient.

\begin{figure*}[tb]
        \centering $
        \begin{array}{cc}
        \includegraphics[trim = 32mm 3mm 34mm 3mm, clip, width=0.47\textwidth]{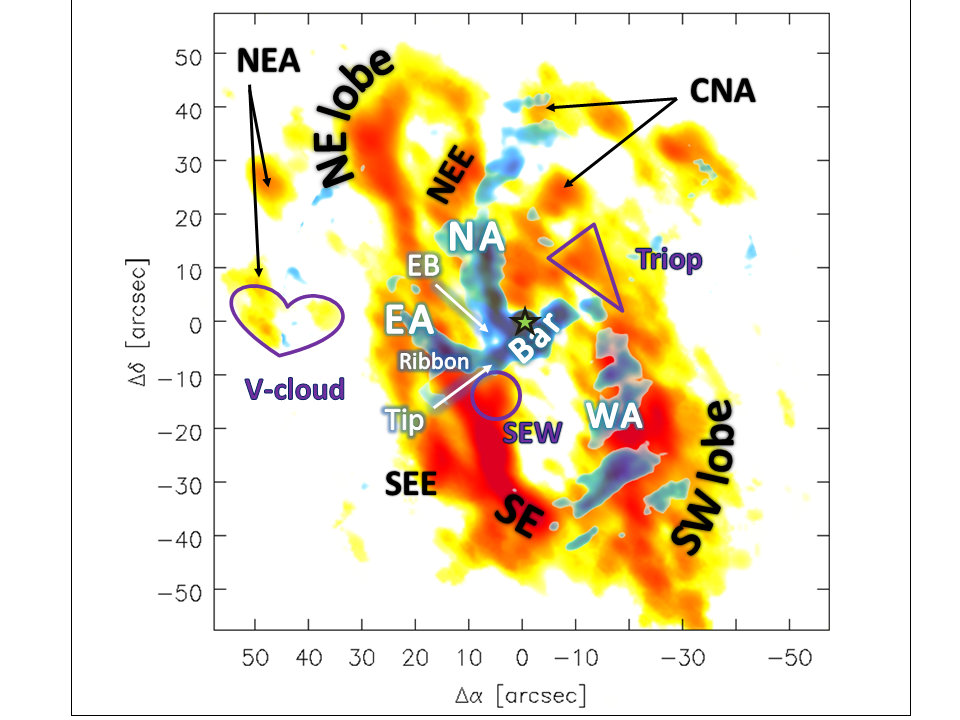}     &
        \includegraphics[trim = 4mm 3mm 7mm 32mm, clip, width=0.46\textwidth]{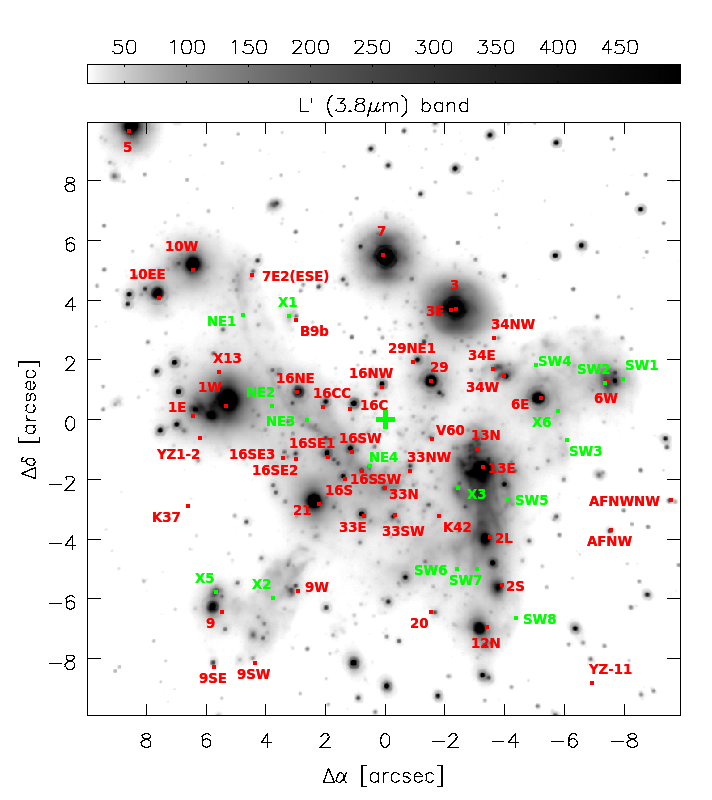}        \\        
        \end{array} $        
        \caption{Left: Overview of the structures in the inner 4.8 pc of the GC. Shown in orange: CND \citep[CN(2--1),][]{Martin2012}; blue: minispiral (100 GHz continuum, this work); and green star: Sgr~A*. 
                The abbreviations are as follows: NE lobe for northeastern lobe, SW lobe for southwestern lobe, SE for southern extension, NEE for northeastern extension, CNA for CND-northern Arm, NEA for       
                northeastern arm, NA for northern arm, WA for western arc, EA for eastern arm, and EB for eastern bridge. The new defined regions are  southern extension west, (SEW) southern extension east (SEE), \textit{Triop}, and V-cloud.
          Right: Overview of the stars and filaments \citet{Paumard2006,Viehmann2006,Muzic2007} mentioned in this work, 
                demonstrated for the inner 20$''$ (0.8 pc) on a Very Large Telescope (VLT/NACO) NIR L' (3.8 $\mu$m) emission map (Sabha, private communication, here: arbitrary units).
        }
        \label{CND-sketch}
\end{figure*}

\section{Results and analysis}
\label{sec:results}

In the following we present the main results of the ALMA data including both continuum and lines. The complexity of the region results in a number of prominent features or objects distributed throughout the region.

In  Fig. \ref{CND-sketch}, we provide a finding chart for the names of the features used in this paper.
Following the CND nomenclature of \citet{Christopher2005} and \citet{Martin2012},
the most dominant regions of the CND are the northeastern (NE) and southwestern (SW) lobe. The southeast side of the ring is called southern extension (SE). The northeastern extension (NEE) is coincident with the eastern edge of the minispiral's northern arm (NA), while the CND northern arm (CNA) is on the west side of the NA. East of the CND is the northeastern arm (NEA) located.

At this occasion, it is useful to define further regions that are discussed in this paper: SEW is a small clump at the northwestern end of the SE; SEE is the large cloud extending east of the middle of the SE; the \textit{triop} is located at the southern tip of the CNA; and the eastern edge of the V cloud coincides with the southern tip of the NEA.

In addition, we use the minispiral nomenclature of \citet{Paumard2004} to describe the features in the ionized gas emission: The brightest part of the minispiral is within the so-called bar extending from west to the south of the SMBH. The NA extends north parallel to the NEE, the western arc (WA) is parallel to the SW CND lobe and the eastern arm (EA) streams toward the center from the east. The latter can be subdivided into the ribbon comprising most of the EA, the tip, a luminous section between the ribbon and the bar, and the eastern bridge (EB), which is visible as vertical connection between the NA and the ribbon.

For the description of the innermost region around Sgr~A* it is helpful to use the IRS sources as reference points (Fig. \ref{CND-sketch}).

\begin{figure*}[tb]
        \centering $
        \begin{array}{ccc}
        \includegraphics[trim = 4mm 0mm 6mm 0mm, clip, width=0.31\textwidth]{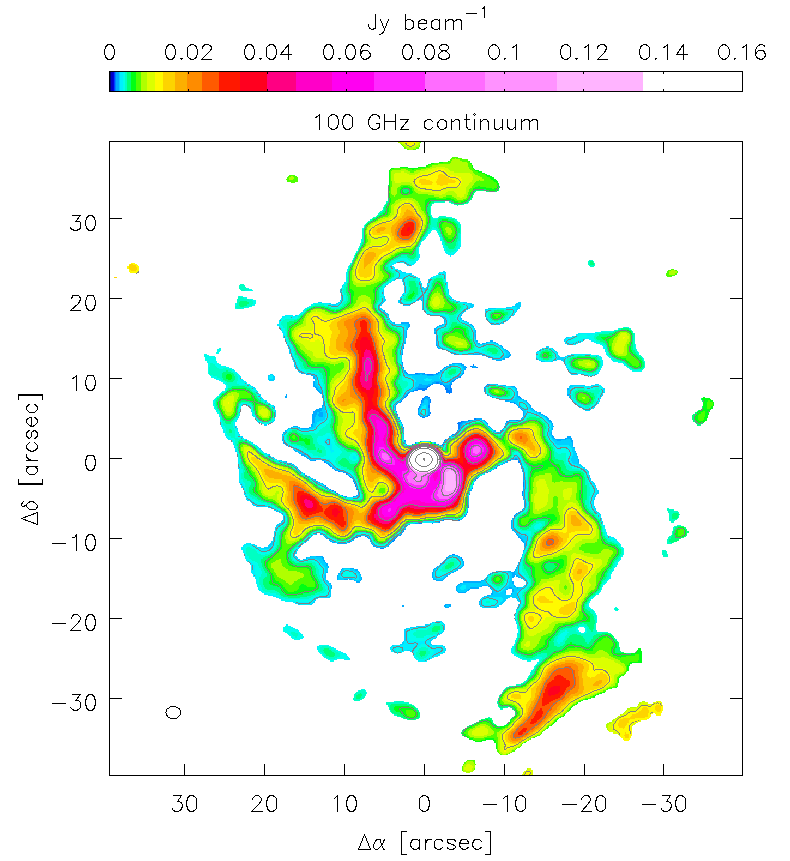}&
        \includegraphics[trim = 5mm 0mm 6mm 0mm, clip, width=0.31\textwidth]{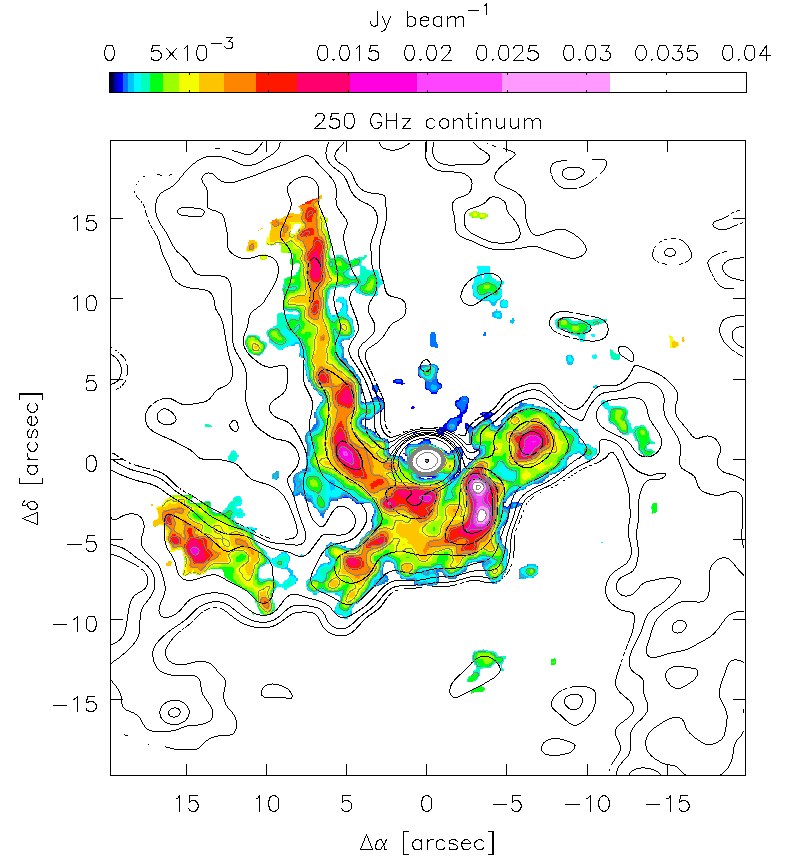}&
        \includegraphics[trim = 7mm 0mm 4mm 0mm, clip, width=0.31\textwidth]{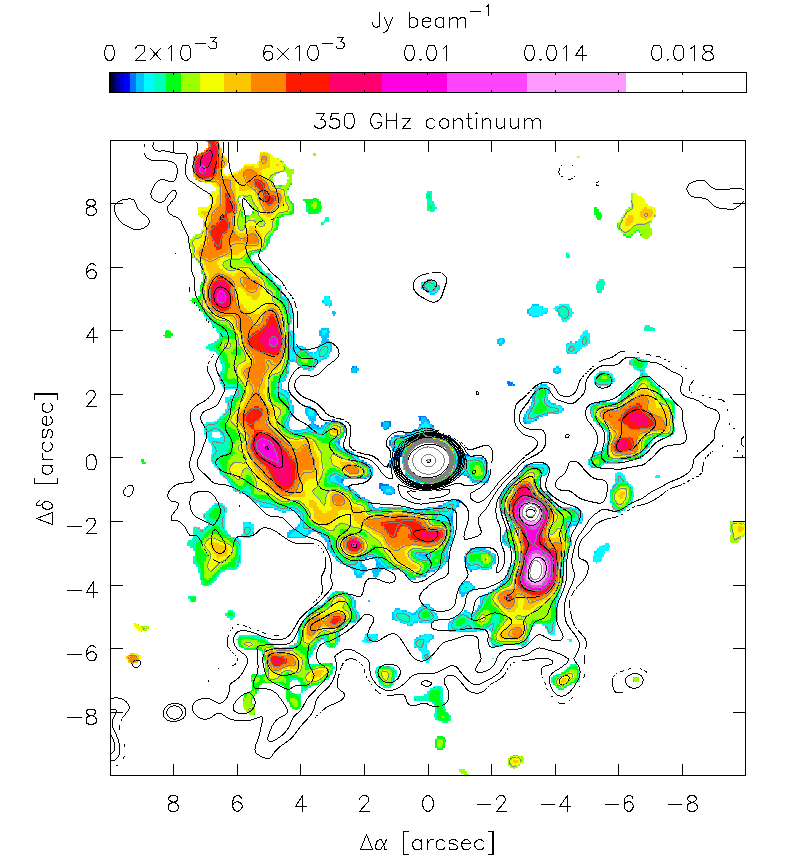}\\                                                                        
        \end{array} $                                                            \caption{Continuum emission images of the inner $\lesssim $ 3 pc.
        	From left to right: central 80$''$ at 100 GHz, central 40$''$ at 250 GHz with 100 GHz contours of [6, 12, 24, 48, 96, 144, 192, 384, 1920, 4800] $\times$ $\sigma$ (= 0.5 mJy~beam$^{-1}$),               and central 20$''$ at 340 GHz with 250 GHz contours of [6, 12, 24, 36, 48, 72, 96, 144, 192, 1920, 19200] $\times$ $\sigma$ (= 0.21 mJy~beam$^{-1}$).
                The beam sizes are 1.8$''$ $\times$ 1.5$''$, 0.7$''$ $\times$ 0.6$''$, and 0.5$''$ $\times$ 0.4$''$, respectively.
        }
        \label{cont-full}
\end{figure*}

\subsection{Continuum}
\label{sec:cont}

Figure \ref{cont-full} shows the continuum emission at 100, 250, and 340 GHz.
Overall, the minispiral is well detected in all three continuum maps.

The 100 GHz continuum emission traces well the arms of the minispiral in their overall extent and prominent clumps with them. 
The resolution of the 100 GHz ALMA map is comparable to the data of, for example, \citet{Lo1983} at 6 cm (Very Large Array; VLA). 
The brightest emission after Sgr~A* comes from the IRS 13/2L complex and the ridge in the IRS 16/21/33 region (see Fig.\ref{CND-sketch} for orientation). Furthermore, the IRS 1W, 5, 6, 7, 8, and 9 regions are discernible.
Nevertheless, we find some differences in the morphology with respect to the 6 cm map; 
there is a southeast extension at the eastern end of the EA to a cloud slightly to the south ($\Delta \alpha \sim 17'', \Delta \delta \sim -15''$). 
While this faint cloud (peak flux $\sim$ 10 mJy~beam) is not visible in the low frequency ($\leq$ 20 GHz) observations, except from a tentatively detected extension \citep[e.g.,][]{Lo1983,Roberts1993,Zhao2009}, it seems to be visible in the 1.3 mm map of \citet{Kunneriath2012a} and the extension toward it is clearly seen in the mid-infrared (MIR). The detection in both emission regimes is suggestive of a dusty nature. 
This clump is located between the northern tip of the CND SE and a CND clump east of the eastern end of the EA.

At 250 GHz the major filaments start to resolve, i.e., in the NA: the filaments X1, NE 1, 3, and 4; in the bar:  SW 6, 7, 8 (south of IRS 2L), 3 (west of IRS 13); and in the tip: X2 \citep[filament nomenclature:][]{Muzic2007}.
The dust features X5 (tip region, north of IRS 9), SW 2, 4, and X6 (IRS 6 region) are located at the edges of the radio continuum emission. Moreover, the IRS 13 cluster and IRS 2L are separated, and the ridge in the IRS 16/21/33 region displays a substructure within which the filament NE4 and IRS 21 are discernible. In the NA, IRS 5 (bowshock to the northeast) and several sources east (most eastern: IRS 17) of the NA emerge (IRS 5S, 5SE1, 5SE2, 17); IRS 10W, 16NE, 16C, and VISIR 60 (K 22) also become visible \citep{Viehmann2006,Zhao2009}-- the latter two are both $<2''$ away from Sgr~A*-- and K 42 \citep{Zhao2009} is detected at the center of the minicavity. 
In addition, we find a positional shift of IRS 7 from the 100 GHz to the 250 GHz map of < 0.5$''$ to the south(west), which is either due to the lower resolution at 100 GHz unable to separate the star or bowshock head from the tail \citep{Serabyn1991,YZ1991,Zhao2009} or to different excitation conditions in the head and the tail.

\begin{figure*}[tb]
        \centering $
        \begin{array}{ccc}
        \includegraphics[trim = 5mm 0mm 7mm 0mm, clip, width=0.31\textwidth]{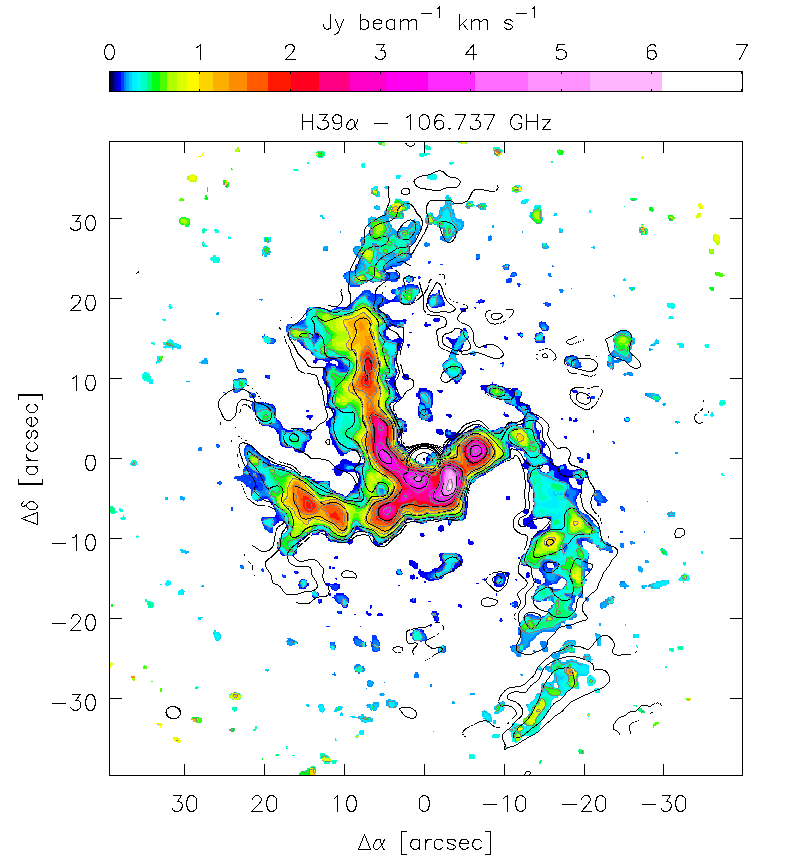}&
        \includegraphics[trim = 5mm 0mm 7mm 0mm, clip, width=0.31\textwidth]{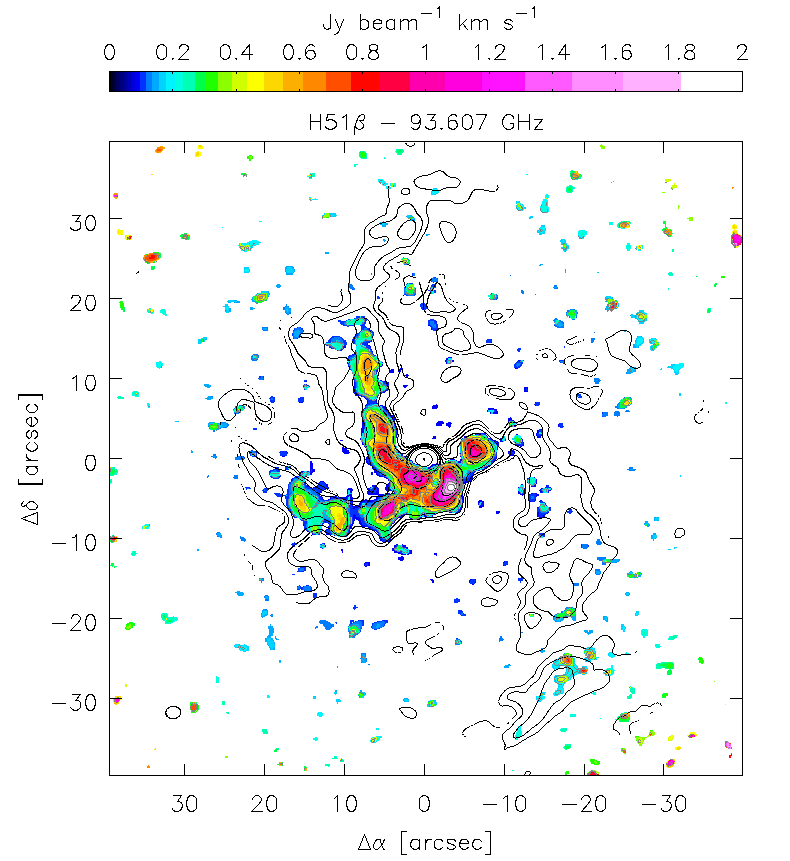}&
        \includegraphics[trim = 5mm 0mm 7mm 0mm, clip, width=0.31\textwidth]{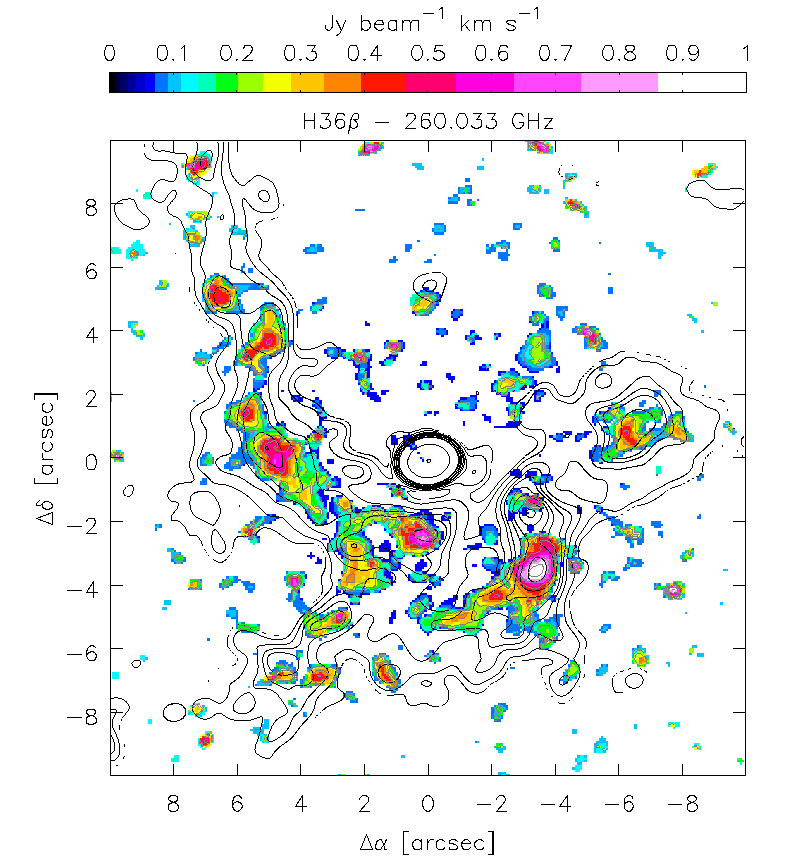}\\                                                                      
        \end{array} $                                                            \caption{Recombination line (RRL) emission images of the inner $\lesssim $ 3 pc. From left to right: H39$\alpha$ and H51$\beta$ in the central 80$''$, 
                both with the 100 GHz continuum contours as in Fig. \ref{cont-full},
                and H36$\beta$ overlayed with the 250 GHz contours as in Fig. \ref{cont-full}. 
                H49$\beta$ and H33$\beta$ can be seen in Fig. \ref{recomb-rest}.
                The beam sizes are 1.9$''$ $\times$ 1.6$''$, 1.8$''$ $\times$ 1.4$''$, and 0.7$''$ $\times$ 0.6$''$, respectively.
        }
        \label{recomb}
\end{figure*}

While the 250 GHz continuum traces the extended gas in the minispiral, the high resolution image at 340 GHz perfectly outlines the compact features as seen in the MIR and in the radio \citep[compare][]{Viehmann2006,Zhao2009}. Additional filaments are visible at 340 GHz are X3 and SW5, which are southeast and southwest of IRS 13, respectively. In addition, the IRS 13 cluster begins to separate into the 13N and 13E clusters at this resolution.

\subsection{Radio recombination lines}
\label{sec:RRL}

Figure \ref{recomb} shows the emission of the brightest RRLs in this data set. Further RRL images can be seen in Fig. \ref{recomb-rest}.
The bright emission in the H39$\alpha$ line very closely mimics the distribution of the 100 GHz continuum. 
The low sensitivity of the primary beam only becomes notable toward the edge of FOV10; neither IRS 8 ($\Delta\alpha \sim 3''$, $\Delta\delta \sim 28''$) nor the southern tip of the WA are properly detected. The clump south of the EA end, which is even fainter than the diffuse WA emission at 100 GHz, is not detected. 

In the following we compare the general brightness distribution in the RRL emission to that in the radio continuum and to the results of corresponding radio cm-wavelength observations.
When comparing our data to the H92$\alpha$ and \ion{Ne}{II} data of \citet{Zhao2009} and \citet{Irons2012}, respectively, we spot some differences. 
In the integrated H92$\alpha$ emission image shown in \citet{Zhao2009}, the region southeast of Sgr~A*, i.e., outlined by IRS 21, 16SW, 33, is fainter close to Sgr~A* and peaks south of IRS 33 (see Fig.\ref{CND-sketch} for orientation). This trend is opposite from what is seen in our data. Furthermore, the middle part of the EA is very weak and the tip is weaker than the eastern end of EA. 
In the \ion{Ne}{II} data of \citet{Irons2012}, the region between IRS 13, 21, 33 is fainter than IRS 1W and 2L, which might simply be related to a different excitation behavior of the Ne ion, but the H30$\alpha$ data of \citet{Zhao2010} looks similar. 
However, our RRL maps might be affected by the large velocity bins given by the observational setup, i.e., $\sim$ 50 km~s$^{-1}$. This is implied by the slight discrepancies between the H49$\beta$ and H51$\beta$ maps. The line transitions are close enough to each other so that opacity changes - if significant at all at 100 GHz - or deviations from local thermal equilibrium (LTE) cannot explain the differences. Therefore, an instrumental bias cannot be ruled out. 
The observations of \citet{Zhao2009,Zhao2010} and \citet{Irons2012} have been conducted at a spatial resolution similar to our data, but with a much higher spectral resolution of $\sim$ 15 km~s$^{-1}$ and $\sim$ 4 km~s$^{-1}$, respectively, suggesting the minispiral clumps are well traced. On the other hand, the region south(west) of Sgr~A* is the second brightest in all of the three RRL at 100 GHz. 
The difference in the velocity ranges for integration could also have an impact on the RRL emission distributions from our data set and theirs.
While we integrate the H39$\alpha$ emission from -400 to 400 km~s$^{-1}$, \citet{Irons2012} use a range of -339 - 299 km~s$^{-1}$ and \citet{Zhao2010} a range of -360 - 345 km~s$^{-1}$ even though faint emission extends to even more extreme negative velocities. This is supported by \ion{Br}{$\gamma$}, \ion{Fe}{III}, and \ion{He}{I} observations discussed in \citet{Steiner2013} and can also be seen in the \ion{Ne}{II} data cube of \citet[][link on online journal page]{Irons2012}, both tracing the ionized gas up to -380 km~s$^{-1}$.
The H36$\beta$ emission outlines the high S/N regions along the minispiral filaments and the H33$\beta$ is too faint to be reliable, except maybe toward the IRS 2L and 13 region.

\begin{figure*}[tb]
        \centering $
        \begin{array}{ccc}
        \includegraphics[trim = 5mm 0mm 7mm 0mm, clip, width=0.31\textwidth]{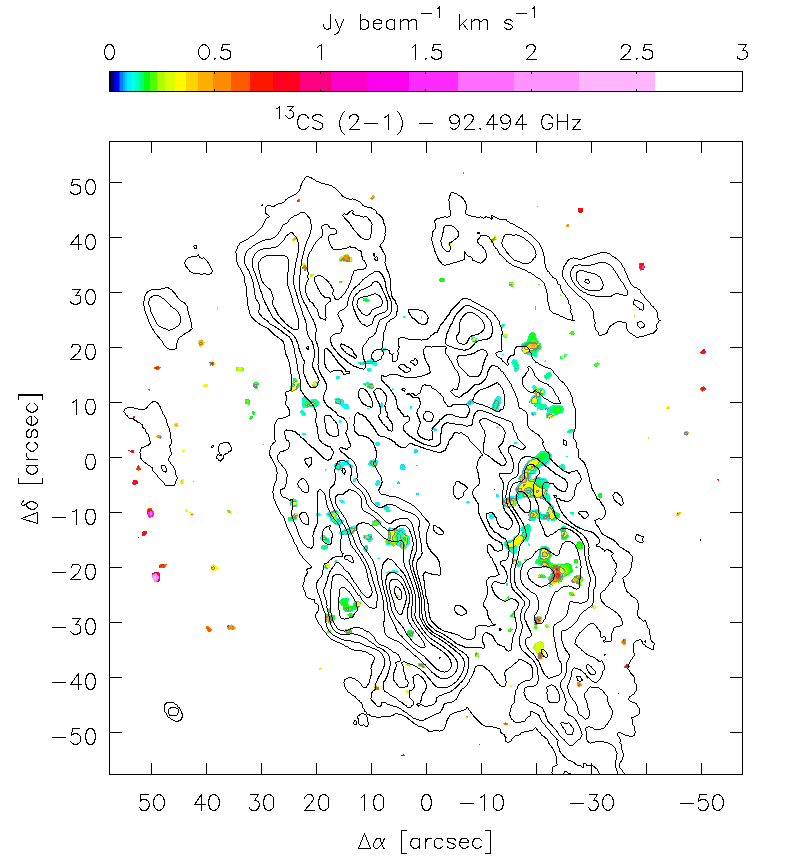}&
        \includegraphics[trim = 5mm 0mm 7mm 0mm, clip, width=0.31\textwidth]{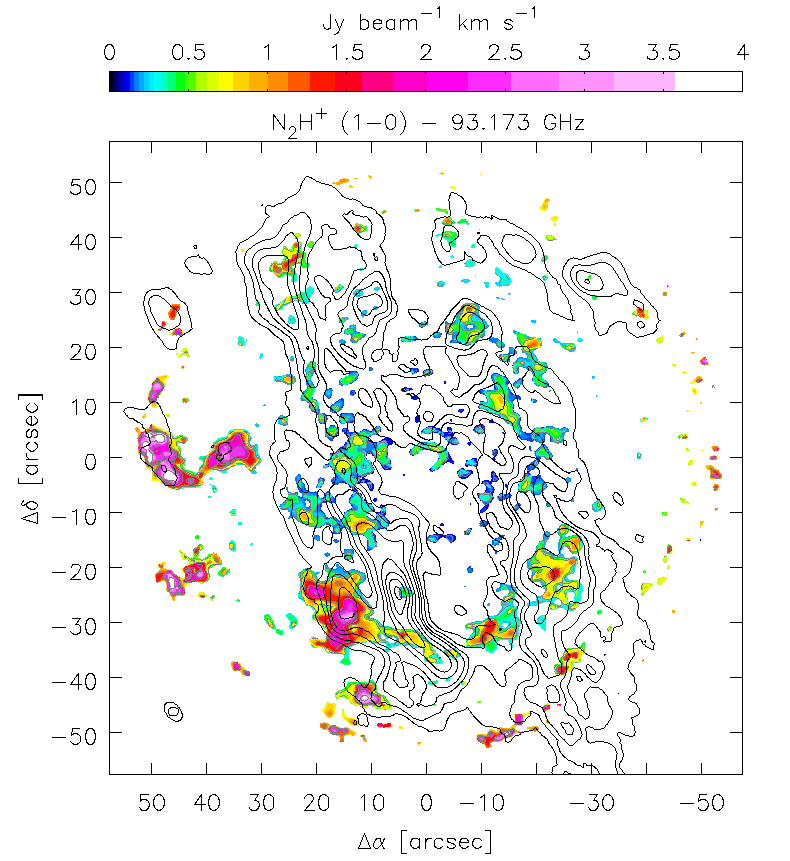}&
        \includegraphics[trim = 5mm 0mm 7mm 0mm, clip, width=0.31\textwidth]{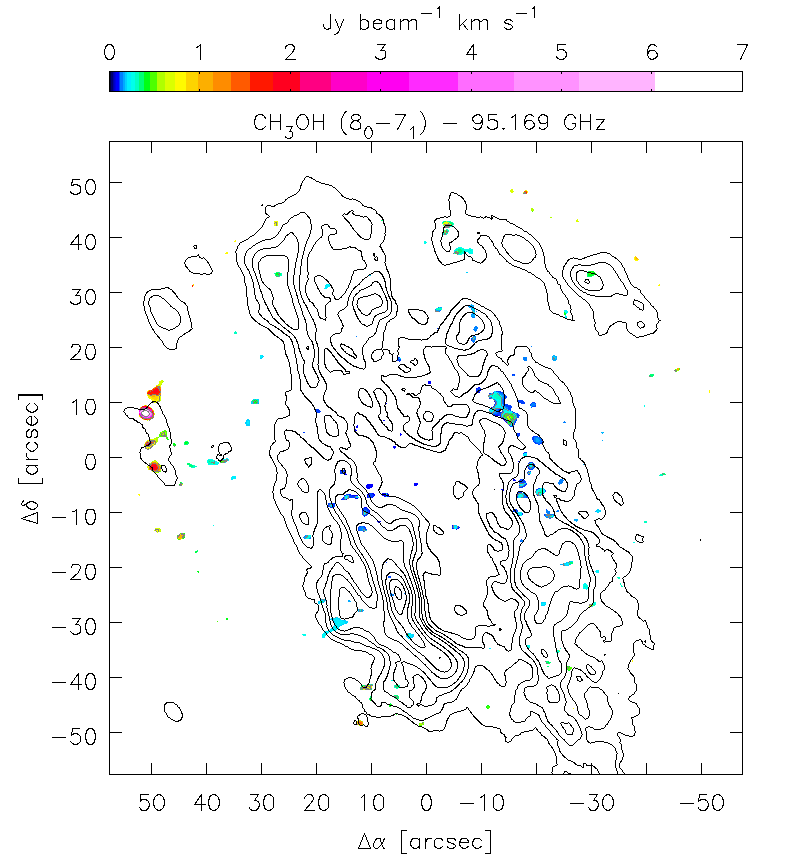}\\
        \end{array} $                                                            \caption{Molecular line emission images of the inner 120$''$ (4.8 pc). From left to right: $^{13}$CS(2--1), N$_2$H$^+$(1--0), and CH$_3$OH(8--7) at a resolution of 1.9$''$ $\times$ 1.6$''$ (see Table \ref{ALMA-lines}). 
                The contours at the levels [8, 16, 24, 32, 48, 64, 80, 96, 112] $\times$ $\sigma$ (= 1.9 Jy~beam$^{-1}$~km~s$^{-1}$)
                show CN(2--1) emission of the CND at a resolution of 4.0$''$ $\times$ 2.6$''$ for orientation \citep[compare Fig. \ref{CND-sketch}; CN data: ][]{Martin2012}.
        }
        \label{outermol}
\end{figure*}

\subsection{Molecular gas in the outer regions}
\label{sec:outer}

The large FOV of the band 3 data enables us to look at the full CND and regions outside of it. 
With outer regions we refer to a distance $r > 40''$ from Sgr~A*.
The emission of $^{13}$CS, N$_2$H$^+$, and CH$_3$OH(8--7) is shown in Fig. \ref{outermol} and compared to the CN data of \citet{Martin2012}.
The regions of $^{13}$CS and CH$_3$OH(8--7) line emission are less extended than in the N$_2$H$^+$ line.
The $^{13}$CS line emission is faint and restricted to the west side of the CND. It is also prominent in the SEW and SEE regions (see Fig.\ref{CND-sketch} for orientation).
N$_2$H$^+$ is found within the CND and partly matches local peaks in CN emission, but the strongest emission resides in regions outside of it, i.e., in the SEE and in the V cloud.
This is consistent with the N$_2$H$^+$ and CH$_3$OH(2--1) distribution obtained from lower resolution ($\theta_\mathrm{beam} \sim 8''$) observations by \citet{Moser2014,Moserinprep}. 
Moreover, the overall distribution resembles very much that of H$_2$CO in \citet{Martin2012}. 
Apart from the molecules just mentioned, the V cloud has never been detected that clearly in any other molecule in the past \citep[cf.][]{Christopher2005,Montero2009,Martin2012}. 
While H$_2$CO has also been well detected in the northern clump of the eastern extension \citep[nomenclature:][]{Christopher2005,Martin2012}, the same clump is faint in N$_2$H$^+$.

CH$_3$OH(8--7) is found in CND regions close to the center, in the clump east of the SE and along the eastern edge of V cloud, that is part of the eastern extension. This is supported by the observations of \citet{Moser2014,Moserinprep} where the CH$_3$OH(2--1) emission is brighter in the eastern part of the V cloud than the western part.
The dominant peak in the V cloud is consistent with a source radiating 36 GHz and 44 GHz class I methanol maser emission \citep{YZ2008,Sjouwerman2010,Pihlstrom2011}. 
The clumps north and south of the V cloud coincide with local peaks in N$_2$H$^+$ and the most southern of the four clumps is situated in a cavity of N$_2$H$^+$ emission at the V-cloud edge.
The other 36 GHz maser \citep{Sjouwerman2010} has no CH$_3$OH(8--7) counterpart.
In the SEE cloud, the position of the CH$_3$OH(8--7) emission coincides with a 36 GHz maser \citep{Sjouwerman2010}, but in contrast to the bright point source in the V cloud, the CH$_3$OH(8--7) emission appears extended and faint.

\begin{figure*}[tb]
        \centering $
        \begin{array}{ccc}
        \includegraphics[trim = 5mm 0mm 7mm 0mm, clip, width=0.31\textwidth]{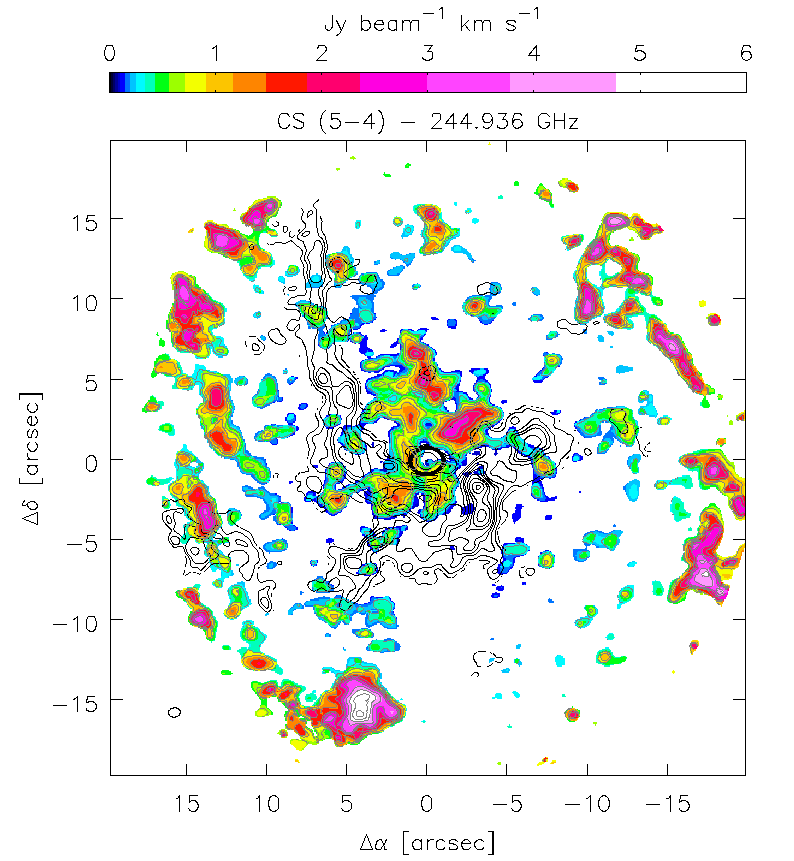}&
        \includegraphics[trim = 5mm 3mm 7mm 0mm, clip, width=0.31\textwidth]{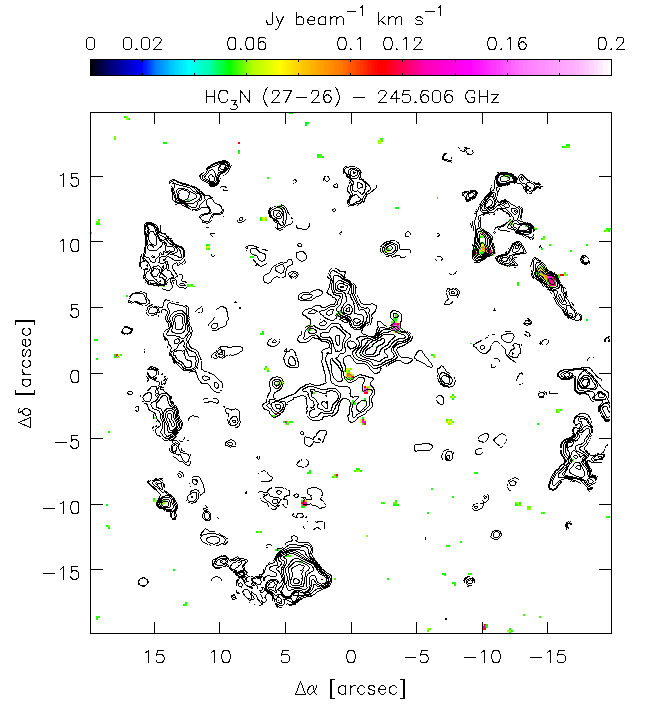}&
        \includegraphics[trim = 5mm 0mm 7mm 0mm, clip, width=0.31\textwidth]{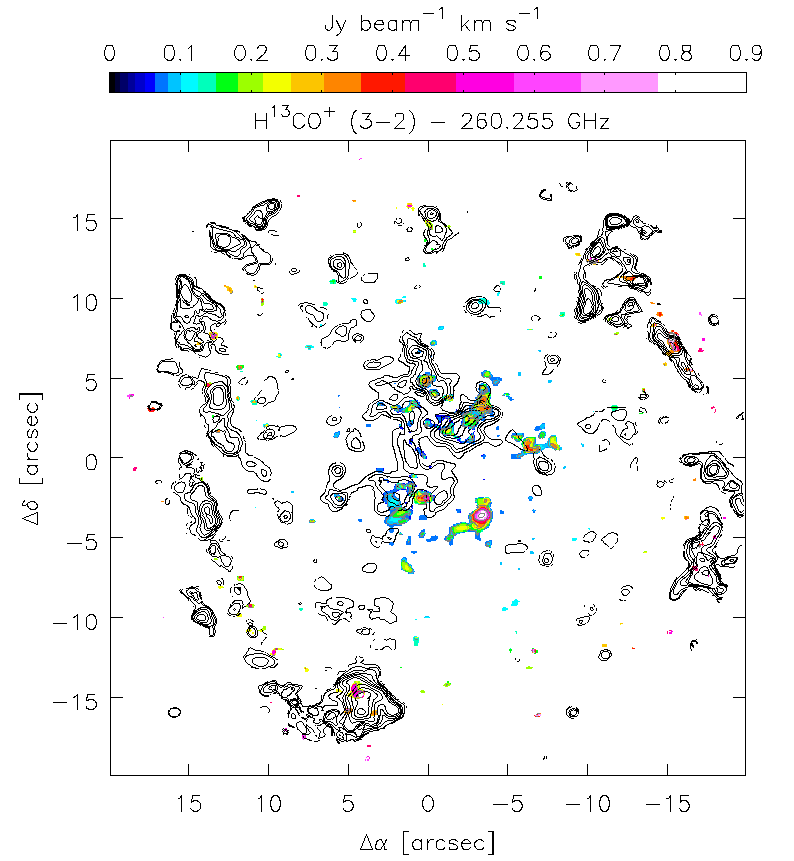}\\
        \includegraphics[trim = 5mm 0mm 7mm 0mm, clip, width=0.31\textwidth]{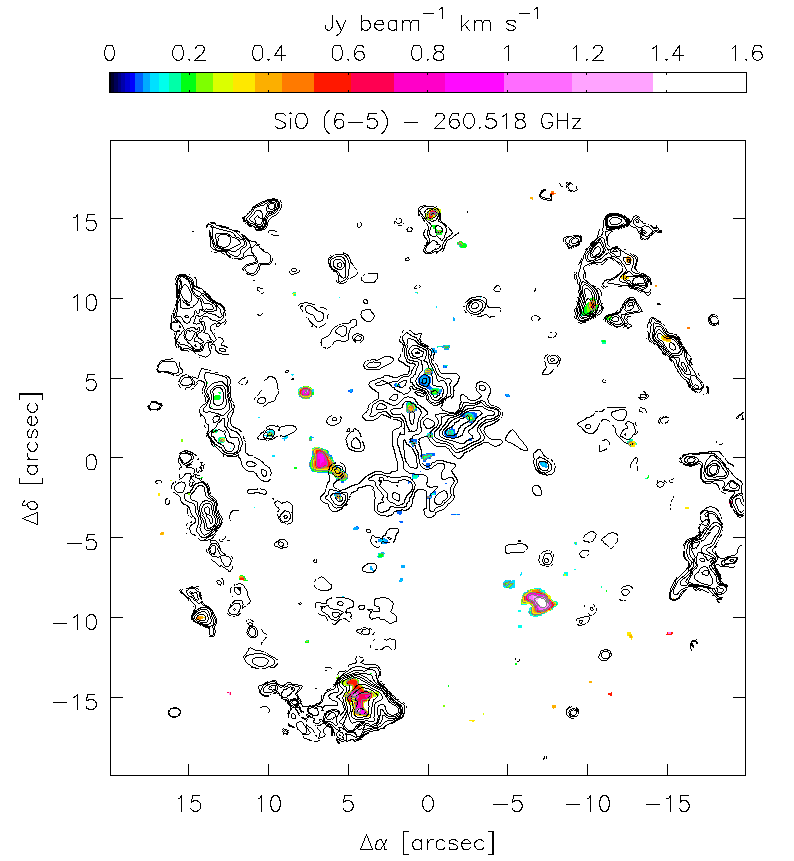}&
        \includegraphics[trim = 5mm 0mm 7mm 0mm, clip, width=0.31\textwidth]{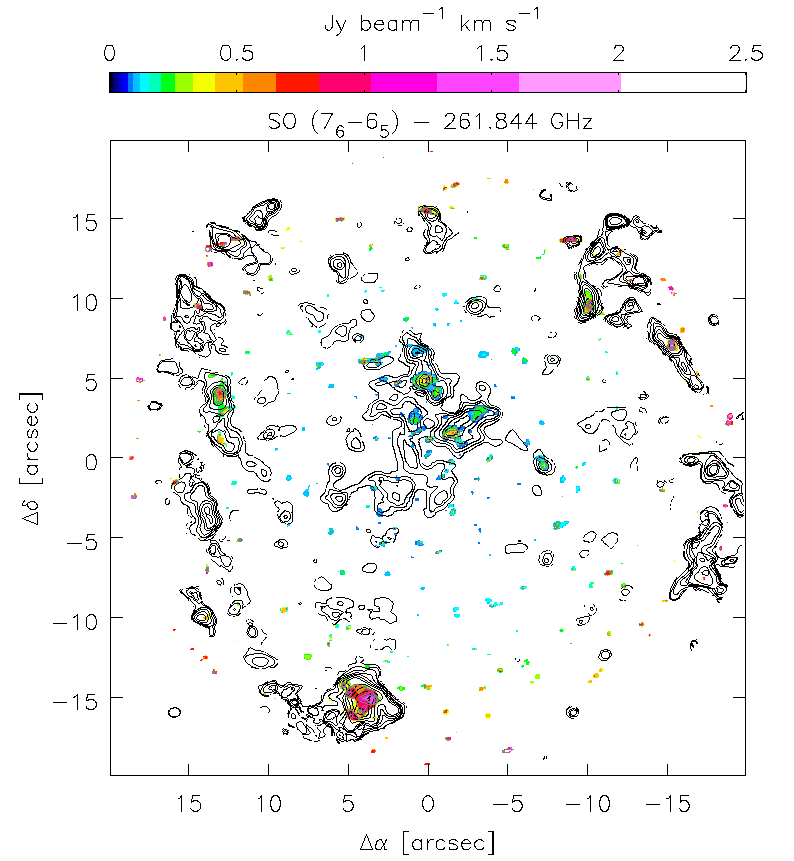}&
        \includegraphics[trim = 5mm 0mm 7mm 0mm, clip, width=0.31\textwidth]{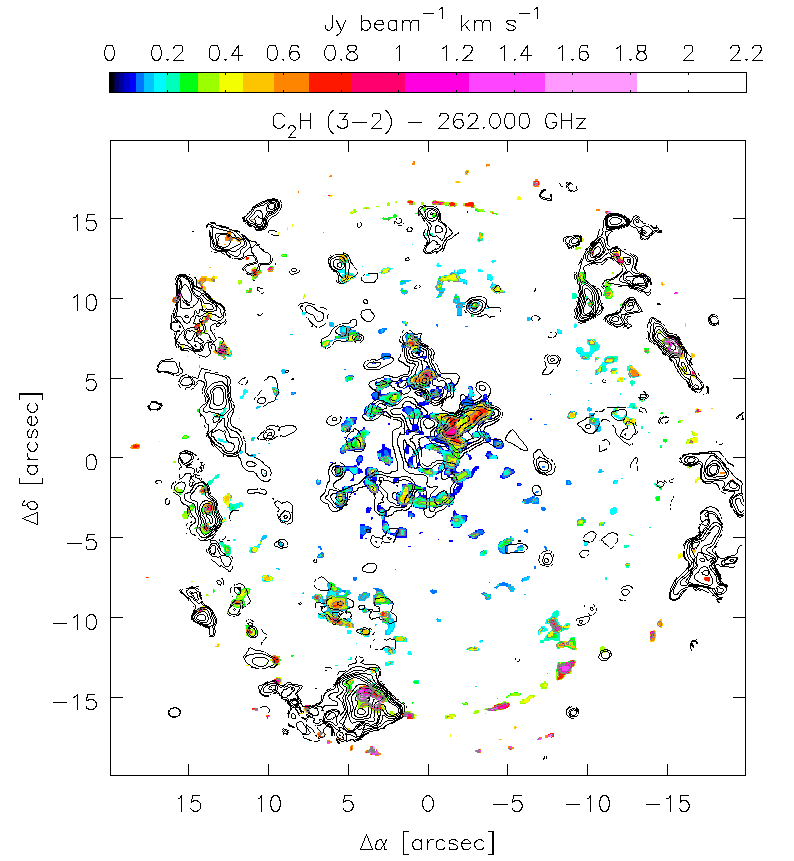}\\                                                                        
        \end{array} $                                                            \caption{Molecular line emission images of the inner 40$''$ (1.6 pc) at a resolution of about 0.7$''$ $\times$ 0.6$''$. 
                Top row (from left to right): CS(5--4), with 250 GHz contours as in Fig. \ref{cont-full}, 
                HC$_3$N(27--26), and H$^{13}$CO$^+$(3--2). 
                Bottom row (from left to right): SiO(6--5), SO(7--6), and C$_2$H(3--2). 
                The contours show the CS(5--4) emission at the levels of 
                [4, 8, 12, 18, 24, 30, 36, 48, 60, 72, 84] $\times$ $\sigma$ (= 0.08 Jy~beam$^{-1}$~km~s$^{-1}$) for comparison (see also Fig. \ref{mols-100-250}).
                Zooms into the inner 20$''$, the \textit{triop} (15$''$ northwest of Sgr~A*), and the SEW clumps (15$''$ south(east) of Sgr~A*) 
                can be found in Figs. \ref{mols-CS-20} and \ref{mols-250-20}, \ref{triop}, and \ref{SEW}, respectively.
        }
        \label{mols-full1}
\end{figure*}

\begin{Contfigure*}[tb]
        \centering $
        \begin{array}{ccc}
        \includegraphics[trim = 5mm 1mm 7mm 0mm, clip, width=0.31\textwidth]{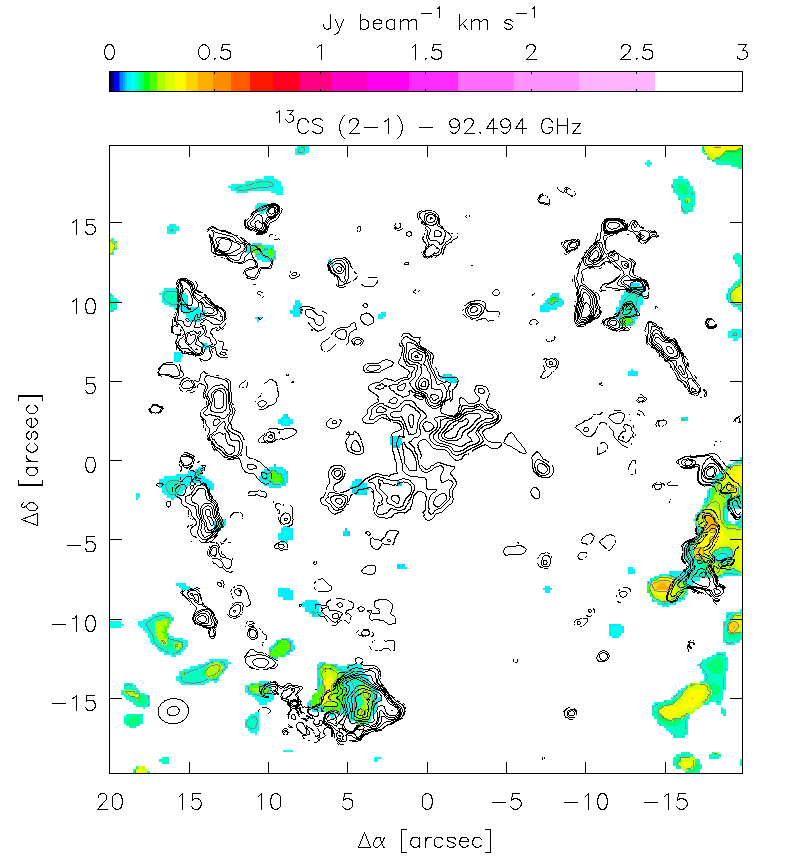}&
        \includegraphics[trim = 5mm 1mm 7mm 0mm, clip, width=0.31\textwidth]{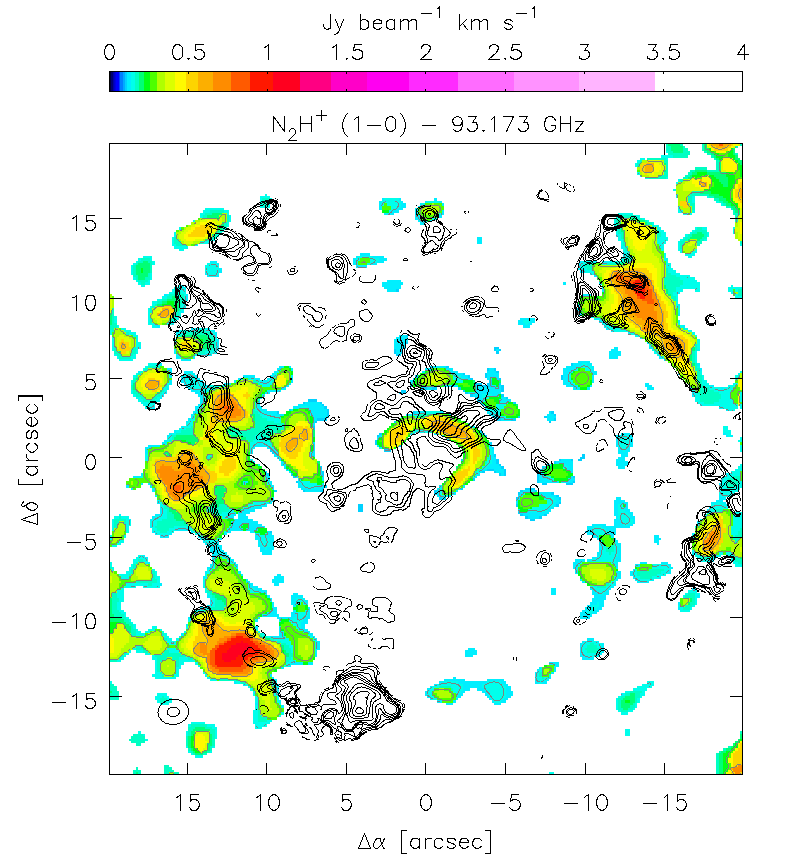}&
        \includegraphics[trim = 5mm 1mm 7mm 0mm, clip, width=0.31\textwidth]{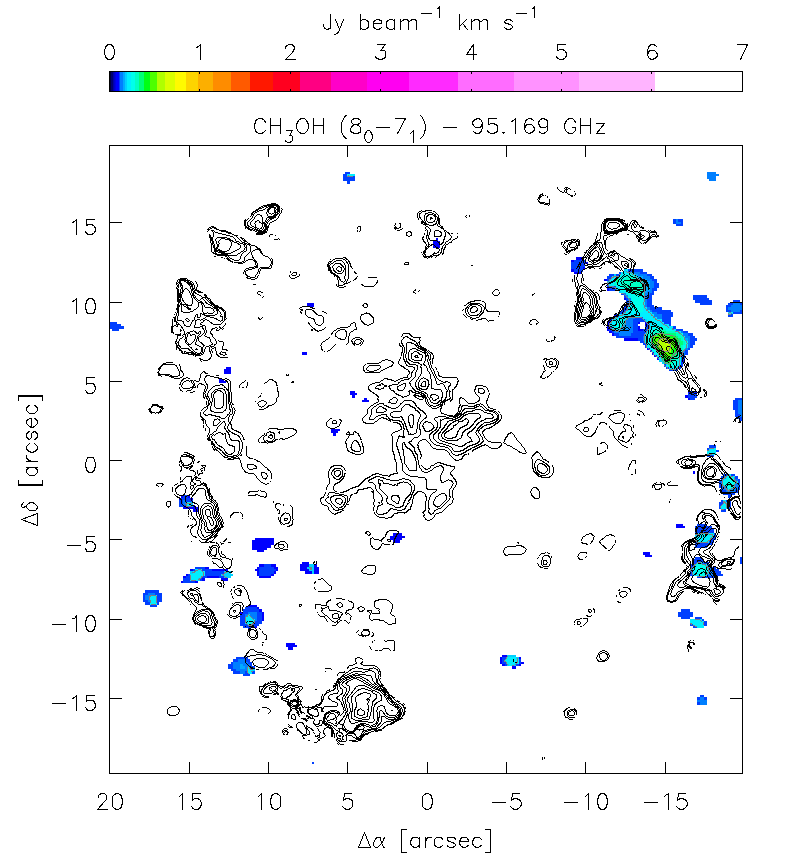}\\
        \includegraphics[trim = 5mm 0mm 7mm 0mm, clip, width=0.31\textwidth]{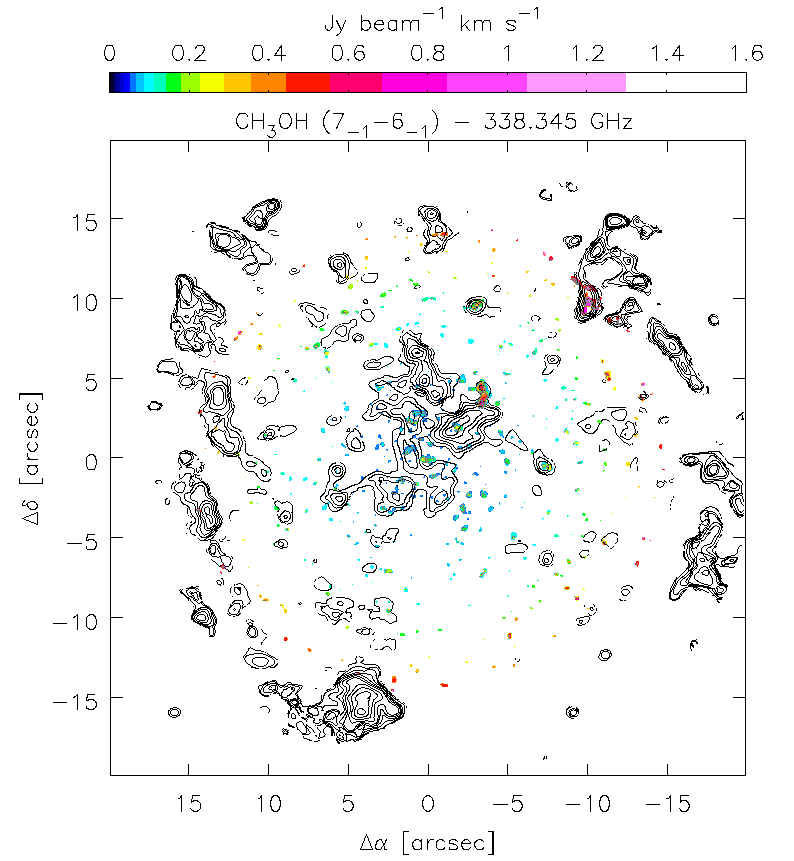}&
        \includegraphics[trim = 5mm 0mm 7mm 0mm, clip, width=0.31\textwidth]{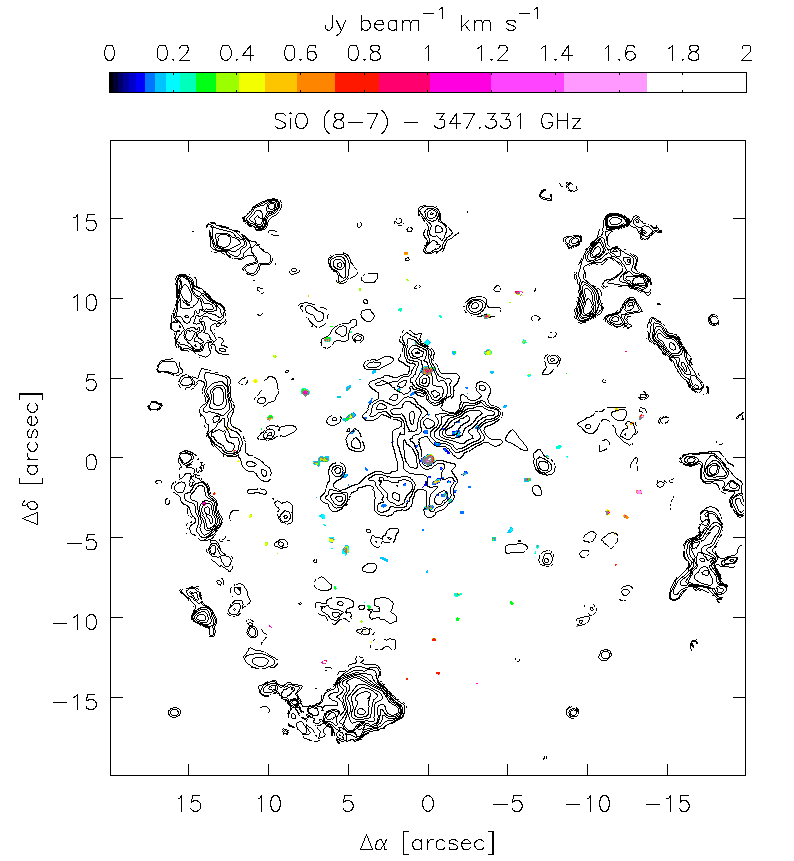}&
        \includegraphics[trim = 5mm 0mm 7mm 0mm, clip, width=0.31\textwidth]{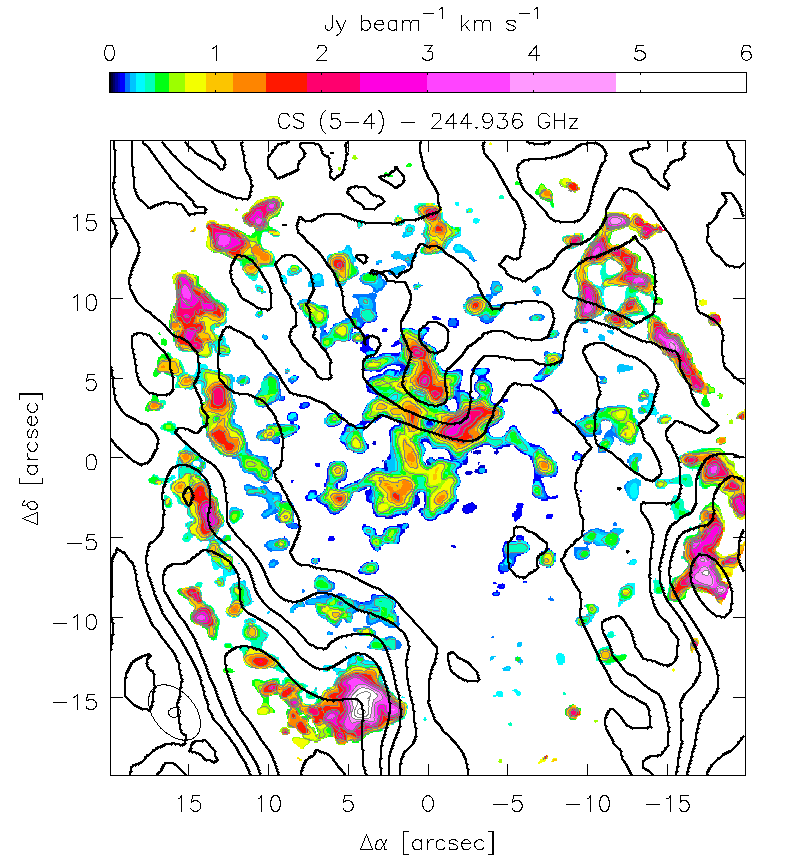}\\                                                                        
        \end{array} $                                                            \caption{Molecular line emission images of the inner 40$''$ (1.6 pc) at a resolution of about 0.7$''$ $\times$ 0.6$''$.
                Top row (from left to right): $^{13}$CS(2--1), N$_2$H$^+$(1--0), and CH$_3$OH(8--7) as in Fig. \ref{outermol}, but zoomed in. 
                Bottom row (from left to right): CH$_3$OH(7--6) and SiO(8--7), both at resolutions of 0.5$''$ $\times$ 0.5$''$, and CS(5--4) as before. 
                The contours show the CS(5--4) emission at the levels of 
                [4, 8, 12, 18, 24, 30, 36, 48, 60, 72, 84] $\times$ $\sigma$ (= 0.08 Jy~beam$^{-1}$~km~s$^{-1}$) for comparison.
                except from bottom right image, which is overlayed with the CN(2--1) emission as in Fig. \ref{outermol}. 
                The inner rim of the CND is detected (cf. Fig. \ref{CND-sketch}) toward the edge of the FOV10 of the band 6 and 7 emission lines.
        }
        \label{mols-full2}
\end{Contfigure*}

\subsection{Molecular gas in the inner 40 arcseconds}
\label{sec:inner40}

In this section, we describe and discuss the features of different molecular emission lines with respect to the line emission of the J=5-4 transition of the excited density tracer CS, which is the brightest line in the central region. The integrated flux images of the lines can be seen in Fig. \ref{mols-full1}.

The CS line emission is strong, widespread, and clumpy within the FOV10 of band 6. 
As expected, CS is found in the CND, i.e., east and west of the center. In fact, it appears at the inner edges of the CND. In the east, the CS emission follows the eastern edge of the extended minispiral emission (e.g., at 100 GHz) and appears intersected by the EA of the minispiral (e.g., at 250 GHz).
The maximum in emission lies in the SEW  (see Fig.\ref{CND-sketch} for orientation).
In the west the emission resides in the \textit{triop} and in the northwestern edge of the WA of the minispiral (see also Fig. \ref{mols-100-250}), where the CS emission is limited to the south by the FOV10 cutoff. 

The emission from H$^{13}$CO$^+$, SiO, SO, and C$_2$H is fainter but typically found, in different fractions, in the strong CS peaks in the center and in the CND, whereas the emission of $^{13}$CS, N$_2$H$^+$, and CH$_3$OH(8--7) is rather found toward the CND, which might also be related to the different angular scales traced in the two ALMA bands.

The CS transition is the brightest in our data set. 
In order to estimate the amount of missing flux, we generated a 5’'-tapered map for the CS transition, which is the brightest transition in out data. The resolution corresponds to the largest angular scales to which the observations are sensitive. This tapered map is compared to our originally 0.75’’ CS map smoothed to 5’’ resolution. Considering these scales, about 20-40\% of the emission is resolved out at 0.75’’.

\subsubsection{Triop}

In our ALMA data, this structure has an unusual triangular shape that is reminiscent of a crustacean shrimp, which is why we gave it the name \textit{triop}.
This region is visible in all molecular emission lines in our ALMA data (see also Fig. \ref{triop}). 
We subdivide the region into the head, which is the wide northeastern part of the \textit{triop}, and the tail, which is the thin southwestern part. Most of the molecular emission lines are predominantly detected toward the edge facing the GC and the nuclear cluster, especially in the center of the tail and in the southeast clump in the head, which is closest to the center. 
The former is the brightest (peak) in all molecules, except from $^{13}$CS and N$_2$H$^+$, which peak rather toward the center or the \textit{triop}. 
The \textit{triop} is visible in all molecules detected in previous studies 
\citep[e.g.,][]{Guesten1987,Jackson1993,Marr1993,YZ2001,Wright2001,Christopher2005,Montero2009,Martin2012}, comprising typical density and photon dominated region (PDR) tracers, such as HCO$^+$, HCN, CN, as well as highly UV-sensitive species, such as HC$_3$N.
While the shape in the previous studies is rather oval or triangular because of the larger beam sizes, these ALMA observations resolve its filamentary structure for the first time.

This region becomes even more extraordinary by its class I methanol maser emission at 44 GHz, possibly indicating an early phase of star formation \citep{YZ2008,Sjouwerman2010}. The maser coincides with the prominent clump in the center of the tail and with the peak in CH$_3$OH(8--7) and is thus the closest class I methanol maser to the center so far.

\subsubsection{Southern extension west}

The SEW cloud stands out by showing emission in the band 6 molecules, except from HC$_3$N, about as bright as in the \textit{triop} but more extended. 
In contrast to the \textit{triop}, the emission of N$_2$H$^+$ and CH$_3$OH(8--7) is not detected in this cloud.
This is consistent with previous studies \citep[compare][]{Christopher2005,Montero2009,Martin2012} and implies a less efficient shielding from the UV field than in the \textit{triop}.
In the CS emission, the SEW shows a diameter of ~4$''$ with a double core in north-south direction and an elongation from the northwestern tip to the southwest that appears as a front perpendicular to the direction to Sgr~A* (see also Fig. \ref{SEW}). Close to this front, i.e., 1$''$ west of the CS peaks, lies the maximum of the C$_2$H emission, which extends to the northern CS peak. The SO emission behaves similar to this but with a larger extent over the CS cores to the east.
The SiO peak is located east of the SO and C$_2$H maxima and slightly east of the southern CS peak, but still on the northern CS peak from where it extends to the north and east. Another SiO peak is on the southern CS maximum itself. Furthest away from the front facing the center is the H$^{13}$CO$^+$ emission displaying a peak between the northern CS and SiO maximum from where it extends to the northeast similar to SiO. 

Additional smaller H$^{13}$CO$^+$ peaks are found 1$''$ east and west of the southern CS core, where the former is consistent with the eastern edge of the SO emission and the latter with southern edge of the C$_2$H peak. The $^{13}$CS emission is strong at the CS maxima and in a region at northeastern edge of SEW as well. For this line, the lower resolution and larger scales traced in band 3 have to be taken into account. The spatial distribution of the line emission from the different molecules with respect to the direction to Sgr~A* and the nuclear stellar cluster suggests some kind of stratification of the emission.

\begin{figure*}[tb]
        \centering
        $
        \begin{array}{ccc}
        \includegraphics[trim = 5mm 0mm 7mm 9mm, clip, width=0.48\textwidth]{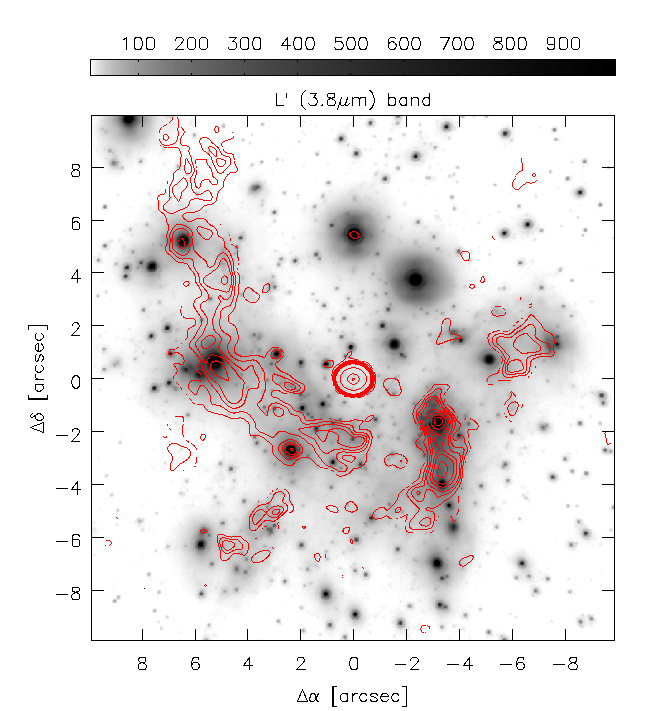}&
        \includegraphics[trim = 5mm 0mm 7mm 9mm, clip, width=0.48\textwidth]{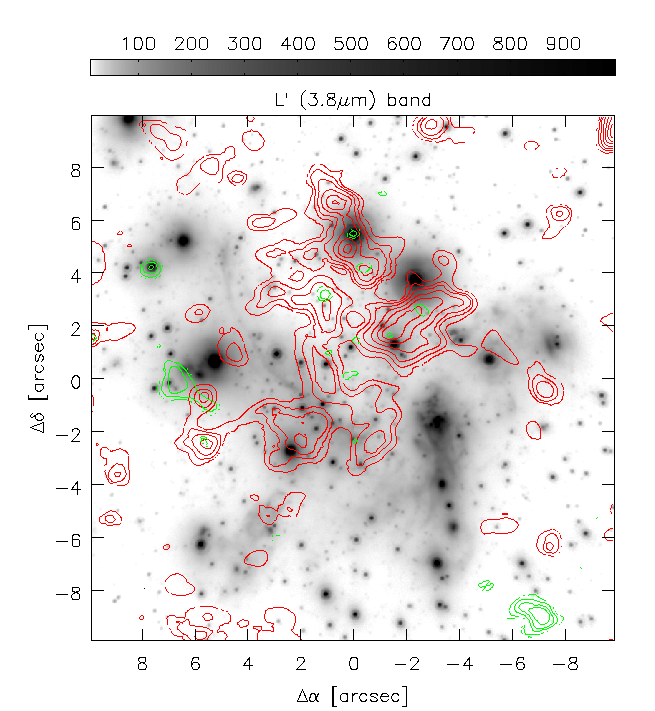}\\
        \includegraphics[trim = 5mm 0mm 7mm  5mm, clip, width=0.48\textwidth]{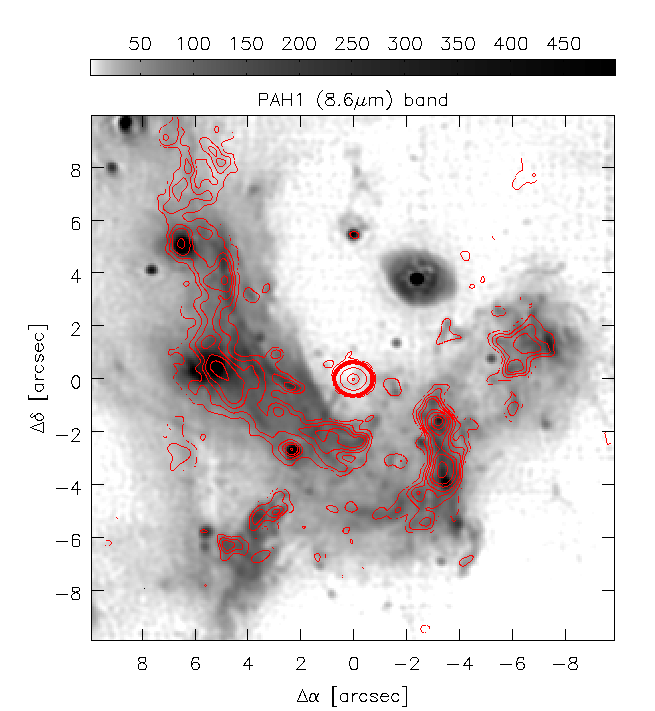}&
        \includegraphics[trim = 5mm 0mm 7mm  5mm, clip, width=0.48\textwidth]{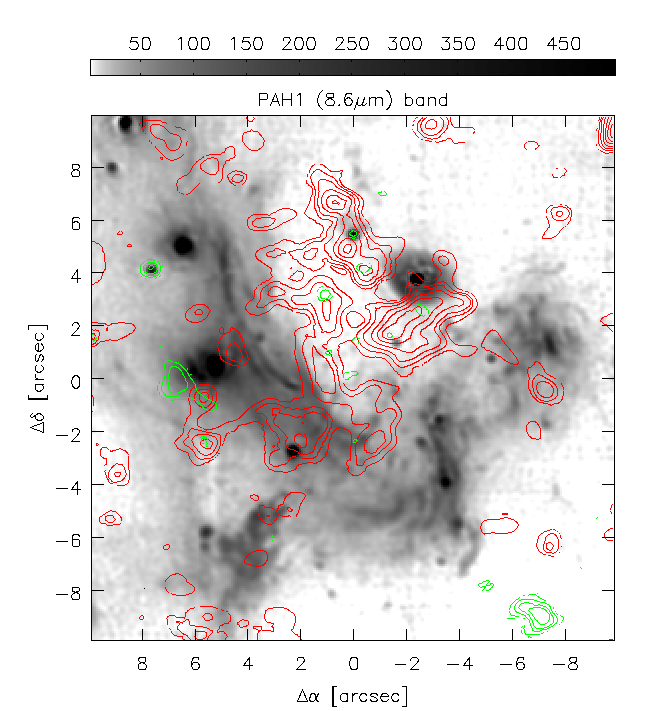}\\      
        \end{array} $  
        \caption{Molecular gas and continuum emission maps in the inner 20$''$ compared to the NIR VLT/NACO L' (3.8 $\mu$m) emission (top, Sabha, private communication)
                and the MIR VLT/VISIR PAH (8.6 $\mu$m) emission images \citep[bottom, ][]{Sabha2016} in arbitrary units.
                The left side shows the 340 GHz continuum
                in red contours of [6, 12, 18, 24, 36, 48, 72, 96, 960, 9600, 17280] $\times$ $\sigma$ (= 0.24 mJy~beam$^{-1}$).
                The right side shows CS(5--4) in red contours as in Fig. \ref{mols-full1} 
                and SiO(6--5) in green contours of [2, 4, 8, 12, 16] $\times$ $\sigma$ (= 0.08 Jy~beam$^{-1}$~km~s$^{-1}$).
        }
        \label{NIR-MIR-CS-350}
\end{figure*}

\subsubsection{The central 20 arcseconds}

It is remarkable that there is CS emission within a (projected) radius of $<$ 8$''$ around the SMBH (see also Figs. \ref{mols-100-250}, \ref{mols-250-20}, and \ref{mols-CS-20}). This central association (CA) of clouds extends over the location where the bar of the minispiral meets the NA to the region north of Sgr~A*, where it appears to be outlined by the inner edge of the bar and the NA facing Sgr~A*. The latter contains two prominent features: one region extends from 2$''$ northwest from Sgr~A* parallel to the bar toward the northwest with a length of about 3$''$ (SE-NW cloud) and the other extends from 4$''$ north of Sgr~A* northward with a length of 4$''$ and a slight tilt to the east (NS cloud). Both regions seem to comprise 2 - 3 clumps. In comparison with the continuum emission, the center of the NS cloud coincides with the ionized emission from IRS 7.
The northwestern part of the CA shows strong similarities to the NIR H$_2$ (hot gas) and extinction maps of \citet{Ciurlo2016}.

The CS emission southeast of Sgr~A* consists of two clumps, one covering a triangular region given by the IRS 16 southern cluster, IRS 16 SE2 and 16SE3 (35), and IRS 21, the other outlining the eastern inner edge of the minicavity. Parts of these clumps are tentatively also detected in C$_2$H, but contaminated by a side lobe. 
South of this configuration the CS emission is found in the EA nearby IRS 9. Furthermore, CS is present in small regions immediately north and south of IRS 1W, which otherwise are only detected in C$_2$H. In addition, the CS emission extends along the EB, i.e., the north-south connection between the NA and the EA (compare MIR images in Fig. \ref{NIR-MIR-CS-350}), where H$^{13}$CO$^+$, SiO, SO, and C$_2$H are also detected. Except from these regions, the CS emission seems to avoid the inner minispiral, especially the IRS 13 to IRS 2 region. 
The SE-NW cloud reaches from the IRS 29 sources to the southwest of IRS 3. These two positions mark two clumps clearly visible in all band 6 emission lines. There are fainter extensions in CS emission on IRS 3 itself and to the northwestern edge of the IRS 3 dust shell. The latter is remarkably bright in HC$_3$N and CH$_3$OH(7--6), neither of which are reliably detected elsewhere in the central 12$''$, and slightly in H$^{13}$CO$^+$, implying a change in the ISM conditions (see Section \ref{sec:ratio} for discussion).
The southern part of the NS cloud originates in a region almost between IRS 3 and 7, passes IRS 7 slightly southwest of it in a second clump, and prolongs to the northeast to a third clump, but not along the tail of IRS 7 (MIR images in Fig. \ref{NIR-MIR-CS-350}). These clumps are also detected in H$^{13}$CO$^+$, SiO, SO, and C$_2$H. Another clump indicated by all band 6 lines is halfway between IRS 16CC and IRS 7. Emission of CS, SiO, SO, C$_2$H, and CH$_3$OH(7--6) is also found at the southwestern edge of the Bar, i.e., south of IRS 6E and 6W.

\begin{figure*}[tb]
        \centering
        \includegraphics[trim = 7mm 10mm 0mm 0mm, clip, width=\textwidth]{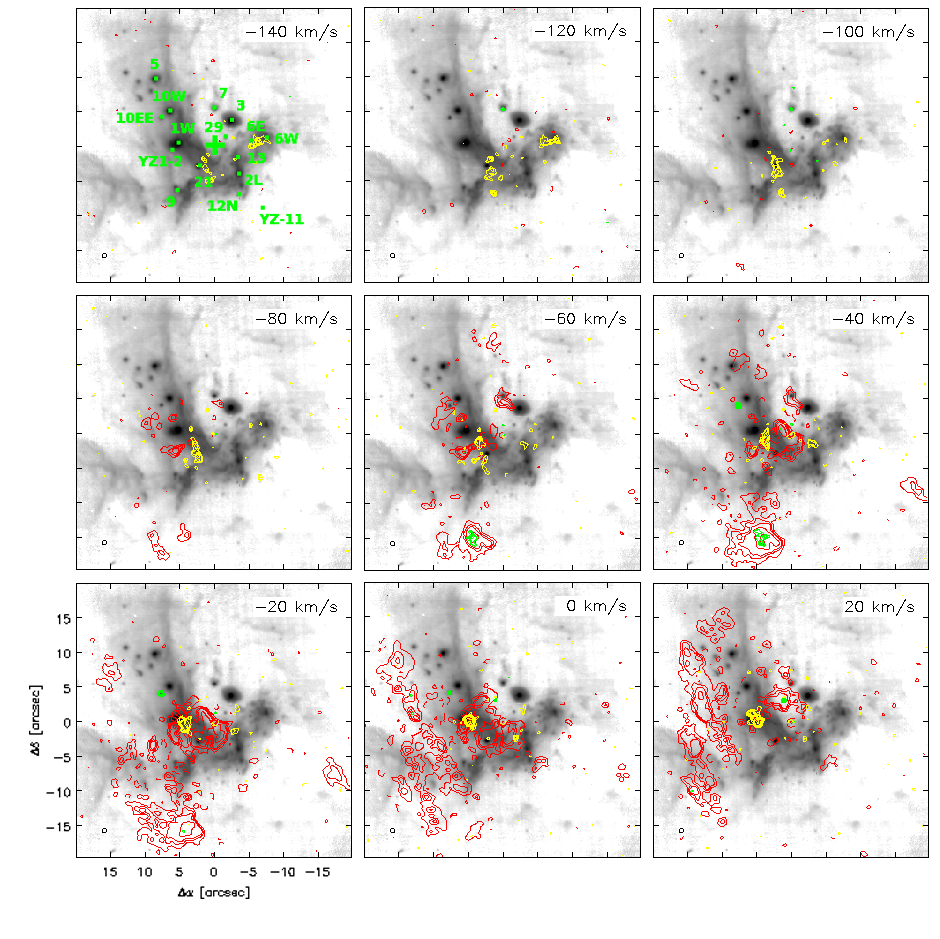}                                                         
        \caption{Velocity distribution maps from -140 to 20 km~s$^{-1}$ in the inner 40$''$
                overlayed onto the MIR (8.6 $\mu$m) PAH image from Fig \ref{NIR-MIR-CS-350}.
                Red contours show the CS(5--4) emission at the levels [4, 8, 12, 24, 48, -3, -45] $\times$ $\sigma$ (= 0.91 mJy~beam$^{-1}$), 
                green contours the SiO(6--5) emission at [4, 5, 6, 7, 8, 10, 12, 14, -4] $\times$ $\sigma$ (= 0.94 mJy~beam$^{-1}$), and
                yellow contours represent the H36$\beta$ emission at [3, 4, 5, 6, 8, 10, -3, -6] $\times$ $\sigma$ (= 1.05 mJy~beam$^{-1}$)
                for comparison to the RRL emission. Green numbers denote the IRS sources and the green cross Sgr~A*.
        }
        \label{mol-vel-ch1}
\end{figure*}

Apart from HC$_3$N and CH$_3$OH(7--6), SiO shows a peculiar deviation from the distribution of the CS detected gas. The two SiO clumps southeast of IRS 1W and southwest of IRS 12, first reported by \citet[][clump 1 \& 2 and clump 11; YZ1-2 and YZ11 hereafter]{YZ2013}, are both detected in SiO(6--5), while only the first is detected in SiO(8--7). There is no other emission line present in the immediate vicinity at the corresponding velocities except from faint CS emission northeast of IRS 1W and south of IRS 21.
Two other prominent SiO point sources seen in both transitions coincide in position and velocities (see Fig. \ref{mol-vel-ch1}) with the SiO maser stars IRS 7 and IRS 10EE \citep{Reid2007,Li2010}. IRS 10EE corresponds to clump 3 (YZ3) in \citet[][]{YZ2013}.

\begin{Contfigure*}[tb]
        \centering
        \includegraphics[trim = 7mm 10mm 0mm 0mm, clip,width=\textwidth]{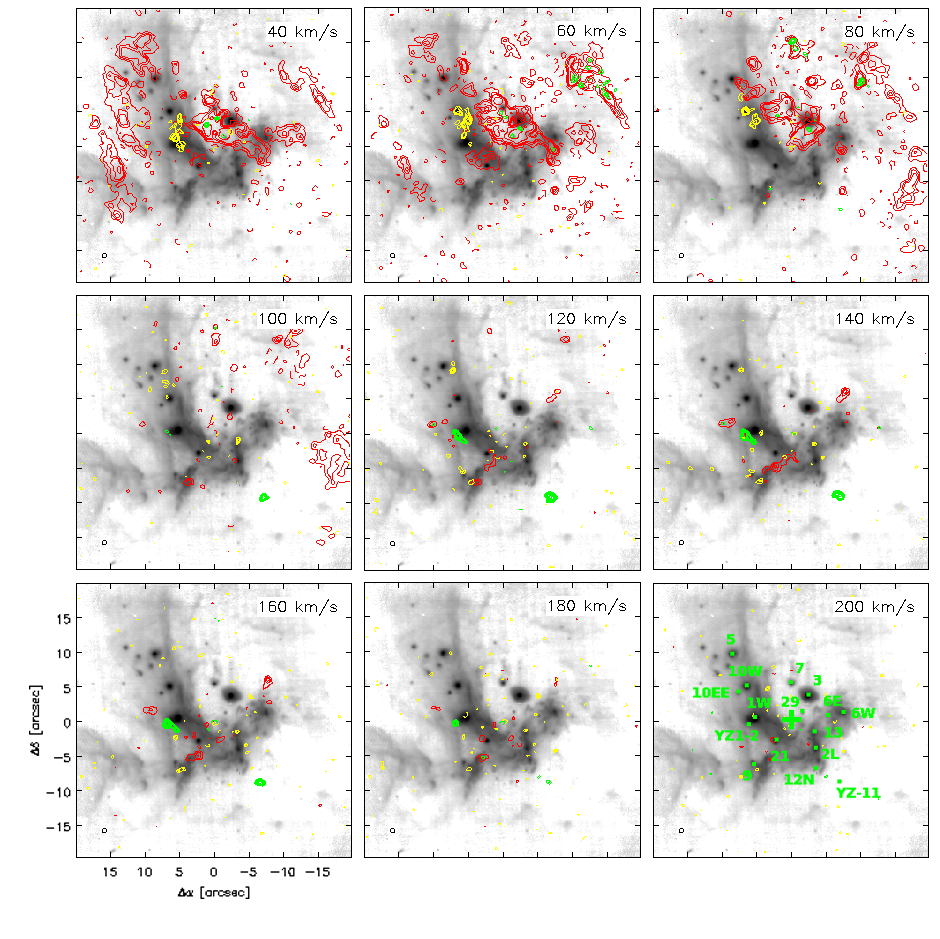}                                                         
        \caption{Velocity distribution from 40 to 200 km~s$^{-1}$ in the inner 40$''$ 
                overlayed onto the MIR (8.6 $\mu$m) PAH image from Fig \ref{NIR-MIR-CS-350}. 
                Red contours show the CS(5--4) emission at the levels [4, 8, 12, 24, 48, -3, -45] $\times$ $\sigma$ (= 0.91 mJy~beam$^{-1}$), 
                green contours the SiO(6--5) emission at [4, 5, 6, 7, 8, 10, 12, 14, -4] $\times$ $\sigma$ (= 0.94 mJy~beam$^{-1}$), and
                yellow contours represent the H36$\beta$ emission at [3, 4, 5, 6, 8, 10, -3, -6] $\times$ $\sigma$ (= 1.05 mJy~beam$^{-1}$)
                for comparison to the RRL emission. Green numbers denote the IRS sources and the green cross Sgr~A*.
        }
        \label{mol-vel-ch2}
\end{Contfigure*}

\subsection{Kinematics in the inner 40 arcseconds}
\label{sec:kin}

In the following we describe the kinematics in a central main velocity range between about -100 km~s$^{-1}$ and +100 km~s$^{-1}$ and higher velocities beyond the limits of this interval. In general, one can expect that the high velocity features can be attributed to either kinetically more active regions (due to outflows or cloud collisions) or the regions that are physically closer ($<$10$''$ or 0.4~pc) to the center.
In order to disentangle the cloud complexes and study the motion of the molecular and ionized gas, we plot the channels maps between -140 km~s$^{-1}$ and 200 (220) km~s$^{-1}$ for CS and SiO in comparison to H36$\beta$ in Fig. \ref{mol-vel-ch1}. The given channel velocities denote the lower edge of the channels, which have a width of 20 km~s$^{-1}$. Details on the spectral properties of the regions can be found in the Tables \ref{mol-src-vel-1} - \ref{mol-src-vel-3}, and \ref{mol-IRS-vel}, maps of the velocity fields in Fig. \ref{mom1}.

\subsubsection{Main velocity range}

The main range of the velocities of about -100 km~s$^{-1}$ and +100 km~s$^{-1}$ in the inner 40$''$ is given by the velocities of the molecular gas clumps within this cutout of the CND. However, in Fig. \ref{mol-vel-ch1}, we start at -140 km~s$^{-1}$ to show the SiO emission in IRS 7, which is visible from channel -140 km~s$^{-1}$ to -100 km~s$^{-1}$. This is consistent with its velocity of -114 km~s$^{-1}$ and a line width of 5-10 km~s$^{-1}$ \citep{Reid2007}, considering our low velocity resolution. 
In this channel range the H36$\beta$ line is located around IRS 21 with an elongation in north-south direction and in the western end of the bar, i.e., between IRS 6E and 6W.
The first CS emission regions appearing between channel -80 to -40 km~s$^{-1}$ are small (r $\sim$ 2$''$) clumps in the NA, i.e., between the EA around IRS 21, and IRS 1W, between IRS 1W, 10 and maybe 5, and south of IRS 7.
In the channel range from -40 to 0 km~s$^{-1}$ the large CS cloud in the minispiral, and south and east of Sgr~A* emerges and extends from the eastern edge of the minicavity to 1W with the thickness of the NA in that region. At the same time IRS 10EE is visible in SiO emission around the expected velocity of -27 km~s$^{-1}$ and a line width of 5 - 10 km~s$^{-1}$ \citep{Reid2007,Li2010}.
The hydrogen RRL emission in the channel range of -80 to 0 km~s$^{-1}$ moves northeast from IRS 21 to 1W and appears in this way mostly at the northeastern edge of the central extended CS emission clump or is surrounded by it. 
These are also the channels in which the SEW arises in CS and SiO, peaking around -40 km~s$^{-1}$. In the channels -40 and -20 km~s$^{-1}$, we make out a faint apparent connection between SEW and the extended central CS clump south of Sgr~A*.

From channel -20 km~s$^{-1}$ onward the SEW emission turns into the eastern CND, passing the EA and extending northward along the NA until the channel of 60 km~s$^{-1}$. We note small CS clumps nearby or at the dusty sources east of IRS 5 \citep{Perger2008} and IRS 5 itself between channels 20 to 60 km~s$^{-1}$. 

The CS emission within the minispiral in channels above 0 km~s$^{-1}$ is not strong or widespread anymore. Instead the bulk of the molecular gas at velocities between 20 to 100 km~s$^{-1}$ is found in the region delineated by the NA, the bar, IRS 3, and 7, with a southwestern extension across the bar, i.e., to the southeast of IRS 6W, and a collection of small clumps between the southwest of the bar and the WA in the velocity range of 40 to 80 km~s$^{-1}$. 
SiO emission appears in several CS peaks in this region and velocity range. Meanwhile, the H36$\beta$ line emission proceeds from IRS 1W northward to IRS 5. It does not overlap with the CS emission anymore but might follow it. We cannot exclude that more extended RRL emission is either too faint or resolved out. The CS emission in the channels 40 to 80 km~s$^{-1}$ seems to be perfectly outlined by the NA edge between IRS 1W and 10W.
From channel 20 to 120 km~s$^{-1}$, or 40 to 80 km~s$^{-1}$ for the weaker lines, the \textit{triop} emission appears, moving from the southwest to the northeast and peaking at 60 km~s$^{-1}$, as well as a larger clump 15$''$ north of Sgr~A*, peaking at 80 km~s$^{-1}$. Both features are also bright in SiO emission.
Furthermore, a clump is visible in the 80 to 100 km~s$^{-1}$ channels southwest of the bar, south of the \textit{triop}, and apparently within the CND.

\subsubsection{High velocities}

Outside these CND typical velocities, i.e., -100 to 100 km~s$^{-1}$ in the CND regions between the lobes, we find some high velocity clumps within 10$''$ around the SMBH (see also Fig. \ref{mom1}).
The two SiO clumps YZ11 and YZ1-2 mentioned before can be seen from 80 to 200 km~s$^{-1}$ and 100 to 180 km~s$^{-1}$, respectively, and peak at $\sim$ 140 to 160 km~s$^{-1}$. This is consistent with the results of \citet[][]{YZ2013}.
In addition, we find three CS emitting clumps: one extending from southwest of IRS 9, i.e., within the tip, past IRS 21 to the eastern edge of the minicavity in the range of 80 to 200 km~s$^{-1}$, another one 5$''$ north of IRS 6W  in the range of 120 to 180 km~s$^{-1}$, and the last one cospatial with faint SiO emission 5$''$ northeast of YZ1-2 in the range of 100 to 180 km~s$^{-1}$.
In the channels above 100 km~s$^{-1}$ the H36$\beta$ line emission is faint and only visible in the form of tiny clumps. Moreover, the noise increases in these channels of the H36$\beta$ image cube. However, the RRL emission appears in the tip at velocities above 100 km~s$^{-1}$, which is in agreement with other ionized emission studies \citep[e.g., \ion{Ne}{II} cube of ][]{Irons2012}. There it partially spatially overlaps with the CS emission at the channels 120 and 140 km~s$^{-1}$.

\begin{figure}[tb]
        \centering 
        \includegraphics[trim = 1mm 0mm 10mm 7mm, clip, width=0.45\textwidth]{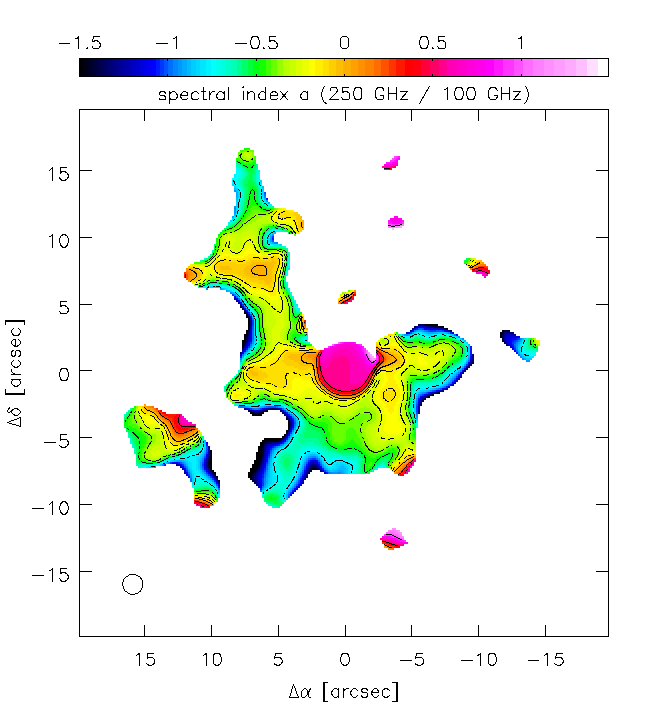}

        \includegraphics[trim = 1mm 0mm  7mm 5mm, clip, width=0.45\textwidth]{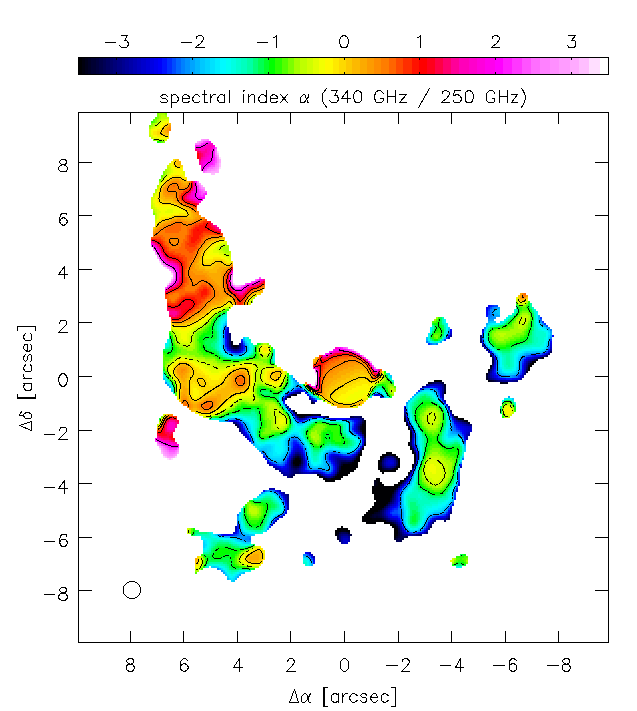}                                                          
        \caption{Images of the continuum spectral index distribution in the inner $\lesssim $ 1.6 pc. 
                Top: Between 100 and 250 GHz (inner 40$''$) tapered to a resolution of 1.5$''$ with contours of 
                [-0.75, -0.5, -0.375, -0.25, -0.125, 0, 0.25, 0.5, 1, 1.5].
                Bottom: Between 250 and 340 GHz (inner 20$''$) tapered to a resolution of 0.65$''$ with contours of [-2,-1, -0.5, 0, 0.5, 1, 2].
                Maps on the uncertainties can be found in Fig. \ref{spex-app}.
        }
        \label{spex}
\end{figure}

\section{Discussion}
\label{sec:Diss}

In this section, we first address the spectral index of the (sub-) mm continuum emission in the inner 40$''$ and the electron temperature of the ionized gas. This is followed by a study of the relation between the gas and the stars. 
The last part of this section deals with the trends in the molecular line ratios with respect to their location in the GC and their possible explanations.

\subsection{Continuum spectral index}
\label{sec:spex}

We computed spectral index maps (see Fig. \ref{spex}) between 100, 250, and 340 GHz defined by a power law with $S_\nu  \propto  \nu^\alpha$, where $\alpha$ is the spectral index. 
To accomplish this, we clipped the tapered (same resolution) continuum maps at 5$\sigma$ before primary beam correction. A map of the uncertainties based on the S/N and the flux calibration uncertainties is given in the Appendix in Fig. \ref{spex-app}. In the high S/N regions $\Delta \alpha$ is dominated by the flux calibration errors (see Sect. \ref{sec:cali}).
We find Sgr~A* to have an inverted spectrum in the 100 to 250 GHz range with $\alpha_{100-250}  \sim  0.58 \pm 0.21$, which is related to the synchrotron emission from the accretion disk. 
The spectral index is in agreement with $\alpha_{100-230}  \sim  0.5$ obtained by \citet{Kunneriath2012a,Kunneriath2012b} and with the results of \citet{Falcke1998}, i.e., $\alpha_{100-150}  \sim  0.76$ and $\alpha_{43-100}  \sim  0.52$ in the 3 - 2 mm  and 7 - 3 mm range, respectively.

The spectral index of Sgr~A* is reliable because of its high S/N detection and its nature as a point source but it may be affected by flux density variability \citep{Kunneriath2012a}. However, the overall spectral trend in this frequency domain remains preserved.

In contrast to the minispiral, Sgr~A* is detected with a high S/N at 340 GHz and displays a flat $\alpha_{250-340} = 0.17 \pm 0.45$. This matches the $\alpha_{230-690}  \sim -0.13$ obtained by \citet{Marrone2006a} and the $\alpha_{217-355} = -0.06 \pm 0.26$ by \citet{Bower2015} very well. The latter suggests the emission at these frequencies to be in a transition between the optically thick and thin regime.

From 345 GHz and 690 GHz SMA measurements, there is evidence that the overall spectrum of Sgr~A* peaks around 345 GHz \citep{Marrone2006a,Marrone2006b,Marrone2006c}. This is also in agreement with \citet{Eckart2012}, who find for the bulk of their synchrotron and synchrotron-self Compton (SSC) models, synchrotron turnover frequencies in the range 300 - 400 GHz.

The spectral indices in the minispiral between 100 and 250 GHz are rather lower limits owing to the flux at larger angular scales being resolved out in the 250 GHz observations, but measured in the 100 GHz observations. Within a region of 10'' around Sgr~A* the spectral indices in the minispiral are negative, similar to the results of \citet{Kunneriath2012b}. 
Toward IRS 13, 2L, 6, 1W, and 10W, $\alpha_{100-250}$ reaches values of $\sim$ -0.1, which is indicative of free-free thermal bremsstrahlung emission. These values might be largely unaffected by resolution and beam filling factor effects because of the brightness and compactness of these sources, consequently dominating the emission from these regions. The filaments south and west of Sgr~A* in the bar and the filaments in the tip are fainter compared to the diffuser emission gas they are embedded in. Missing the extended 250 GHz flux could be the reason why the spectral index value drops to $\sim$ -0.4 and even less, i.e., $\sim$ -0.7 in the tip. 

The spectral index map of \citet{Kunneriath2012b} shows similarly steep values toward the compact sources, but their results suffer from the lower angular resolution of the CARMA observation. 
The reliability of the spectral index for faint extended emission depends heavily on the $uv$ coverage, which is superior for ALMA.
At ALMA resolutions, the spectral index of point source is solely dominated by the corresponding point source fluxes.       
Positive spectral indices prevail in the region south of IRS 5 (at $\Delta \delta \sim 7''$ from Sgr~A*) and in the IRS 1 to IRS 16NW region of the NA with $\alpha_{100-250}  \gtrsim  0$ , in the ribbon of the EA with $\alpha_{100-250}  \gtrsim  0.1$, and in a few single clouds in the field with $\alpha_{100-250}  \gtrsim  0.5$. These indices imply a growing importance of dust for the continuum emission at frequencies $\gtrsim$ 230 GHz \citep[see discussion in][]{Kunneriath2012a}.

This trend continues to the next higher band at 340 GHz, where an $\alpha_{250-340} \gtrsim 0$ between 250 and 340 GHz is even found downstream the NA toward the south of IRS 1W. 
The bar and the tip show a steep $\alpha_{250-340}  \sim  -1 - 0$. 
The 340 GHz data is not only influenced by resolution effects but also by a lower S/N in this region.
An $\alpha_{250-340}  \sim  -0.3$ is only found in the brightest sources. 
It is very likely that diffuse emission, although within the angular scales of the 340 GHz observation, is too faint to be detected at this noise level, which results in even more extreme apparent spectral indices than for the 100 - 250 GHz range.

\begin{figure}[tb]
        \centering
        \includegraphics[trim = 5mm 0mm 7mm 9mm, clip, width=0.46\textwidth]{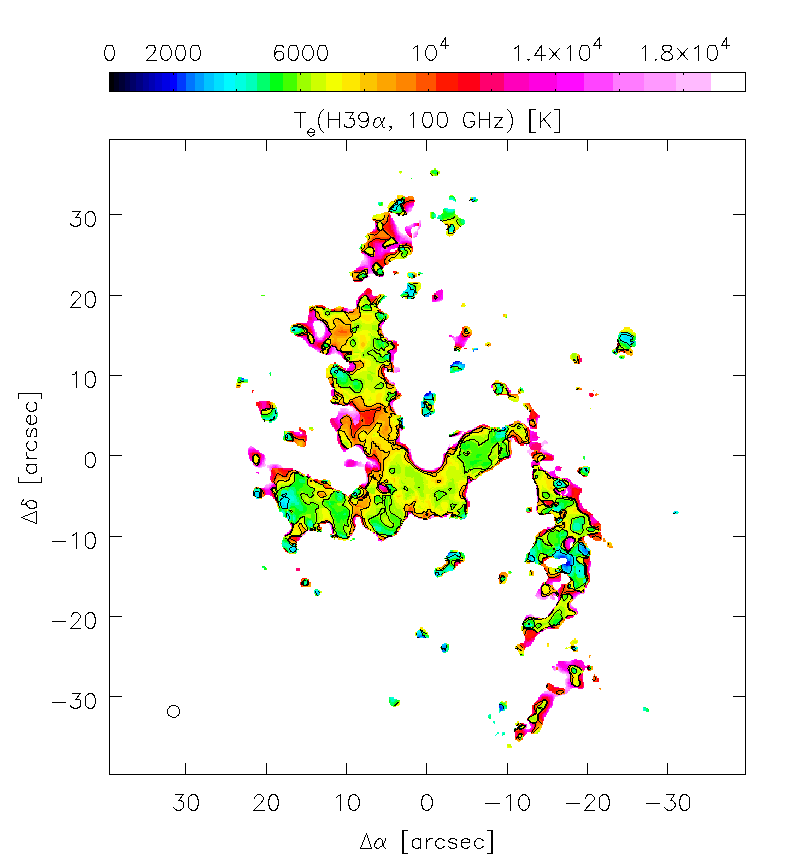}

        \includegraphics[trim = 5mm 0mm 7mm 6mm, clip, width=0.46\textwidth]{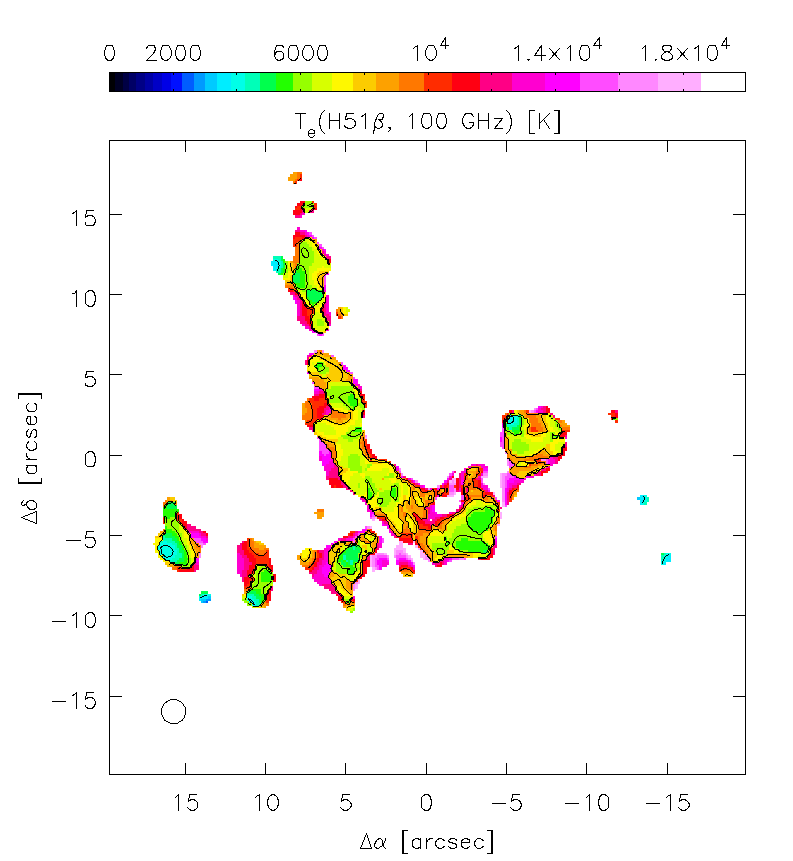}
        \caption{Images of the electron temperature distribution in the inner $\lesssim $ 1.6 pc.
                Top: Based on H39$\alpha$ (inner 40$''$) and tapered to a resolution of 1.5$''$ with contours of [4, 6, 8, 10] $\times$ 1000 K.
                Bottom: Same as top, but based on H51$\beta$ (inner 20$''$).
                Maps on the uncertainties can be found in Fig. \ref{Te-app}.
        }
        \label{Te}
\end{figure}

\subsection{Electron temperature}
\label{sec:Te}

The electron temperature can be used as a tool to investigate the strength and location of the ionization in the central region. This allows us then to speculate on the cause for the ionization.
Under the assumption of optically thin RRL and continuum emission and LTE conditions for the ionized gas, we use the formula given in \citet{Zhao2010} to derive the LTE electron temperature from the H39$\alpha$ and the continuum emission. 
To accomplish this, we clipped the tapered (same resolution) continuum map and RRL cube at 3$\sigma$ and 3$\sigma$, respectively, before primary beam correction. 
Since the line width in our data cannot be properly measured due to the low velocity resolution, we replaced $S_L \Delta V_{FWHM}$ with the integrated flux $\int S_L dV$. 
Moreover, we adjusted the correction factor for the power-law approximation to the value for 100 GHz, i.e., $\alpha(\nu=100$ GHz, $T_e \sim 10^4$ K)$ \sim $ 0.904 \citep{Mezger1967}. 
In Figure \ref{Te} we show that the LTE electron temperature is relatively uniform over the whole minispiral with $\sim$ 6000 K. In the NA and the central part of the bar it approaches 7000 K, around IRS 1W and 10W 9000 K, and drops in the western bar, the EA, and the WA to 5000 K. Extreme values at the edges of the minispiral are nonphysical, i.e., S/N related, and can be neglected. 
In general, the temperature range of 6000 $\pm$ 1000 K is in agreement with the results of H76$\alpha$, H92$\alpha$, and H30$\alpha$ observations by \citet{Schwarz1989}, \citet{Roberts1993}, \citet{Roberts1996}, and \citet{Zhao2010} reporting temperatures of $\sim$ 7000 K, especially in the arms. 
The rise in $T_e$ in the IRS 1W to 10W region in our data might result from an improper H39$\alpha$ flux detection due to the wide channels as mentioned in Section \ref{sec:RRL}. We derive $T_e$ from the H51$\beta$ line (Fig. \ref{Te}), which is the better detected of the two band 3 H$\beta$ lines, by scaling its flux to the H$\alpha$ emission using the LTE H$\alpha$/H$\beta$ ratio of 3.55 \citep[e.g.,][]{Gordon1990}; this yields $\sim$ 6000 K for this region and looks very similar to the H$\alpha$ based LTE $T_e$ in all other regions in the minispiral.

However, toward the center of the bar, i.e., the region between IRS 21, 16, and 33, \citet{Zhao2010} find temperatures well above 10000 K. There could be an indication for this behavior in the H51$\beta$ based temperature map, but the high temperature region does not reach up the NA as far as in the observations of \citet{Zhao2010}. 
Based on the $T_e$ (H51$\beta$) distribution, we attribute the higher temperatures to a low S/N effect; a similar behavior can also be seen toward the edges of the minispiral. 
In contrast to the H51$\beta$ data, our H39$\alpha$ data is sensitive enough to reliably detect the very broad (in velocity) but faint ionized gas component close to and in the minicavity \citep[see also spectra and line cube in][]{Roberts1996,Zhao2010,Irons2012} resulting in an electron temperature of 6500 $\pm$ 500 K throughout this region, which is consistent with the $\sim$ 6850 K found by \citet{Roberts1996}.

Owing to the moderate significance of our RRL line cubes (spectral resolution), we can only reason that our data reproduces the electron temperature well for regions with single velocity components, i.e., the arms, but deviates from former results in the crowded regions, where the minispiral arms and, potentially, the until now unidentified and unassigned velocity components overlap in projection.

\subsection{Stellar sources}
\label{sec:IRS}

In order to investigate the relation between the molecular gas, the ionized gas, and the stars, we compare the line centroids of these features (see Table \ref{IRS-vel-sum}). We find that even for a generous velocity offset of 70 km~s$^{-1}$ between two features, the list of matches is still not long. 
The line of sight (LOS) velocity of the RRL emission of the ionized gas agrees within the offset range with the stellar velocities for the IRS 1W, 1E, 7, 9SW, 9SE, 10W, 13W, 13E2, 16NE, 33N, 34E, 34NW, W11b, and W13b (see Figs. \ref{CND-sketch} and \ref{find-IRS-app}).
This agreement in velocity suggests that the stellar motion might be related to the gas motion, but one has to keep in mind that the stellar IR spectra might have been blended by the bright ionized gas emission in some cases, which then dominate the resulting spectrum.
In contrast to the RRL emission, the motion of the CS traced molecular gas agrees with stellar velocities only in the case of IRS 3E and W13b. Hence, an origin of the molecular gas in the stellar atmospheres appears unlikely, despite the large amount of O/B and Wolf-Rayet stars. However, the interaction of the winds from all the stars in the cluster might have dispersed the hint for a common velocity and origin. 
\\
As already indicated in Section \ref{sec:kin}, the molecular and ionized gas have common velocities in the NA/bar crossing region of the minispiral, i.e., in IRS 5, 6E, 16SSW, 16CC, 16SE2, 33SW, 33NW, 33N, W13b, W10b, W7b, and W14b from our selection. We revisit this coincidence in more detail in Section \ref{sec:Loc}.
In the following, we discuss the most famous IRS sources in the light of the emission lines.

\begin{table*}[ptb]
        \centering
        \caption{The gas radial velocities in RRL and CS emission at the position of the nuclear cluster stars, their stellar radial velocities and spectral type}
        \tabcolsep=0.05cm
        \tiny
        \begin{tabular*}{0.99\textwidth}{@{\extracolsep{\fill}} lccccc}
                \toprule
                sources                         & \multicolumn{3}{c}{gas radial velocity $v_g$}          &    stellar radial velocity $v_s$                                      & type            \\
                & \multicolumn{3}{c}{[km~s$^{-1}$]}                     &    [km~s$^{-1}$]                                                        &               \\
                & H39$\alpha$   &  H36$\beta$   & CS(5--4)              &                                                                         &               \\
                \midrule                                                                                                                                                                                                                                                                                                                                            
                IRS 1W          \tfm{P06}       & 32            & 5             & -                       & 35    \tfm{P06}                                                       & Be?             \\              
                IRS 1E          \tfm{P06}       & 34            & -             & -                       & 18    \tfm{P14}                                                       & B1-3 I  \\                      
                IRS 2L          \tfm{V06}       & -251, 22      & -284          & -                       & -                                                                     & -               \\              
                IRS 2S          \tfm{V06}       & -237          & -289          & -                       & 107   \tfm{G00}                                                       & L               \\
                IRS 3           \tfm{V06}       & -             & -             & 61                      & -                                                                     & E               \\
                IRS 3E          \tfm{P06}       & -             & -             & 63                      & 107   \tfm{P06}                                                       & WC5/6           \\              
                IRS 4           \tfm{V06}       & 175           & -             & -20                     & -                                                                     & -               \\              
                IRS 5           \tfm{V06}       & 125           & -             & 66                      & -                                                                     & -               \\              
                IRS 6W          \tfm{V06}       & -128          & -147          & -                       & -150  \tfm{G00}                                                       & E               \\
                IRS 6E          \tfm{V06}       & -134          & -67, 21       & 28, 99          & -                                                                     & -               \\
                IRS 7           \tfm{V06}       & -110          & -118          & -130 \tfm{SiO}, -53     & -114  \tfm{R07}                                                       & L               \\
                IRS 7E2(ESE)    \tfm{P06}       & 102           & -             & -                       & -80   \tfm{P06}                                                       & Ofpe/WN9        \\                      
                IRS 9           \tfm{V06}       & 173           & -             & -                       & -342  \tfm{R07}                                                       & -               \\              
                IRS 9N          \tfm{V06}       & 170           & 160           & -                       & -110  \tfm{G00}                                                       & -               \\              
                IRS 9W          \tfm{P06}       & -70, 300      & -             & -                       & 140   \tfm{P06}                                                       & WN8             \\      
                IRS 9SW         \tfm{P06}       & 202           & -             & -                       & 180   \tfm{P06}                                                       & WC9             \\      
                IRS 9SE         \tfm{P06}       & 198           & -             & -31                     & 130   \tfm{P06}                                                       & WC9             \\                      IRS 10W         \tfm{V06}       & 91              & 65            & -                     & 7     \tfm{G00}                                                       & -               \\              
                IRS 10EE        \tfm{V06}       & 102           & -             & -31 \tfm{SiO}           & -55   \tfm{G00},      -27     \tfm{R07,L10}                           & L               \\
                IRS 12N         \tfm{V06}       & -207          & -             & -                       & -96   \tfm{G00},      -63     \tfm{R07,L10}                           & L               \\
                IRS 13E         \tfm{V06}       & -200, -40     & -             & -                       & 45    \tfm{G00}                                                       & E               \\
                IRS 13W         \tfm{V06}       & -220, -30     & -             & -                       & -74   \tfm{G00}                                                       & L               \\              IRS 13N$\eta$   \tfm{V06}       & -40           & -33             & -                     & 40    \tfm{P06}                                                       & B V/III \\      
                IRS 13E1        \tfm{P06}       & -200, -30     & -45           & -                       & 71    \tfm{P06}                                                       & B0–1 I        \\                      
                IRS 13E4        \tfm{P06}       & -180, -30     & -             & -                       & 56    \tfm{P06}                                                       & WC9             \\              
                IRS 13E2        \tfm{P06}       & -200, -30     & -55           & -                       & -2    \tfm{P14}                                                       & WN8             \\              
                IRS 16NW        \tfm{P06}       & -             & -             & 46                      & -30   \tfm{G00},      -44     \tfm{P06},      17      \tfm{P14}       & Ofpe/WN9        \\                      
                IRS 16C         \tfm{P06}       & -             & -             & -41                     & 125   \tfm{P06},      186     \tfm{P14}                               & Ofpe/WN9        \\                      
                IRS 16SW        \tfm{P06}       & -130          & -155          & -36                     & 320   \tfm{P06},      460     \tfm{P14}                               & Ofpe/WN9        \\                      
                IRS 16SSW       \tfm{P06}       & -157          & -222, 37      & -36, 45         & 206   \tfm{P06},      221     \tfm{F15}                               & O8–9.5 I      \\                      
                IRS 16CC        \tfm{P06}       & -52           & -             & -41                     & 241   \tfm{P06},      145     \tfm{P14},      256     \tfm{F15}       & O9.5–B0.5 I   \\
                IRS 16NE        \tfm{P06}       & -19           & -             & -                       & 17    \tfm{G00},      -10     \tfm{P06},      53      \tfm{P14}       & Ofpe/WN9        \\
                IRS 16SSE2      \tfm{P06}       & -122          & -137          & -34                     & 286   \tfm{P06},                                                      & B0-0.5 I        \\      
                IRS 16SSE1      \tfm{P06}       & -117          & -133          & -33                     & 216   \tfm{P06},      229     \tfm{F15}                               & O8.5–9.5 I    \\      
                IRS 16SE1       \tfm{P06}       & -93           & -30           & -                       & 450   \tfm{G00},      366     \tfm{P06}                               & WC8/9           \\
                IRS 16S         \tfm{P06}       & -128          & -155          & -34                     & 100   \tfm{P06},      123     \tfm{P14},      149     \tfm{F15}       & B0.5–1 I      \\      
                IRS 16SE2       \tfm{P06}       & -40           & -72           & -38                     & 327   \tfm{P06}                                                       & WN5/6           \\
                IRS 16SE3       \tfm{P06}       & -30           & -54           & -                       & 281   \tfm{P06}                                                       & O8.5–9.5 I    \\                      
                IRS 17          \tfm{V06}       & -             & -             & 16                      & 185   \tfm{G00},      73      \tfm{R07}                               & L               \\
                IRS 20          \tfm{V06}       & -199          & -             & -                       & 17    \tfm{G00}                                                       & L               \\
                IRS 21          \tfm{V06}       & -84           & -96           & -9                      & -                                                                     & -               \\
                IRS 29NE1       \tfm{P06}       & -             & -             & 46                      & -130  \tfm{G00},      -99     \tfm{P14}                               & WC8/9           \\      
                IRS 29          \tfm{P06}       & -             & -             & 46                      & -190  \tfm{P06}                                                       & WC9             \\              
                IRS 33SW        \tfm{V06}       & -220, 29      & -             & -27, 138                & -                                                                     & -               \\
                IRS 33NW        \tfm{V06}       & -300, 18      & -312          & -30                     & -                                                                     & -               \\
                IRS 33N         \tfm{P06}       & -235, 19      & -295          & -29                     & 68    \tfm{P06},      93      \tfm{P14},      105     \tfm{F15}       & B0.5–1 I      \\      
                IRS 33E         \tfm{P06}       & -150          & -167          & -                       & 160   \tfm{G00},      170     \tfm{P06},      214     \tfm{P14}       & Ofpe/WN9        \\      
                IRS 34E         \tfm{P06}       & -200, -30     & -             & 63                      & -154  \tfm{P06}                                                       & O9–9.5 I      \\              
                IRS 34W         \tfm{P06}       & -180, -50     & -             & 40                      & -215  \tfm{G00},      -290    \tfm{P06},      -184    \tfm{P14}       & Ofpe/WN9        \\              
                IRS 34NW        \tfm{P06}       & -248          & -200          & 55                      & -150  \tfm{P06}                                                       & WN7             \\      
                AFNW            \tfm{P06}       & -             & 74            & -                       & 150   \tfm{G00},      70      \tfm{P06}                               & WN8             \\      
                AFNWNW          \tfm{P06}       & -             & 84            & -                       & 30    \tfm{P06}                                                       & WN7             \\              
                W11b            \tfm{P06}       & -295, 22      & -             & -34                     & -364  \tfm{P06}                                                       & OB              \\              
                W13b            \tfm{P06}       & -301, 20      & -             & -33                     & -24   \tfm{P06}                                                       & OB I?           \\              
                W10b            \tfm{P06}       & 27            & -             & -30                     & -434  \tfm{P06}                                                       & O8–9.5 III/I  \\                      
                W7b             \tfm{P06}       & 31            & -             & 79                      & -344  \tfm{P06}                                                       & O9–9.5 III?   \\                      
                W14b            \tfm{P06}       & 11            & -             & -26                     & -224  \tfm{P06}                                                       & O8.5–9.5 I?   \\                      
                B9b             \tfm{P06}       & 145           & 130           & 3                       & -150  \tfm{P06}                                                       & WC9             \\              
                \bottomrule                                                                      
                
        \end{tabular*}
        \tablefoot{Spectra were obtained from a beam sized aperture centered on the source position (see Table \ref{mol-IRS-vel}, and Fig. \ref{CND-sketch} and \ref{find-IRS-app} for details).
                References for the source positions and velocities are 
                \tablefoottext{P06}{\citet{Paumard2006},}
                \tablefoottext{V06}{\citet{Viehmann2006},}
                \tablefoottext{P14}{\citet{Pfuhl2014},}
                \tablefoottext{G00}{\citet{Genzel2000},}
                \tablefoottext{R07}{\citet{Reid2007},}
                \tablefoottext{L10}{\citet{Li2010},} and
                \tablefoottext{F15}{\citet{FeldmeierK2015}}.
                The spectral type is taken from \citet{Paumard2006}, except from the classes E and L that are denoting the early-type and late-type stars as given in \citet{Genzel2000}.
                \tablefoottext{SiO}{velocities from SiO(6--5) emission of SiO maser stars.}
        }
        \label{IRS-vel-sum}
\end{table*}

\subsubsection{IRS 1W}
In the NIR, IRS 1W is a dust embedded bowshock star \citep{SanchezB2014} at the lower tip of the minispiral NA close to the location at which it touches the minispiral bar. It shows spectral line features in CS(5-4) and RRL at its radial velocity of 20 km~s$^{-1}$.
The line maps show two CS(5-4) emission clumps to northwest and southeast of the source position.
The overall velocity pattern follows that of the NA. The gas has a velocity of 0 km~s$^{-1}$ to 15 km~s$^{-1}$ southeast of the source, then splits to a northwest and southeast component in the 15 km~s$^{-1}$ to 30 km~s$^{-1}$ interval, leaving out the exact position of IRS 1W, to finally present itself as a northwest component in the 30 km~s$^{-1}$ to 45 km~s$^{-1}$ interval. The reason for this behavior might be a combination of absorption and excitation.

The 340 GHz and 250 GHz continuum emission peak on the bright minispiral ridge about 0.2$''$ to 0.3$''$ southwest of the stellar source position at 8 mJy~beam$^{-1}$ and $\sim$13 mJy~beam$^{-1}$, respectively. As the compact source blends into the minispiral in the lower resolution 100 GHz map, IRS 1W is bright with peak emission of $\sim$34 mJy~beam$^{-1}$ and more extended flux density toward the southwest along 0.5$''$ to 1$''$.

\subsubsection{IRS 2L}
The RRL velocity at IRS 2L is about -270 km~s$^{-1}$ \citep[e.g.,][]{Zhao2010} in consistency with a H36$\beta$ line peak at -280 km~s$^{-1}$. 
The source is in a void of molecular gas emission. There is no clear line identification right at the position of IRS 2L. RRL line flux from material north of IRS 2L (see IRS 13E) is smeared into the aperture. At 340 GHz and 250 GHz, there is a continuum flux peak at the position of IRS 2L with flux densities of 19 mJy~beam$^{-1}$ and 27 mJy~beam$^{-1}$, respectively. At 100 GHz, we find a continuum flux of about $\sim$100 mJy~beam$^{-1}$ toward IRS 2L.

\subsubsection{IRS 3}
This is a dust-enshrouded star. Very Large Telescope Interferometer (VLTI) observations in the NIR have shown that it has an extended shell that contains also a central compact source \citep{Pott20083}. Its radial velocity is not known. Toward the position of the central star the ALMA spectroscopy data show a CS(5--4) and a C$_2$H(3--2) line at 60 km~s$^{-1}$ as well as somewhat less well-defined RRL at 80 km~s$^{-1}$. The line maps reveal that the CS(5--4) and C$_2$H(3--2) line detections are associated with a bright bar extending about 1$''$ to the north and south with line flux toward IRS 7. 
In addition, we find a mysterious clumps 2$''$ west of IRS 3 extending 2$''$ in north-south direction. It appears as a northern extension of northwest tip of the SE-NW cloud in CS(5--4) and H$^{13}$CO$^+$(3--2) emission at around 50 to 80 km~s$^{-1}$. It is the brightest source in CH$_3$OH(7--6) and HC$_3$N(27--26) emission in the CA.
The RRL emission of the H36$\beta$ line represents itself as a bright bar extending about 1$''$ to southwest with a velocity of about -200 km~s$^{-1}$. In the continuum, we detect a weak point source with a 340 GHz peak flux density of 1.4 mJy~beam$^{-1}$. There is also low level extended continuum emission to the south and west. At 250 GHz, there is a ridge of continuum emission with IRS 3 and IRS 7 at the southwest and northeast tips. The ridge has a peak brightness of 1.5 mJy~beam$^{-1}$ with a $\sim$ 2 mJy~beam$^{-1}$ (peak) unresolved source at the position of IRS 3. At 100 GHz, the entire continuum structure is speared out.

\subsubsection{IRS 5}
The RRL radial velocity at the dust embedded bowshock star IRS 5 is 110 km~s$^{-1}$ \citep[e.g.,][]{Zhao2010,SanchezB2014}. It shows no strong emission line that could be attributed to the source. Line emission of the NA occurs about 1.3" to the west. No continuum emission can be attributed to the source at 340 GHz, 250 GHz, and 100 GHz. Only two sources to the southeast, i.e., IRS 5S and 5SE1, have counterparts in the continuum emission (see Table \ref{cont-src}).

\subsubsection{IRS 6}
The RRL emission of IRS 6E and IRS 6W occurs at velocities of about -120 km~s$^{-1}$. The RRL and continuum emission avoid the IRS sources in that region and fill the gap between them as a filaments \citep[K13, K14, K17, K19;][]{Zhao2009}. The CS(5--4) emission passes IRS 6E 1$''$ north and extends 3$''$ to the southwest. At the latter position, also SiO(6--5), SO(7--6), and  CH$_3$OH(7--6) peak locally.

\subsubsection{IRS 7}
At a radial velocity of about -120 km~s$^{-1}$ IRS 7 shows spatially unresolved maser emission in SiO(6--5), and a bit stronger in SiO(8--7) \citep[compare with][]{Reid2007}. This corresponds to the RRL velocities given the channel sizes. The CS(5--4) line peaks around -50 km~s$^{-1}$ and +50 km~s$^{-1}$. 
C$_2$H(3--2) shows a behavior that is very similar to CS(5--4). Maps of the CS(5--4) line show that the source is part of a NS ridge that peaks in molecular line emission about 0.5$''$ south-southeast of the source. The C$_2$H(3--2) line emission is unresolved with an offset if about 0.2$''$ to the south.
In the 340 GHz and 250 GHz continuum, we have a point source right at the source position with peak flux densities of about 2 mJy~beam$^{-1}$ and 2.8 mJy~beam$^{-1}$, respectively. In the 100 GHz continuum, we confirm a 1.5$''$ to 2$''$ long tail to the north \citep{Serabyn1991,YZ1991,Zhao2009}. This extended component peaks at 0.65" north of the stellar position. At the position of IRS 7, we find a peak flux density of 1 mJy~beam$^{-1}$ and on the tail of about 1.6 mJy~beam$^{-1}$.

\subsubsection{IRS 9}
The stellar radial velocity for IRS 9 is -340 km~s$^{-1}$ \citep[][]{Reid2007}. Toward its position, there is weak CS(5-4) line emission at 0 km~s$^{-1}$. Otherwise there is no conspicuous line emission to be detected neither on IRS 9 nor on the NIR source 0.5$''$ to the north. 
On IRS 9 itself, there is no continuum detection above 1 mJy~beam$^{-1}$ at 100 GHz to 340 GHz. The dusty source IRS 9N \citep[X5 in ][]{Muzic2007} 0.3$''$ to the north shows a faint 2.5 mJy~beam$^{-1}$ emission peak at 340 GHz and $\sim$5 mJy~beam$^{-1}$ at 250 GHz.

\subsubsection{IRS 10}
For IRS 10W, the radial stellar velocity is about 10 km~s$^{-1}$. We find emission in the RRL H36$\beta$ line at about 70 km~s$^{-1}$.
In the 340 GHz and 250 GHz continuum, it is a marginally extended source with total fluxes of $\sim$23 mJy and peak flux brightnesses of $\sim$7 mJy~beam$^{-1}$ and  $\sim$8 mJy~beam$^{-1}$, respectively. In the lower resolution 100 GHz map, the continuum emission blends in with the minispiral flux with a peak brightness of $\sim$50 mJy~beam$^{-1}$.

IRS 10EE is bright in the masing SiO lines at about -30 km~s$^{-1}$ \citep[compare with][]{Reid2007,Li2010}. For IRS 10EE, we also find a bright compact component in the H36$\beta$ emission
at 70 km~s$^{-1}$ within the stream of minispiral NA. There are no other lines and no continuum emission detected for IRS 10EE.

\subsubsection{IRS 12N}
The stellar radial velocity of the prominent NIR source IRS 12N lies at about -60 km~s$^{-1}$. In the ALMA data, it has no prominent submillimeter line or continuum emission.

\subsubsection{IRS 13E}
The stellar radial velocity of the compact 0.5$''$ diameter cluster \citep{Eckart2013} is about 45 km~s$^{-1}$ \citep{Paumard2006}. In the RRL H36$\beta$ line emission map at the position of IRS 13E, we find peaks at -190 km~s$^{-1}$ and -40 km~s$^{-1}$ both about 75 km~s$^{-1}$ broad. 
In line maps, the source is devoid of in molecular gas emission. The RRL H36$\beta$ is locally brightest between IRS 13E and IRS 2L. It peaks about 0.5$''$ north of IRS 2L at -300 km~s$^{-1}$ at the position of a faint NIR L-band source. We also find a component at -160 km~s$^{-1}$ peaking on IRS 13E and a component at -60 km~s$^{-1}$ and  -43 km~s$^{-1}$ in the general IRS 13E and IRS 13N region.
The behavior in the H39$\alpha$ RRL is consistent with the findings for the H36$\beta$ line except that the spatial resolution is lower.
The 340 GHz and 250 GHz continua peak on the center of IRS 13E with a peak flux of 23 mJy~beam$^{-1}$ and $\sim$30 mJy~beam$^{-1}$, respectively. In the lower angular resolution 100 GHz band, we find a peak brightness of about 110 mJy~beam$^{-1}$.

\subsubsection{IRS 13N}
IRS 13N \citep{Eckart2004,Eckart2013} is a 0.3$''$ diameter stellar cluster disjunct from IRS 13E at a IR radial velocity of about 40 km~s$^{-1}$ \citep{Paumard2006}. It is rather inconspicuous in its line emission.
There is a detection in the  CS(5-4) line at 90 km~s$^{-1}$ that is possibly not associated with the source.
In the H39$\alpha$ emission, we find possible extensions in line emission at velocities of 44 km~s$^{-1}$ and from -90 till -170 km~s$^{-1}$.
In the H36$\beta$ emission, the southern edge of the source is bright over a velocity range of -43 km~s$^{-1}$ till -60 km~s$^{-1}$.
In the 0.5$''$ resolution 340 GHz map, we find continuum flux peak on all 13N members at about 7.5 mJy~beam$^{-1}$ with some a diffuse component about 0.5$''$ to the northwest. 
In the 250 GHz data, the cluster has a point flux density of 18 mJy~beam$^{-1}$. In the lower resolution 100 GHz map, IRS 13N appears only as an extension of the brighter IRS 13E source.

\subsubsection{IRS 16NE}
The stellar radial velocity of IRS 16NE is 50 km~s$^{-1}$ \citep{Pfuhl2014}. At its position, we see in RRL the H33$\beta$ peaking at a velocity of -30 km~s$^{-1}$ and -80 km~s$^{-1}$ with a width of 100 km~s$^{-1}$. The source appears as part of the minispiral flow. The H36$\beta$ line emission peaks 0.5$''$ to the east at -80 km~s$^{-1}$.
There is RRL emission in the range of -100 km~s$^{-1}$ to +100 km~s$^{-1}$ east of source. We find a similar behavior for H33$\beta$ line except that the emission at -80 km~s$^{-1}$ peaks weakly upon source itself. In the velocity range -40 km~s$^{-1}$ until -10 km~s$^{-1}$, we find emission line flux 0.5$''$ east of the source. Other than that, IRS 16NE is clear of line emission in all line maps.
At 340 GHz and 250 GHz, IRS 16NE is a point source with $\sim$5 mJy~beam$^{-1}$ and $\sim$4 mJy~beam$^{-1}$. At 100 GHz, the source flux is smeared out and blends in with the
minispiral.

\subsubsection{IRS 16SW}
IRS 16SW has a radial velocity of 460 km~s$^{-1}$  \citep{Pfuhl2014}. The ALMA data indicates that spectra toward its position are free of strong emission lines except from faint CS(5-4) emission at -40 km~s$^{-1}$. The emission line map shows that at this velocity, there is a 0.5$''$ wide valley between two CS(5-4) line emitting regions. There is a possible continuum detection at 340 GHz and 250 GHz with fluxes of 1 mJy~beam$^{-1}$ and 2.5 mJy~beam$^{-1}$, respectively.

\subsubsection{IRS 21}
IRS 21 is a dust-enshrouded star \citep{SanchezB2014} located in the minispiral bar.
The ALMA data shows an emission peak on source in the H36$\beta$ line at -96 km~s$^{-1}$, which is consistent with the minispiral flow \citep[e.g.,][]{Zhao2010}. There is also RRL emission to north and south of the source.
Comparably extended CS(5-4) emission at the source position appears at velocities from -40 to 60 km~s$^{-1}$.
The  340 GHz and 250 GHz continuum peaks about 0.1$''$ to the northwest giving the appearance of a relatively isolated compact source on the minispiral ridge with a peak flux density 7  mJy~beam$^{-1}$ and 10 mJy~beam$^{-1}$, respectively. Because of the lower angular resolution at 100 GHz the source is smeared out and cannot easily be identified as a point source within the minispiral. The corresponding region appears to be brighter to the northwest of the IRS 21 source position.

\subsubsection{IRS 29}
IRS 29 and IRS 29NE have radial velocities of -190 km~s$^{-1}$ \citep{Paumard2006} and -100 km~s$^{-1}$ \citep{Pfuhl2014}, respectively. The ALMA spectra toward their position show CS(5-4) and C$_2$H(3-2) line emission at 50 km~s$^{-1}$. In fact, the molecular emission belongs to the clump between IRS 29 and IRS 29NE 0.4$''$ northeast of IRS 29. This clump is part of an emission line ridge (SE-NW cloud) south of IRS 3. There is a possible weak continuum detection at 340 GHz with a flux of $\le$2 mJy~beam$^{-1}$. At 250 GHz and 100 GHz, no continuum detection can be claimed.

\subsubsection{IRS 33}

There are a few NIR sources to the west and southwest of IRS 16SW, which are listed under IRS 33. Here, we consider IRS 33NW, N, (S)E, and SW.
Radial velocities are $\sim$100 km~s$^{-1}$ for 33N and 214 km~s$^{-1}$ for 33E \citep{Pfuhl2014,FeldmeierK2015}. The spectra toward IRS 33SE reveal no prominent emission lines. 
IRS 33SW, N, and NW show CS(5-4) line emission at -30 km~s$^{-1}$. 
The maps show that the CS(5-4) line emission toward IRS 33SW and NW lies at the bridge or tips of emission clumps not associated with the sources.
As part of the minispiral flow, hydrogen RRL emission passes 0.1$''$ south of IRS 33N at -330 km~s$^{-1}$. No clear continuum emission can be attributed to the IRS 33 sources. Extended continuum flux passes though the cluster probably as part of the minispiral.

\subsection{The molecular line ratios}
\label{sec:ratio}

Molecular line ratios allow us to probe the molecular excitation and abundance. 
In Table \ref{mol-rat} we list the molecular line ratios for different molecules for all regions identified in Sect. \ref{sec:specrat-how} (details on computation therein).
As an overview we roughly summarize the ratios by regions with similar ratios in Table \ref{ratio-sum} and sort them by the general trend of the CS ratios.

\begin{table*}[tb]
        \centering
        \caption{Molecular line ratio trends on larger regional scales}
        \tabcolsep=0.05cm
        \small
        \begin{tabular*}{0.995\textwidth}{@{\extracolsep{\fill}} lccccccccc}
                \toprule
                &  \multicolumn{9}{c}{Regions} \\
                \cmidrule{2-10}
                Ratio                                   & center                                & edge    & mHVC  & Triop                         & SEW                   & CND N\&E                & CND W         & V-cloud       & SEE   \\ 
                \midrule                                                                                                                                                                                                                                      
                CS / C$_2$H                             &  \qquad\;\, 5 - 7 \tfm{\, 5,12,20}      & 2 - 4         &  -            &  1 - 2                        &\qquad\:\! 5 - 7 \tfm{\, SEW2} &\qquad\;\; 1 - 3 \tfm{\, CND-10}   &  -            &  -              &  -    \\
                CS / SO                                 &  \quad\, 5 - 15 \tfm{\, 4}              & 2             &  2 - 3        &  \quad\:\! 2 - 4  \tfm{\, T1} &  5 - 8                         &  2 - 4                                &  -              &  -            &  -    \\
                CS / H$^{13}$CO$^+$                     &  \,\, 5 - 16                           & 5             &  -            &  2 - 4                         &  \; 8 - 14                    &  -                                    &  -              &  -            &  -    \\
                CS / SiO                                &  \,\, 5 - 17                           &  -            &  2            &  \quad\:\! 3 - 5  \tfm{\, T9} &  7 - 9                         &  2 - 5                                &  -              &  -            &  -    \\
                CS / HC$_3$N                            &  6 - 9                                 & 3             &  -            &  \quad\:\! 6 - 7  \tfm{\, T4} &  -                             &  -                                    &  -              &  -            &  -    \\
                CS / CH$_3$OH(7--6)                     &  \,\, 4 - 11                           & 4             &  -            &  -                            &  -                              &  -                                    &  -              &  -            &  -    \\
                C$_2$H / SO                             &  1.3 - 2.1                             & 1.2           &  -            &  1.5                          &  1.0 - 1.6                      &  -                                    &  -              &  -            &  -    \\
                C$_2$H / H$^{13}$CO$^+$                 &  \quad\!\! 1.8 - 3.0 \tfm{\, 3}        &  -            &  -            &  1.5                          &  2.1 - 2.6                      &  -                                    &  -              &  -            &  -    \\
                C$_2$H / SiO                            &  \quad\:\! 2.2 - 2.6 \tfm{\, 14}       &  -            &  -            &  -                            &  1.1 - 1.9                      &  -                                    &  -              &  -            &  -    \\
                SO / SiO                                &  1.0 - 2.4                             &  -            &  -            &  1.1                          &  1.2 - 1.7                      &  1.4                                  &  -              &  -            &  -    \\
                SO / H$^{13}$CO$^+$                     &  0.7 - 1.9                             &  -            &  -            &  1.2                          &  1.7 - 2.3                      &  -                                    &  -              &  -            &  -    \\
                SiO / H$^{13}$CO$^+$                    &  0.6 - 1.7                             &  -            &  -            &  1.2                          &  1.6                    &  -                                    &  -            &  -              &  -    \\
                H$^{13}$CO$^+$ / HC$_3$N                &  1.3 - 1.9                             &  -            &  -            &  1.6                          &  -                              &  -                                    &  -              &  -            &  -    \\
                H$^{13}$CO$^+$ / CH$_3$OH(7--6)         &  1.0 - 1.6                             & 0.7           &  -            &  -                            &  -                              &  -                                    &  -              &  -            &  -    \\
                CS / N$_2$H$^+$                         &  -                                    &  -      &  -            &  2 - 4                        &  -                            &  1 - 7                          &  2 - 8        &  -            &  -    \\
                CS / CH$_3$OH(8--7)                     &  -                                    &  -      &  -            &  4 - 8                        &  -                            &  8                              &  10 - 15      &  -            &  -    \\
                CS / $^{13}$CS                          &  -                                    &  -      &  -            &  5 - 7                        &  28 - 33                      &  4 - 8                          &  \; 8 - 13    &  -            &  -    \\
                N$_2$H$^+$ / H$^{13}$CO$^+$             &  -                                    &  -      &  -            &  0.6 - 1.3                    &  -                            &  1 - 2                          &  0.8          &  -            &  -    \\
                N$_2$H$^+$ / CH$_3$OH(8--7)             &  -                                    &  -      &  -            &  1.3 - 2.5                    &  -                            &  2.3                            &  1.3 - 2.1    &  1.0 - 1.7    &  4.0  \\                                                                                                                                                                                                                                                        
                N$_2$H$^+$ / $^{13}$CS                  &  -                                    &  -      &  -            &  2.3 - 2.5                    &  -                            &  1.9 - 3.2                              &  1.0 - 1.6    &  -            &  3.9    \\
                $^{13}$CS / CH$_3$OH(8--7)              &  -                                    &  -      &  -            &  0.8                          &  -                            &  -                                      &  1.1 - 1.7    &  -            &  -      \\
                \bottomrule
        \end{tabular*}
        \tablefoot{Ratios are obtained from a beam-sized aperture of 0.65$''$ centered onto the average position of all integrated emission line peaks in the corresponding region. For ratios with N$_2$H$^+$, CH$_3$OH(8--7), and $^{13}$CS a beam-sized aperture of 1.5$''$ is used. SiO refers to the J = 6--5 transition of SiO. The regions comprise the following clumps (see Fig. \ref{find-app}) listed in the Table \ref{sec:app-ratio} in the Appendix:
                center: the bulk of the CA, clumps 1 - 16, 20; edge: edge  of the CA, clumps  19, 21, 22; mHVC: high velocity molecular clouds, clumps 24 - 27; \textit{Triop}: clumps T1 - T11; SEW: clumps SEW1 - SEW4; CND N\&E: northern and eastern edge of the CND, clumps SEW8 - SEW13, CND1 - CND10; CND W: western CND, clumps CND-W1 - CND-W3; V-cloud: clumps V1 - V12; SEE: clumps SEB1 - SEB6.      
                \tablefoottext{annotations}{regions that do not fall into the range of ratios}
        }
        \label{ratio-sum}
\end{table*}

We find that all CS/X (X: any other observed molecule) ratios are significantly elevated in the center and the SEW clump, i.e., more than three times higher than in the \textit{triop} and single clumps further out of the center, which are rather related to the CND. 
A similar behavior can be seen in the CS(7-6)/HCN(4-3) intensity ratio map of \citet{Montero2009}: Around Sgr~A* the intensity ratio, and with it the luminosity ratio due to the similar frequencies, is about 10 times higher than in the CND.
The C$_2$H/X and SO/X listed in Table \ref{ratio-sum} appear to be marginally enhanced by a factor of 1.5 in the center and the SEW clump.
In ratios with C$_2$H, the C$_2$H flux might be overestimated in the channels -20 and 0 km~s$^{-1}$ because of strong side lobe artefacts. Furthermore, the deviation of the integrated flux ratios from the channel flux ratios is here about a factor of 2 since the integrated emission contains both C$_2$H fine structure lines.
The SiO(6--5)/H$^{13}$CO$^+$ ratio seems to remain constant, slightly decreasing toward the center. 

The reason for the above described behavior of the line ratios can be manifold.
The GC is a region of extreme conditions in terms of intense IR to UV radiation from the nuclear cluster of massive stars and the X-ray emission from a population of stellar remnants and the SMBH \citep[e.g.,][]{Serabyn1985,Krabbe1991,Baganoff2003,Perez2015,Mori2015}.
Furthermore, the molecular emission, maybe itself an infalling CND clump, is located in a turbulent region, where the three minispiral gas streamers meet and where stellar winds and gravitational shear impact the environment. Consequently, shocks and magnetic fields may also play an important role. 

In a scenario where the molecular gas is very close to the center, one can assume the winds from the stellar cluster to sweep away a large amount of the gas so that only the densest cores and a diffuse intercloud medium is left over. This affects the densities, temperatures, and overall chemistry in a complicated way.

\subsubsection{Excitation}

In the following we speculate on how the molecular line ratios and in particular the molecular excitation is linked to variations of the physical properties in the central region.

From the emission lines in band 6, H$^{13}$CO$^+$ is the easiest to excite with an upper state energy of $E_u/k_B$ $\sim$ 25 K and a critical density of $n_c$ $\sim$ 3 $\times ~ 10^6$ cm$^{-3}$. Despite having the same $E_u/k_B$, C$_2$H requires a density that is twice as high as for H$^{13}$CO$^+$ to thermalize. CS, the molecule with the strongest emission in the center, requires a slightly lower density but has an upper state temperature of 35 K. The transitions of SO and SiO correspond to energies of  $\sim$  45 K, where SO thermalizes at densities similar to the density of H$^{13}$CO$^+$ but SiO has one of the highest critical densities in band 6, i.e., 8 $\times ~ 10^6$ cm$^{-3}$. The next group of higher upper state energies comprise CH$_3$OH (around 70 and 80 K and densities of 0.6 - 1 $\times ~ 10^6$ cm$^{-3}$) and SiO (at 75 K with the highest critical density of the sample, i.e., 2 $\times ~ 10^7$ cm$^{-3}$). Such high excitation conditions obviously only occur in few regions. 
The molecule with the highest upper state energy and found in the same regions as the former is HC$_3$N with an energy of 165 K. 

Because of the similar critical densities of about 5 $\times ~ 10^6$ cm$^{-3}$ the excitation might be rather determined by the temperature for CS, C$_2$H, and H$^{13}$CO$^+$, which trace the overall emission. 
Indeed, non-LTE radiation transfer models of H$_3^+$ and CO in the cloud toward IRS 3 suggest a temperature and density of $T_\textrm{k}$ = (300 $\pm$ 50) K and $n_\textrm{H$_2$}$ $\geq$ $10^4$ cm$^{-3}$ \citep{Goto2014}, which is consistent with the estimate for a small filling factor ensemble of irradiated dense clumps and clouds of \citet{Goicoechea2013}. 
Minispiral dust temperatures of $T_\textrm{d}$ = 150 - 300 K obtained from MIR and far-infrared (FIR) dust observations \citep{Gezari1985,Cotera1999,Lau2013} give a lower limit for the kinetic temperature of the gas there. 
Compared to this, recent estimate ranges for CND clumps, especially in the SW lobe, are $T_\textrm{k}$ $\sim$ 100 - 500 K, $n_\textrm{H$_2$}$ $\sim$ $10^{4.5}$ - $10^{6.5}$ cm$^{-3}$ and $T_\textrm{d}$ = 50 - 90 K \citep{Mills2013,RT2012,Lau2013}.

CS can be abundant enough to be affected by radiative trapping so that it appears thermalized at far lower densities, i.e., a factor of 25 less than the critical density \citep{Shirley2015}. Considering abundance ratios found in the GC of SiO/H$^{13}$CO$^+$ $\sim$ 2-3 and CS/H$^{13}$CO$^+$ $\sim$ 25-70 \citep{AmoBaladron2011} H$^{13}$CO$^+$ is expected to be faint. SiO and SO might trace the denser cloud interior, but not only owing to excitation, as we discuss later in this section. 

Another effect that can skew the level population distribution of molecules is IR pumping, i.e., the excitation of the lowest vibrational states and a consequent decay to higher purely rotational levels than expected for a given collisional excitation rate. \citet{Carroll1981} found the CS molecule to be the most efficiently pumped species of those tested in their study, followed by SiO, and, somewhat less effectively, HCN.
They obtain a minimum dust temperature needed to turn on the CS IR pumping of $T \geq 114$ K, which results in a maximum distance from a dust-embedded star ($L \sim  10^5 L_\odot$) at which pumping still occurs of $r \lesssim  0.04$ pc $=$ 1$''$. 
The environment of Sgr~A* comprises several dust-embedded stars. In fact, the CS emission covers regions at and around IRS 1W, 3, 6, 7, 21, 29, which are known as dust-embedded sources with temperatures of T > 200 K \citep{Gezari1985,Tanner2002,Tanner2005,Moultaka2004,Viehmann2006,Pott20087,Pott20083}.

In addition, the 19$\mu$m/37$\mu$m color temperature map from \citet{Lau2013} shows dust temperatures of $T = 120 - 150$ K within r  $\sim  5''$ of the minispiral and $T = 105 - 135$ K in the region between IRS 3 and 7.
In the case that the CS is indeed engulfing the minispiral and the stars, the impact of IR pumping by hot dust on the CS line flux cannot be neglected.

Vibrationally excited HCN has already been detected in the dense, shielded region in the SW lobe \citep{Mills2013}, where a possible stellar heating origin is not discernible in the NIR HST-NICMOS images, which is most likely due to high extinction. Since the conditions in this clump are sufficient to excite the vibrational transition of HCN, the conditions in the center should be ideal. In fact, \citet{Goto2014} had difficulty reproducing the CO excitation ladder observed toward IRS 1W and IRS 3 by radiative transfer modeling and suspect the IR pumping to affect the level population.
Observations of vibrationally excited molecules (CS, HCN, SiO) in these regions could clarify the impact of this effect.

\subsubsection{Abundance} 

CS has been found to be the most abundant S-bearing molecule in edges of PDRs and diffuse ISM \citep[PDR shell tracer; e.g.,][]{LucasL2002,Goicoechea2006} because of the high abundance of ionized sulphur in the gas phase \citep[e.g.,][]{Lepp1988}. However, in dense clouds sulphur is locked up in the progenitor molecules depleted onto grains \citep{Charnley1997,vdT2003} and therefore the CS abundance is low \citep[e.g.,][]{Bergin2001,DiFrancesco2002}.
Chemical modeling by \citet{Benz2007} suggests that the presence of X-ray emission enhances the gas phase abundances of CS and SO.
However, neither SO appears to be significantly elevated in the center, nor does SiO appear to be elevated for the assumption of IR pumping conditions. 
Either X-rays and/or IR pumping do not play an important role or another effect depletes these molecules relative to CS.
Considering the photodissociation rates, it becomes evident that SiO and SO are very sensitive to UV radiation and are only outmatched by HC$_3$N.
On the other end of the ladder, H$^{13}$CO$^+$ is the most UV-resistant species, followed by C$_2$H and CS at a separation of < 2 magnitudes. CH$_3$OH lies in between the two extremes. Therefore, finding the emission of the UV-sensitive species rather localized at the CS peaks might not only relate to excitation effects but also to the shielding properties of the gas.
CS, SO, SiO, and CH$_3$OH are related to grains. The CH$_3$OH and the progenitor molecules of CS and SO are formed on grains, which often contain silicates (i.e., SiO). These molecules or their progenitors can be released to the gas phase at a molecule specific temperature (evaporation) or by shocks (sputtering). 
Therefore, while SiO is rather accepted to be a shock tracer, the other molecules can also indicate a warmer environment enhancing their abundances,  
which might be the case for the region west of IRS 3. This region shows strong HC$_3$N and CH$_3$OH emission and moderate H$^{13}$CO$^+$ emission, but faint CS emission. The lack of SiO in this obviously UV-shielded, warm region suggests a temperature-related excitation and/or abundance enhancement rather than a shock chemistry.

\section{Nature of the molecular gas}
\label{sec:molgas}

\begin{figure*}[tb]
        \centering 
        \includegraphics[trim = 0mm 0mm 0mm 0mm, clip, width=0.98\textwidth]{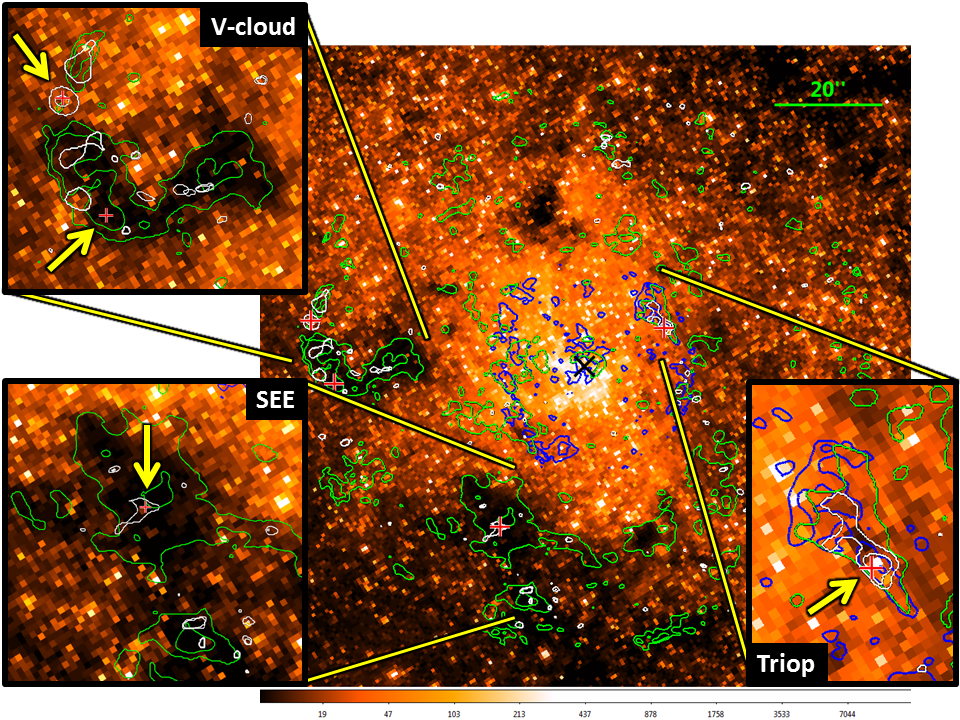}
        \caption{Schematic view on the central $\lesssim$ 5pc. N$_2$H$^+$(1--0) is indicated by green contours of [4, 32] $\times \sigma$ (= 0.07 Jy~beam$^{-1}$~km~s$^{-1}$), CS(5--4)
                shown as blue contours of [8, 48] $\times$ $\sigma$ (= 0.08 Jy~beam$^{-1}$~km~s$^{-1}$), and CH$_3$OH(8--7) indicated by white contours of [3, 6, 96] $\times$ $\sigma$ (= 0.06 Jy~beam$^{-1}$~km~s$^{-1}$) on a NIR HST NICMOS (1.87 $\mu$m) image (HST archive). 
                Red crosses in the \textit{triop}, SEE, and the V cloud show the class I methanol masers (see text) and the black cross Sgr~A*.}
        \label{dark}
\end{figure*}

\subsection{Infrared dark clouds and methanol masers}
\label{sec:dark}

When comparing to the NIR emission in Figure \ref{dark}, it is evident that the large N$_2$H$^+$ clouds, i.e., the V cloud and SEE, correspond to large dark clouds in the foreground of the center absorbing the NIR radiation. In fact, this is not unexpected; N$_2$H$^+$ is mainly destroyed by reactions with CO, which is depleted onto dust grains at temperatures of $T_\textrm{k} \lesssim 25$ K \citep{Vasyunina2012} such as those found in IRDCs. The absence of relatively strong emission from other molecules suggests a similar fate for them as for CO. Therefore, these clumps seem to be much cooler than the CND. A lower temperature limit is given by the freezeout of N$_2$ at $T_\textrm{k} \sim 15$ K \citep{Vasyunina2012} prohibiting the formation of N$_2$H$^+$ from it.
Nevertheless, the connection between dark clouds and N$_2$H$^+$ emission is not universal as is seen in -- among many more -- for example, the region 10$''$ east of Sgr~A*, where there is N$_2$H$^+$, but no NIR absorbing dust cloud, and in the dark clumps $\sim 30''$ north of Sgr~A*, which do not show N$_2$H$^+$ emission. Obviously, the excitation and abundance evolution are more complicated here. \citet{Martin2012} find the velocity centroids of the V cloud to be at $\sim 0$ km~s$^{-1}$ in the west and $\sim 60$ km~s$^{-1}$ in the east. The velocity centroid of SEE is stated to be at $\sim 10$ km~s$^{-1}$ and the line FWHMs are typically around $\sim 20$ km~s$^{-1}$. This is in agreement with our low spectral resolution data. 
The proximity of these two clouds in velocity and space to the $20$ km~s$^{-1}$ and $50$ km~s$^{-1}$ GMCs in the southeast and east of the GC rather suggests a relation to them than to the CND, where N$_2$H$^+$ is comparably faint. 

These dark clouds are discussed above along with the \textit{triop}, which actually contains a tiny dark cloud, and are also sources of class I CH$_3$OH maser emission at 36 and 44 GHz (see Section \ref{sec:outer}) and of CH$_3$OH(8--7) emission. The latter transition can also be excited to amplified stimulated emission (95 GHz class I maser, \citep[e.g.,][and references therein]{Voronkov2012}. Indeed, the CH$_3$OH(8--7) emission peaks at three of the four masers, i.e., the two 44 GHz masers in the eastern edge of the V cloud and in the \textit{triop} and the 36 GHz maser in SEE, where the latter two are much fainter compared to the first and surrounded by extended emission. Because of the extremely narrow line widths of $ \sim 1 - 2 $ km~s$^{-1}$ \citep{YZ2008,Sjouwerman2010,Pihlstrom2011}, the maser nature of CH$_3$OH(8--7) line emission in these peaks cannot be assessed with our observations, but the very bright and point source-like emission of the first peak strongly implies an amplification by a maser process.

Class I methanol masers are believed to be collisionally pumped and hence tracing shocks, while Class II methanol masers are pumped by FIR-radiation relating them to star-forming regions.
The 36, 44, and 95 GHz class I masers are often found together, suggesting the same underlying material and conditions, where the 95 GHz line emission is intrinsically fainter \citep[e.g.,][]{Fontani2010,McEwen2014,Kang2015}.
Modeling of the methanol maser transitions results in the favored conditions for 36 GHz and 44 GHz maser line emission overlapping with $ T_\textrm{k} \gtrsim 50$ K and $n \sim 10^{5} - 10^{6}$ cm$^{-1}$, where in star-forming regions the 36 GHz masers occurs in less dense (and cooler gas) than the 44 GHz masers and vice versa for supernova remnants \citep[SNR;][]{Pratap2008,McEwen2014}.
On the one hand, the eastern edge of the V cloud and the SEE region appear not only cold and dense enough to form stars, but they are also cospatial with bright SiO emission, that is only surpassed in the southern lobe, which, among others, could imply an interaction of the SNR Sgr A East with the $20$ km~s$^{-1}$ and $50$ km~s$^{-1}$ - GMCs producing large-scale shocks. \citep{Sato2008,Martin2012,Moser2014,Moserinprep}. On the other hand, shocks from outflows from YSOs are also able to excite the maser.
In fact, the 44 GHz maser in the \textit{triop} and at the prominent CH$_3$OH(8--7) peak in the V cloud as well as the 36 GHz maser in SEE are located next ($r < 0.5''$) to a star or weak NIR sources in projection. Consequently, more observations are needed to understand and disentangle the conditions and their causes in these mixed regions better.

\begin{figure}[tb]
        \centering
        \includegraphics[trim = 35mm 0mm 35mm 0mm, clip, width=0.45\textwidth]{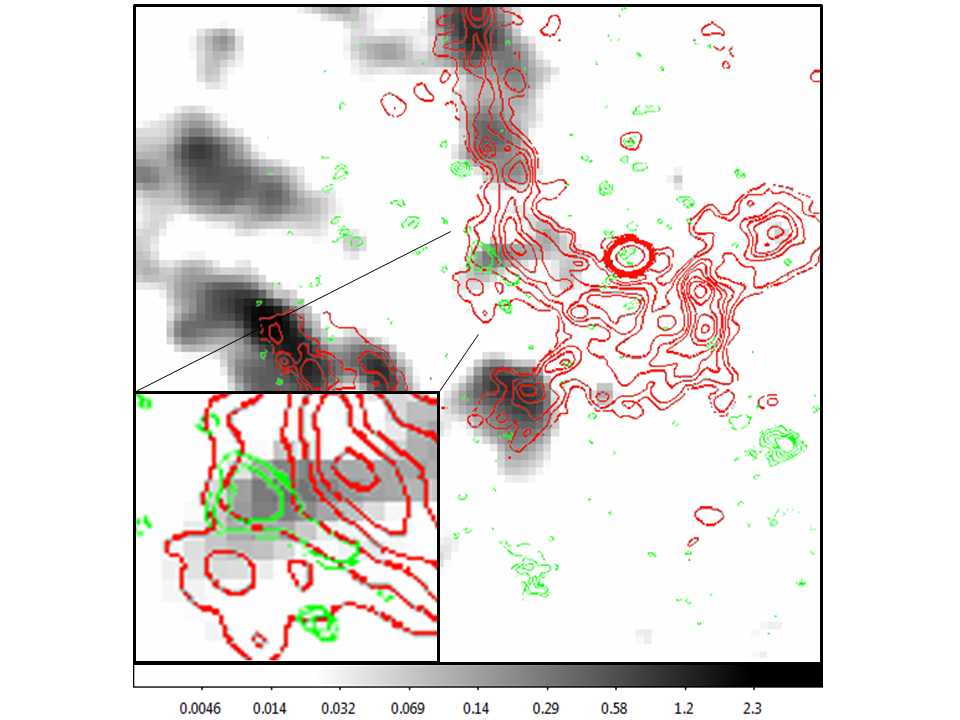}
        \caption{130 km~s$^{-1}$ channel image of the Ne II line emission cube from \citet{Irons2012}. 
                Ionized emission appears at the high velocity SiO cloud YZ1-2 southwest of IRS 1W (arbitrary units). SiO(6--5) emission is shown in green contours as in Fig. \ref{mols-full1} and the 250 GHz continuum in red contours as in Fig. \ref{cont-full}.
        }
        \label{HVC-SiO}
\end{figure}

\subsection{High velocity clouds}
\label{sec:HVC}

Here we consider clouds as high velocity clouds that move faster than the typical ranges for molecular gas in the CND, i.e., -100 to 100 km~s$^{-1}$ (see Fig. \ref{mol-vel-ch1} and \ref{mom1}).    

Based on SiO(5--4) emission, \citet[][]{YZ2013} measure a line center and width of $v \sim$ 148 km~s$^{-1}$ and FWHM $\sim$ 47 km~s$^{-1}$ for clump YZ1-2 and $v \sim$ 136 km~s$^{-1}$ and FWHM $\sim$ 56 km~s$^{-1}$ for clump YZ1. They interpret these properties and their line transition modeling results of $n_\textrm{H$_2$} \sim 10^5 - 10^6$ cm$^{-3}$ and $T_\textrm{k} \sim 100 - 200$ K as an indication of highly embedded protostellar outflows.  

In contrast to the earlier data, the SiO clumps YZ1-2 and YZ11 are partly resolved in this observation showing an elongated structure pointing at each other (see Fig. \ref{NIR-MIR-CS-350}). 
On the one hand, the configuration is reminiscent of a double lobed outflow from a hypothetical source in the middle of the two SiO clumps. But such a configuration can be ruled out owing to the similar velocities. Instead, it could be indeed possible that we are seeing the single-sided lobes from two independent outflow sources as suggested by \citet[][]{YZ2013}.

On the other hand, YZ1-2 looks especially like a boundary due to its elongated or filamentary shape closely following the NA streamer. In fact, it is located in the region where the EB, i.e., the minispiral component bridging the NA and the EA, overlaps with the NA and their proper motions are opposite each other: The NA has an x- and y-velocity components of $v_\alpha \sim$ -188 km~s$^{-1}$ and $v_\delta \sim$ -560 km~s$^{-1}$ and the EB shows $v_\alpha \sim$ 32 km~s$^{-1}$ and $v_\delta \sim$ 360 km~s$^{-1}$ \citep[clumps X13 and K37, respectively, in][]{Zhao2009}. In addition, the EB covers radial velocities of $\sim$ 40 - 140 km~s$^{-1}$ and shows a velocity component at 130 km~s$^{-1}$ at the discussed position, whereas the NA moves with $\sim$ 20 km~s$^{-1}$ in this region (Fig. \ref{HVC-SiO}). 
The main filament in the EB, apparently connecting the tip with the IRS 1W region and containing the K37 \citep{Zhao2009} component, is about as thick as the YZ1-2 is long. While a head-on collision of the NA streamer and the EB filament would surely destroy the dust grains and molecules at such high velocities, scraping past each other could generate shock velocities in the layers further inward of the clouds that would be ideal for grain sputtering and the release or production of gas phase SiO. 

Apart from the known SiO clumps YZ1-2 and YZ11, we discover three more clumps in CS emission at similar velocities (see Fig. \ref{mol-vel-ch1} and \ref{mom1}). 
One of these clumps seems to be associated with YZ1-2, showing faint CS emission at the location of the SiO peak and slightly brighter CS emission 5$''$ northeast of it, cospatial with faint SiO emission in the same channels. This extension of YZ1-2 to the northeast parallel to the elongation of YZ1-2 can also be seen in the SiO data of \citet[][]{YZ2013}.
The other two CS clumps, one between the tip and the minicavity, the other north of the IRS 6 region in the western bar, are elongated and in line with each other. The first clump overlaps in space and velocity with the ionized gas emission, while the second appears as unassociated as the YZ11 cloud, where no other emission is detected in the corresponding channels. The occurrence of several clumps with similar velocities and FWHM within a radius of 10$''$ suggests a relation between them that possibly even includes the SiO clumps. They could be the denser leftovers of a tidally disrupted cloud that is falling toward the center similar to what has been proposed for the high negative velocity cloud around -180 km~s$^{-1}$ detected in, for example, OH absorption and NH$_3$ emission \citep{Zhao1995,Karlsson2003,Donovan2006}. In contrast to the latter, large-scale emission of the 150 km~s$^{-1}$ clumps has not been detected in previous observations \citep{Wright2001,Martin2012} except from a tentative feature in the channel maps of \citet{Christopher2005}.

\subsection{Origin of the molecular gas in the central 20 arcseconds}
\label{sec:Loc}

\begin{figure}[tb]
        \centering
        \includegraphics[trim = 5mm 15mm 3mm 20mm, clip, width=0.49\textwidth]{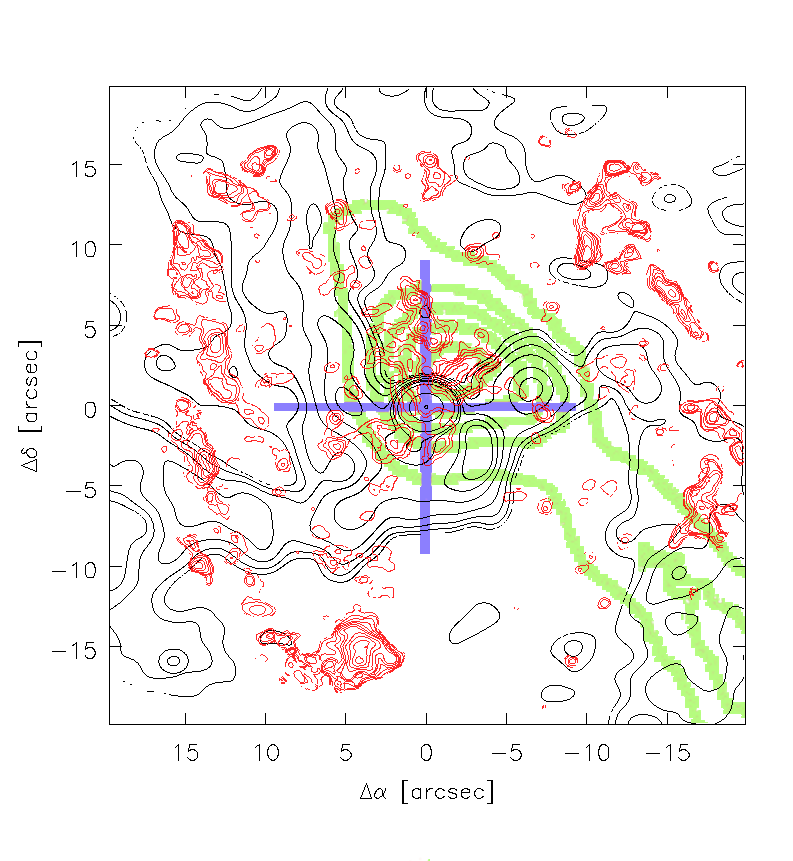}
        \caption{Schematic view on the OH streamer in OH emission at 50 km~s$^{-1}$ in green contours with lowest level at$\sim$ 90 mJy~beam$^{-1}$ ($\sim 3.5\sigma$) and steps of $1\sigma$ from \citet{Karlsson2015}. Black contours show the 100 GHz continuum emission as in Fig. \ref{cont-full}, red contours indicate the CS(5--4) emission as in Fig. \ref{mols-full1},
                and the blue cross indicates Sgr~A*.
                 The shape of the streamer head resembles the distribution of the CS clumps.
        }
        \label{OH}
\end{figure}

In the following we discuss the nature and possible origin of the CA gas in light of mainly two hypotheses: the CA gas may be linked to the triop or to the OH streamer (Fig. \ref{OH}).

\textit{General considerations on the CA:}
The presence of molecular gas in the central parsec has been indicated by several observations in the past. 
Molecular gas is expected to be dissociated into ionized elements if the conditions are as harsh as they are found in the very center close to the super massive black hole Sgr~A* and well within the cluster of young and hot He-stars. However, the presence of molecular gas or even ice in the central parsec has been indicated by several observations in the past. The reason for this lies in the crossing timescale of freshly infalling or orbiting gas streams through the central stellar cluster. This timescale is short compared to the dissociation timescale in the presence of shielding and the cloud clump evaporation timescale \citep[discussion in][]{Moultaka2004,Moultaka2005,Moultaka2015H2O,Moultaka2015CO}.

\textit{Summary of previous findings:}
\citet{Geballe1989} discovered strong CO absorption toward IRS 3 and 7 and even water and CO ice absorption is detected in the minispiral and toward the mass losing stars \citep{Moultaka2004,Moultaka2005,Moultaka2015H2O,Moultaka2015CO}. 
Furthermore, the presence of molecular gas at and around Sgr~A* is evident in the HCN(4--3) and CS(7--6) maps of \citet{Montero2009} and the CN map of \citet{Martin2012}.
From the up to now largest line study of the cavity, comprising CO transition lines from J=4--3 to J=24--23, \citet{Goicoechea2013} inferred that the ISM can be described either by a single, hot ($T_\textrm{k}$ > 1000 K), low-density ($n_\textrm{H$_2$}$ $\leq$ $10^4$ cm$^{-3}$) component or by multiple rather compact components at a lower temperature and higher density. 
The ALMA data discussed here supports the latter case. In addition, hot gas in the CA has been detected in several H$_2$ transitions in the NIR whose level populations suggest a strong impact by UV radiation \citep[i.e., UV pumping and dissociation,][]{Ciurlo2016}.

Recent NIR and MIR data on the distribution of H$_2$O ice, hydrocarbons, CO ice, and gaseous CO reveal that molecular gas and ices are ubiquitous in the whole central minispiral and around IRS 3, 7, and 29 \citep{Moultaka2015H2O,Moultaka2015CO}. The H$_2$O ice absorption extends over the whole minispiral, whereas the hydrocarbons and gaseous CO absorption are rather restricted to the dust filaments and the stars in the case of CO, but the optical depth of all three species is large around IRS 3, 7, 29, and the western part of the bar. Solid CO, which requires even lower temperatures \citep[T < 25 K,][]{Vasyunina2012} than H$_2$O ice to exist, shows strong absorption only in the dust filaments between IRS 1W and the minicavity, IRS 21 and in the IRS 2L region, and faint absorption toward IRS 3, 7, and 29.

The strong absorption features of molecular gas and ice toward IRS 3, 7, and 29 and the western bar are accompanied by strong IR absorption features of dust grains: 
\citet{Pott20083,Pott20087} detected silicate absorption in the spectra of IRS 3 and IRS 7 and they point out that the dust is of interstellar nature and is not related to the circumstellar regions of IRS 3 and IRS 7. 
In fact, there is a silicate dust veil extending between IRS 7 over IRS 3 and IRS 29 across the western part of the bar \citep{Viehmann2006}. 
This dust could be a likely reservoir for SiO molecules in the gas phase when these molecules are released into it via shocks, evaporation, UV, or X-rays.
The CS emission at positive velocities covers the IR absorption regions at IRS 3, 7, and 29, and partly the western Bar.
Although they are strongly mass-losing stars, the shifts in the peaks of the distributions of gas and ice and the location of IRS 3 and IRS 7 suggest the IRS sources play a marginal role in the enrichment of the gas phase molecules. Only IRS 29 coincides with a CS peak. A relation with the silicate dust appears more likely, but needs to be probed.
The blueshifted CS emission is more puzzling; this emission coincides with the CO gas and ice absorption in the dusty region between IRS 1W, IRS 21, and the minicavity, but not in the IRS 2L region. On the one hand, the coincidence of the molecular gas emission with the RRL in projection and velocity, i.e., -40 to 0 km~s$^{-1}$, might indicate a common molecular cloud with a limited ionized surface or an interaction of a molecular gas streamer, not yet ionized, with the NA. On the other hand, the ambiguous behavior of CS emission compared to the IR ice and gas absorptions suggests that the features are rather not related.

\textit{The velocity range of the CA and its possible link to the triop: }From the spectroscopic view, \citet{Goto2014} reported strong absorption in the higher excitation lines of H$_3^+$ and CO in the NIR toward IRS 3 around 50 and 60 km~s$^{-1}$, respectively, and weaker absorption around 45 km~s$^{-1}$. In contrast to this, the spectrum of IRS 1W shows only the 45 km~s$^{-1}$ component. Similar results of 50 km~s$^{-1}$ were found by \citet{Geballe1989}, but at a much lower spectroscopic and spatial resolution. In their observed spectra, it becomes evident that the IRS 1 region covers a range of 0 - 40 km~s$^{-1}$, whereas the broad absorption between 0 and 60 km~s$^{-1}$ is not only seen in IRS 3, but also toward the IRS 6 region and IRS 7. This behavior is recognizable in the CS emission channel maps (Fig. \ref{mol-vel-ch1}). \citet{Martin2012} report a velocity component at 46 km~s$^{-1}$ for clump 18, which covers the bulk of the positive velocity CS clumps between IRS 3, 7 and Sgr~A* and whose peak is almost cospatial with IRS 7. Furthermore, they determined a second component at -77 km~s$^{-1}$, which is most likely the same component detected in absorption at -72 km~s$^{-1}$ in the lower excitation CO emission towards IRS 1, IRS 3, and IRS 16NE \citep{Goto2014}. This is consistent with the ALMA data showing emission at or nearby these sources in the -80 to -40 km~s$^{-1}$ channels. 

The H$_3^+$ and CO absorption and the sub-mm molecular emission originate from the same cloud, which is likely located in front of or around IRS 3 and 7. Moreover, the silicate dust absorption could take place in the same cloud.
Its LOS position with respect to the minisprial remains unclear, since the LOS position of IRS 3 and 7 relative to minispiral is not clear. 
In the case that the molecular material is not a product of mass ejection from the evolved stars in the nuclear cluster, this clump might be linked to the CND. In fact, the velocity range matches the CND velocities in the \textit{triop} region, which appears at 40 - 100 km~s$^{-1}$ and peaks at 60 - 70 km~s$^{-1}$. In this way our data supports the suggestion of \citet{Goto2014} who considered the clump to be an extension of the CND.

\textit{Is the CA linked with the OH streamer?}
Another intriguing aspect is given by \citet{Karlsson2003,Karlsson2015}, who detect an additional streamer in OH absorption that has not been noticed in other molecular lines so far (Fig. \ref{OH}). 
The absorption feature is visible from -30 to 70 km~s$^{-1}$ and extends from the center, where it peaks northwest from Sgr~A* at 50 km~s$^{-1}$ ("head"), to the SW lobe and beyond, where it peaks at $\sim$ 70 km~s$^{-1}$ ("tail"). One can find faint hints of it in the CS(1--0) channel maps of \citet{Liu2012} and the denser parts are also detected as single clumps in, for example, CN \citep{Karlsson2015}. 
We see that the ALMA CS emission extends slightly across the western bar in the same direction as the streamer (Fig. \ref{OH}).
Between velocities of -30 to 15 km~s$^{-1}$ the OH absorption of the head is centered on Sgr~A*. The same is visible for the CS emission. The OH and CS detected gas also shows an apparent connection to the southeast around velocities of 0 km~s$^{-1}$.

\citet{Karlsson2015} constrain the streamer head to have a molecular gas mass of 65 $M_\odot$, based on its extent and the assumption of an ellipsoidal structure and a density of $n_\textrm{H$_2$}$ $\sim$ $10^5$ cm$^{-3}$. Applying the same method for all (deconvolved) clumps in the CA, we obtain a total gas mass of the CA of about 5 $M_\odot$. The results match well because the observed OH transition (1665 and 1667 MHz) is already sensitive to densities of $\sim$ 10$^2$ cm$^{-3}$ so that the adopted density is overestimated for the traced OH gas. A difference in the average densities traced by the OH transition and the transitions in our data of an order of magnitude is highly likely. This approach for a mass estimate is of limited accuracy since the density and structure in these clumps/cloud are unknown and therefore simplified. However, the assumption of virialization yields unplausible masses of several 10$^3$ $M_\odot$ per clump and few 10$^5$ $M_\odot$ for the whole CA. The masses of atomic hydrogen inside the cavity (dominated by the minispiral NA) range between 50 - 300 $M_\odot$ \citep[from FIR and \ion{O}{I} observations;][]{Telesco1996,Latvakoski1999,Jackson1993}, while the molecular gas mass is most likely only a fraction of this.

The OH streamer seems to be a viable explanation at least for the redshifted emission in the center. The observed OH transition traces the rather diffuse material, so that the tidally stretched tail is only tentatively detected by the high density tracers (e.g., HCN, HCO$^+$, CS) toward higher density clumps within the streamer, which then appear to be unrelated at first sight. 
The density of the head of the OH streamer seems to be higher than that of the tail gas. Either the head is the dense core of the initial cloud or the gas is compressed by stellar winds and  other gas streamers, or, most likely, it is a mixture of both. Several CS emission peaks within the region between IRS 3, IRS 7 and Sgr~A* are cospatial with SiO emission peaks, which could indicate shocks, although a significant impact of the X-ray and UV radiation field in the gas phase SiO cannot be ruled out. 

The blueshifted molecular gas in the center could be part of the OH-streamer orbit. Such a case raises several questions on the trajectory of the clump, the eccentricity, and the periapse distance of the orbit, and intimately related to this, the excitation and lack of ionization.
Furthermore, the streamer could interact with the NA, which could entrain the streamer toward the center.
On the one hand, this would explain the coincidence in projected location and velocity. On the other hand, no blueshifted SiO emission is detected in the NA except from the SiO maser stars, but this could be due to an insufficient S/N. The integrated CS emission in the blueshifted peaks in the center is only half as bright as in the redshifted peaks. If the SiO emission scales accordingly, it might be too weak to be detected in this observation. In a case where the OH streamer is interacting with, but not entrained by the NA, the gas might be slowed down enough to form a disk feeding the central 10$''$ (0.4 pc) and might be even related to the counterclockwise disk \citep{Paumard2006,Bartko2009,Bartko2010,Lu2009}. 

In order to understand the dynamical processes and the relations between the sources in the center, proper motions are mandatory to complete the picture. Based on few data sets within a total time range of $\sim$ 15 yr, \citet{Zhao2009,Zhao2010} established a plausible 3D model of the minispiral streamers as Keplerian orbits around Sgr~A*. A similar campaign needs to be conducted for the molecular gas in the CND and the cavity.

\section{Summary}
\label{sec:sum}

The interferometer ALMA was used to observe Sgr~A*
and its environment at 100, 250, and 340 GHz with a spatial resolution of 0.5 - 1.5$''$ and a spectral resolution of 15-50 km~s$^{-1}$. We found, in addition to the continuum in the three bands, 11 molecular line transitions and 5 RRL transitions as well as many interesting regions and features. The main results are summarized in the following:

\paragraph{\bf Continuum and RRL emission:} 

The minspiral is well detected in its whole extent in the 100 GHz continuum and the H39$\alpha$ emission yielding an almost uniform electron temperature around $T_e \sim 6000$ K. The 250 and 340 GHz continuum emission trace the minispiral filaments and IRS sources at an up to now unprecedented resolution (< 0.75$''$) in the sub-mm domain. The spectral index of Sgr~A* is $\sim$ 0.5 at 100 - 250 GHz and $\sim$ 0.0 at 230 - 340 GHz. The compact and high S/N regions in the center show spectral indices around -0.1 implying Bremsstrahlung emission, while indications that dust emission gain importance for the continuum emission is seen in the NA and EA in the form of a positive spectral index.

\paragraph{\bf Molecular line emission:}

The most striking result from this data is the for the first time resolved view on sub-mm molecular line emission of the inner 20$''$. It shows a clumpy distribution in CS line emission with the bulk of the emission at positive velocities and in a region confined by the minispiral NA, bar, and the sources IRS 3 and 7. Although partly spatially overlapping with the RRL emission at the same negative velocities, the relation to the minispiral remains unclear. Other molecules, such as SO, SiO, and H$^{13}$CO$^+$, are confined to the CS emission peaks. It is possibly an infalling clump, as suggested by earlier OH observation, which might consist of denser cloud cores embedded in diffuse gas. 
The CA of clouds shows three times higher CS/X (X: any other observed molecule) luminosity ratios than the CND, suggesting a combination of higher excitation, by a temperature gradient and/or IR-pumping, and abundance enhancement due to UV and/or X-ray emission. We conclude that the association is closer to the center than the CND is.

We find emission at unexpectedly high velocities. 
Between 100 and 200 km~s$^{-1}$, we detect few SiO and CS clumps, where the first group has previously been interpreted as outflows from YSO. Nevertheless, the common velocity ranges around 150 km~s$^{-1}$ might also hint at a connection between SiO and the CS clumps, such as coming from another infalling tidally sheared cloud.

Moreover, we identify two further intriguing regions. One of them is the \textit{triop} in the northwest of the center and located in the CND. It is detected in almost all molecules observed previously and the ALMA observations presented here resolve its filamentary structure for the first time. The fact that it harbors a class I methanol maser and the detection of transitions with high upper state energies (e.g., HC$_3$N(26--25), $T_u \sim 160$ K) point at the possibility of hot core conditions and early stages of star formation. 

The other region is the northwestern tip of the CND SE. It shows higher CS/X ratios than the CND, but not as high as the center. The molecular species detected appear to form a layered structure perpendicular to sightline to the center. Seemingly, it becomes affected by the radiation and or the winds from the nuclear stellar cluster. 

Outside the CND, we find the traditionally quiescent gas tracer N$_2$H$^+$ coinciding with the largest IRDCs in the field. Methanol emission is found at and around previously detected methanol class I masers, which are assumed to be evoked by large-scale shocks from the Sgr A East shell. Since dark clouds are likely to contain prestellar cores, these clouds make another ideal candidate to investigate the earliest stages of star formation.

\paragraph{\bf Outlook:}

The observation yielded a collection of interesting regions with properties distinct from the majority of CND clouds, which deserve further investigation. 

Our data clearly show a trend of more extreme conditions for the molecular emission toward the center.
In order to constrain the density and temperature distribution, the excitation ladder of specific tracers (e.g., density) has to be probed by observations of several transitions per molecule spanning a wide range of J levels. In addition, the distribution of vibrationally excited CS, SiO, or HCN emission needs to be tested to infer the effect of IR pumping on the excitation. 
Observation of atomic line emission could mark the transition or link from molecular gas to ionized gas and  helps to understand the relations between the features. For all of this, a higher spectral resolution than given in this observation is mandatory. 
In the long term, a better constraint on the distance of the molecular and neutral clouds to Sgr~A* and their fate can be obtained when studying the proper motions. 

The presence of, for example, CH$_3$OH, HC$_3$N, and H$_2$CO in the warmer \textit{triop} and the colder IRDCs calls for observations of more complex hydrocarbons, and maybe deuterated species, which are characteristic for the temperature and density dependent core chemistry. 
The assessment of their properties might give hints about the open questions on (recent) star formation in this turbulent region. 

The observations discussed have already proven the high capabilities of ALMA in early science cycle 0 so that we can look forward to full array observations of the GC with ALMA.

\begin{acknowledgements}
        This paper is based on the following ALMA data: ADS/JAO.ALMA\#2011.0.00887.S. ALMA is a partnership of ESO (representing its member states), NSF (USA) and NINS (Japan), together with NRC (Canada) and NSC and ASIAA (Taiwan) and KASI (Republic of Korea), in cooperation with the Republic of Chile. The Joint ALMA Observatory is operated by ESO, AUI/NRAO, and NAOJ. Furthermore, the paper makes use of observations made with ESO Telescopes at the La Silla Paranal Observatory under programme IDs 089.B-0145 and 085.C-0047 and of observations made with the NASA/ESA Hubble Space Telescope, obtained from the data archive at the Space Telescope Science Institute. STScI is operated by the Association of Universities for Research in Astronomy, Inc. under NASA contract NAS 5-26555.
        This work was supported in part by the Deutsche Forschungsgemeinschaft (DFG) via the Cologne Bonn Graduate School (BCGS), the Max Planck Society through the International Max Planck Research School (IMPRS) for Astronomy and Astrophysics, as well as special funds through the University of Cologne and SFB 956 – Conditions and Impact of Star Formation. L. Moser is funded by the SFB 956. L. Moser, B. Shahzamanian, and A. Borkar are members of the IMPRS.  
\end{acknowledgements}

\bibliographystyle{aa} 
\bibliography{LMoser_GC-ALMA_final}                                                      


\begin{appendix} 

\section{Additional images}
\label{app:img}

\begin{figure*}[htbp]
        \centering $
        \begin{array}{cc}
        \includegraphics[trim = 0mm -5mm 0mm 0mm, clip, width=0.32\textwidth]{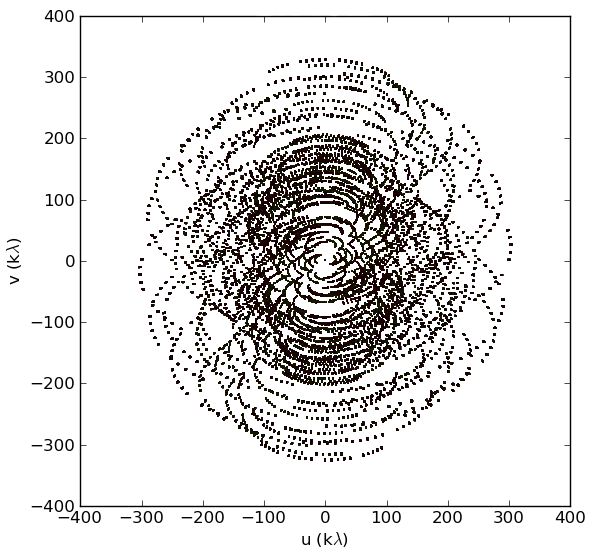}&
        \includegraphics[trim = 5mm 0mm 7mm 0mm, clip, width=0.31\textwidth]{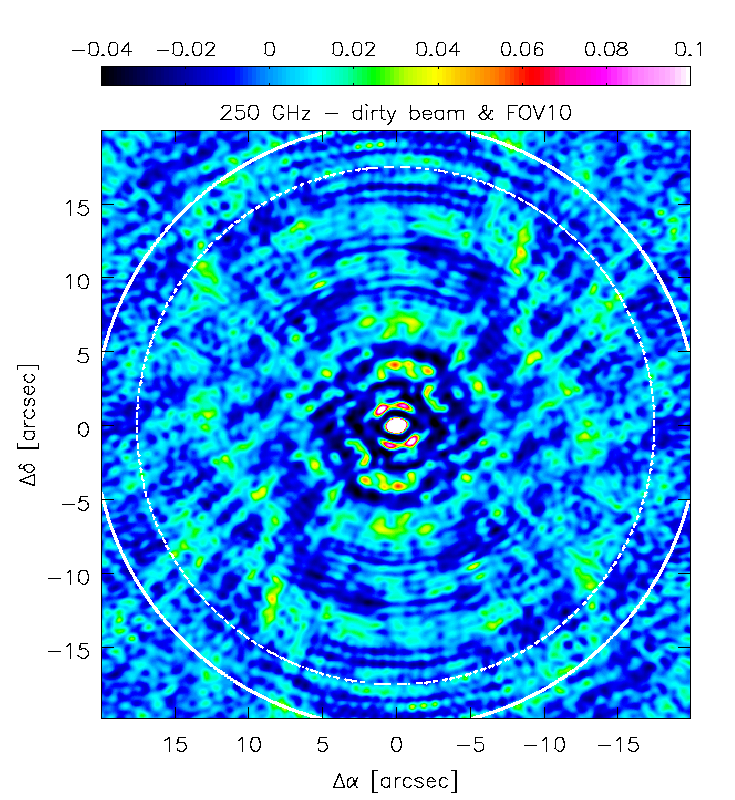}\\
        \end{array} $                                                            \caption{Left: $uv$ coverage in a single channel at 250 GHz. Right: Dirty beam for the 250 GHz continuum. Color scale shows the intensity or sensitivity normalized to 1.0 and is clipped to 0.1 to show the side lobe structure. First side lobe maximum and minimum are at 0.12 and -0.05, respectively, and located in the inner 5''. White contour outline the CS(5-4) (solid) and 250 GHz continuum (dashed) primary beam at 10\% sensitivity (FOV10).      }
        \label{psf-uv}
\end{figure*}

\begin{figure*}[htbp]
        \centering $
        \begin{array}{cc}
        \includegraphics[trim = 5mm 0mm 7mm 0mm, clip, width=0.31\textwidth]{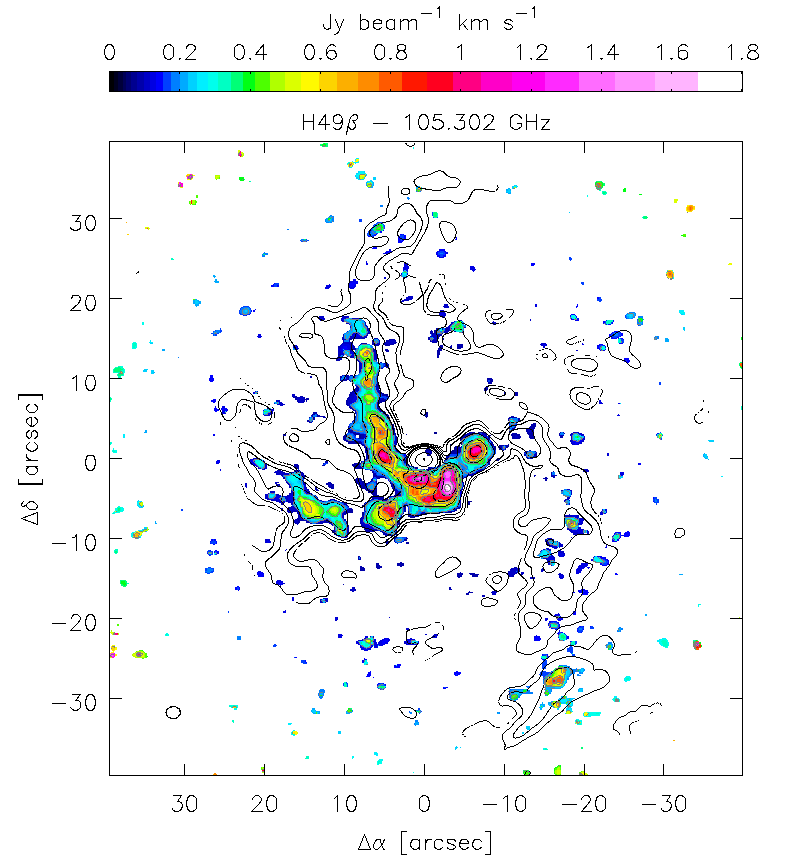}& 
        \includegraphics[trim = 5mm 0mm 7mm 0mm, clip, width=0.31\textwidth]{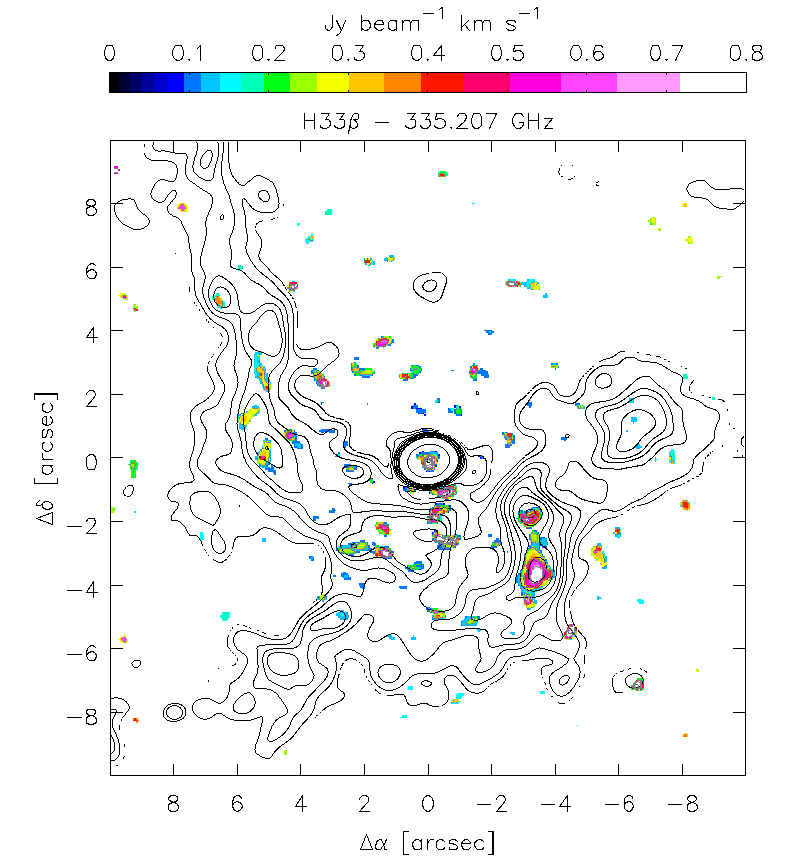}\\                                                                       
        \end{array} $                                                            \caption{Recombination line (RRL) emission images of the inner $\lesssim $ 3 pc. Left: H49$\beta$ with the 100 GHz continuum contours as in Fig. \ref{cont-full}.
                Right: H33$\beta$ with the 250 GHz continuum contours as in Fig. \ref{cont-full}.
        }
        \label{recomb-rest}
\end{figure*}

\begin{figure*}[htbp]
        \centering $
        \begin{array}{ccc}
        \includegraphics[trim = 5mm 0mm 7mm 0mm, clip, width=0.31\textwidth]{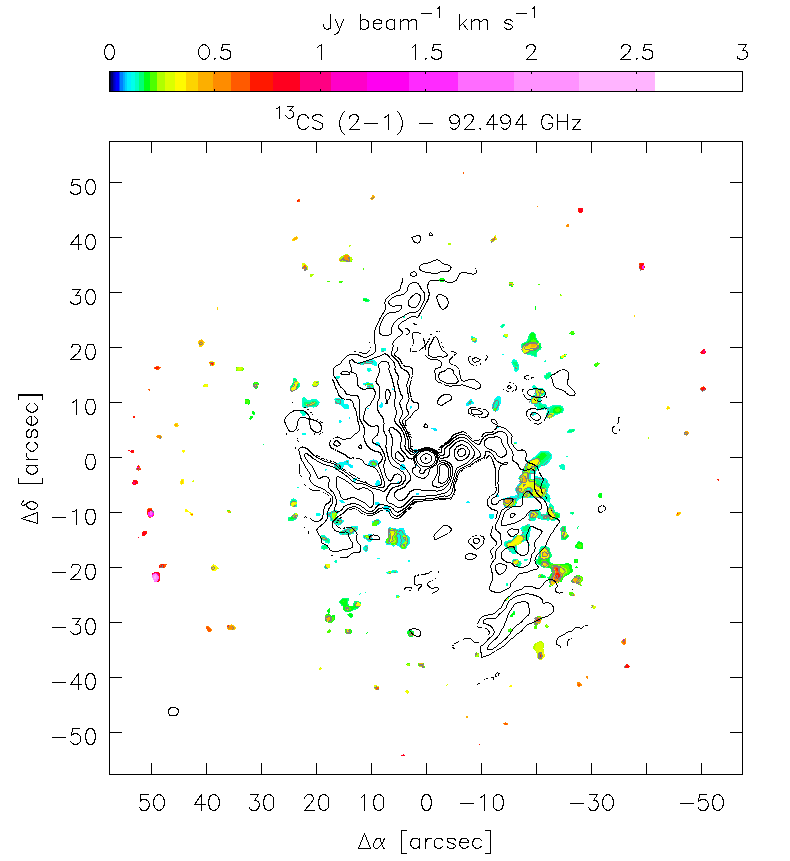}&
        \includegraphics[trim = 5mm 0mm 7mm 0mm, clip, width=0.31\textwidth]{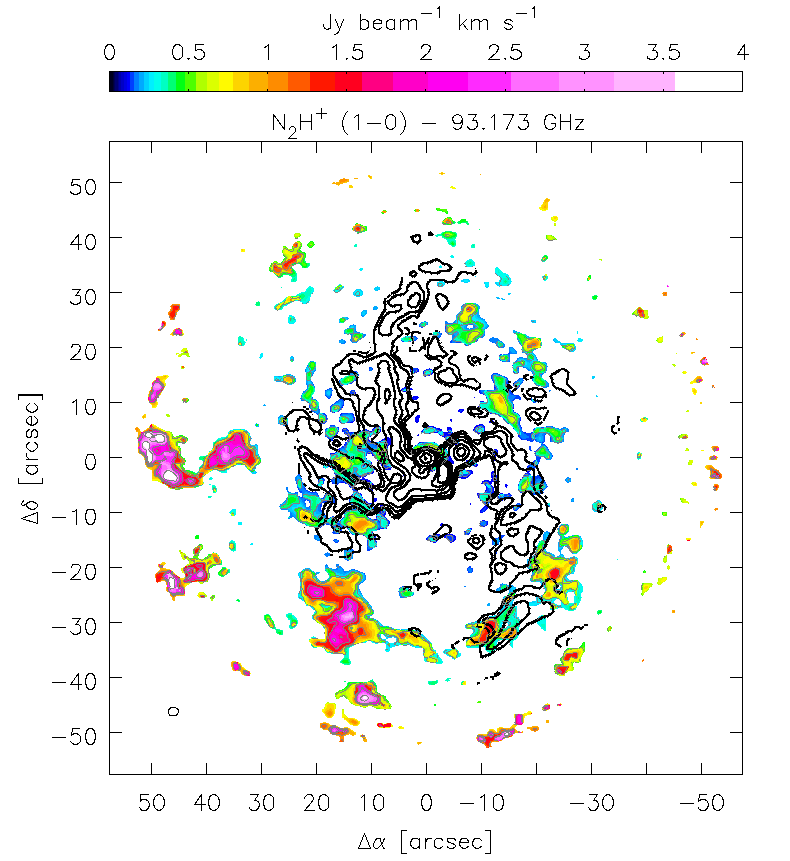}&
        \includegraphics[trim = 5mm 0mm 7mm 0mm, clip, width=0.31\textwidth]{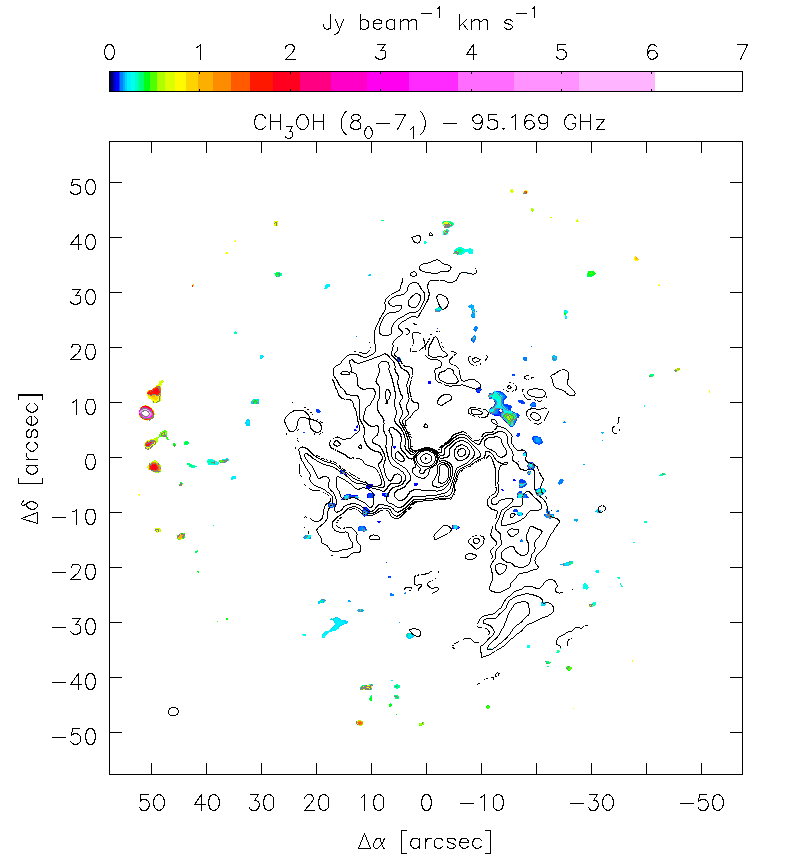}\\
        \end{array} $                                                            \caption{Molecular line emission images of the inner 120$''$ (4.8 pc) from left to right: $^{13}$CS(2--1), N$_2$H$^+$(1--0), and CH$_3$OH(8--7). 
                The contours at the levels of [6, 12, 24, 48, 96, 144, 192, 384, 1920, 4800]$\times \sigma$ (= 0.5 mJy~beam$^{-1}$)
                show the 100 GHz continuum emission.}
        \label{outermol-100}
\end{figure*}

\begin{figure*}[htbp]
	\centering $
	\begin{array}{cccc}
	\includegraphics[trim = 5mm 0mm 7mm 0mm, clip, width=0.23\textwidth]{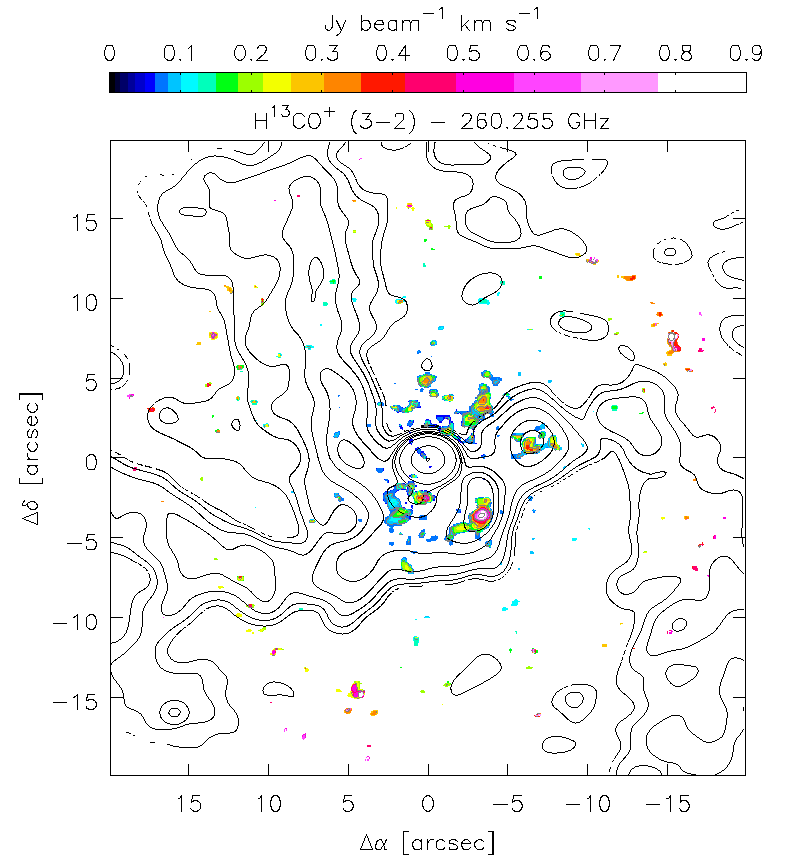}&
	\includegraphics[trim = 5mm 0mm 7mm 0mm, clip, width=0.23\textwidth]{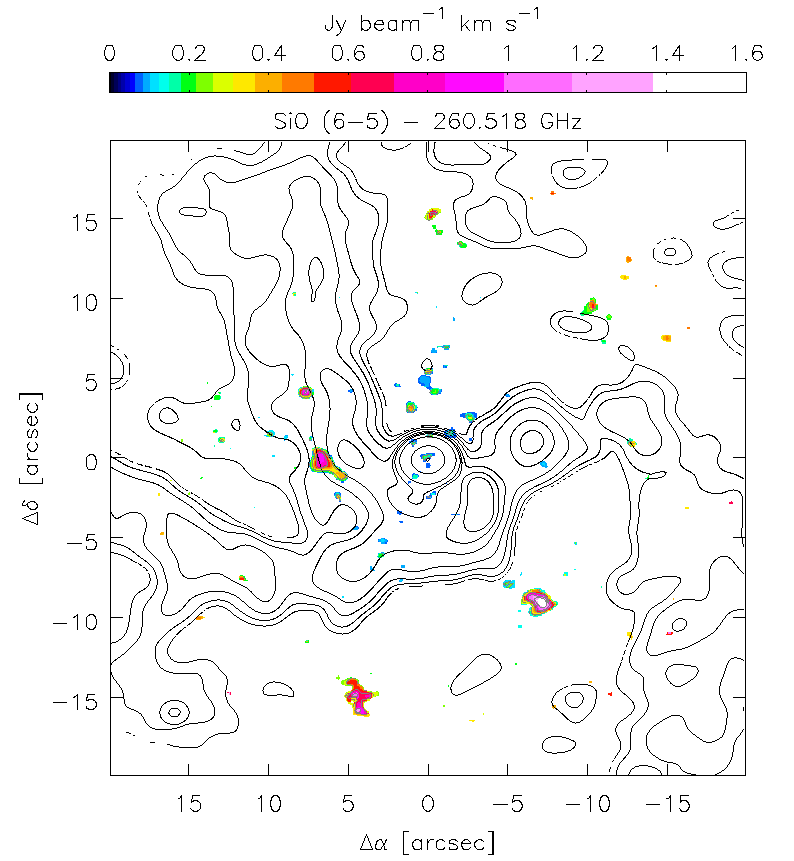}&
	\includegraphics[trim = 5mm 0mm 7mm 0mm, clip, width=0.23\textwidth]{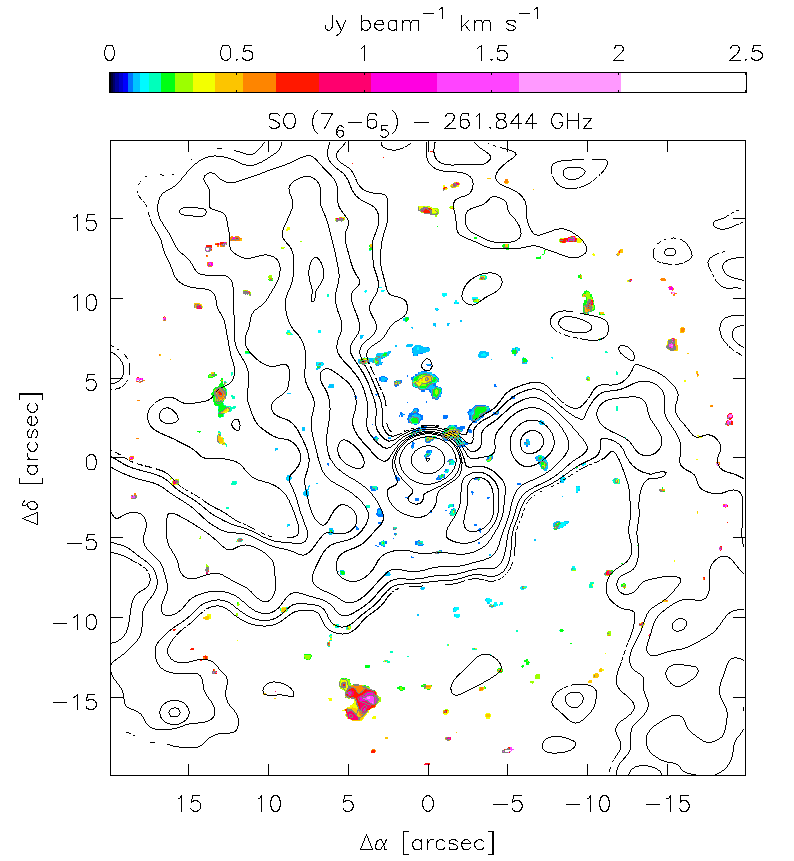}&
	\includegraphics[trim = 5mm 0mm 7mm 0mm, clip, width=0.23\textwidth]{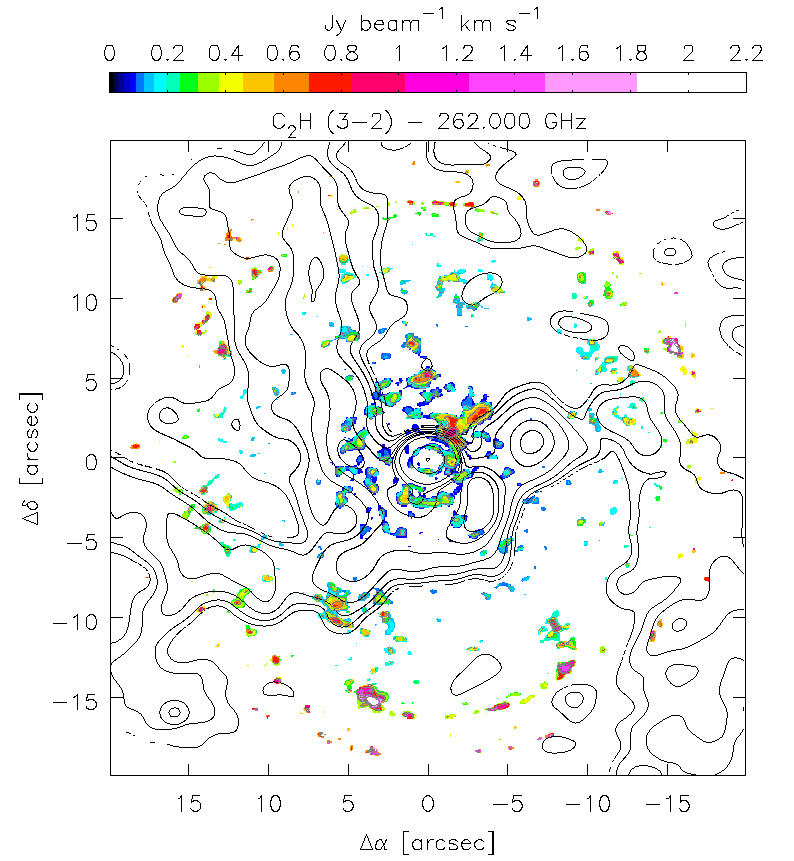}\\
	\includegraphics[trim = 5mm 0mm 7mm 0mm, clip, width=0.23\textwidth]{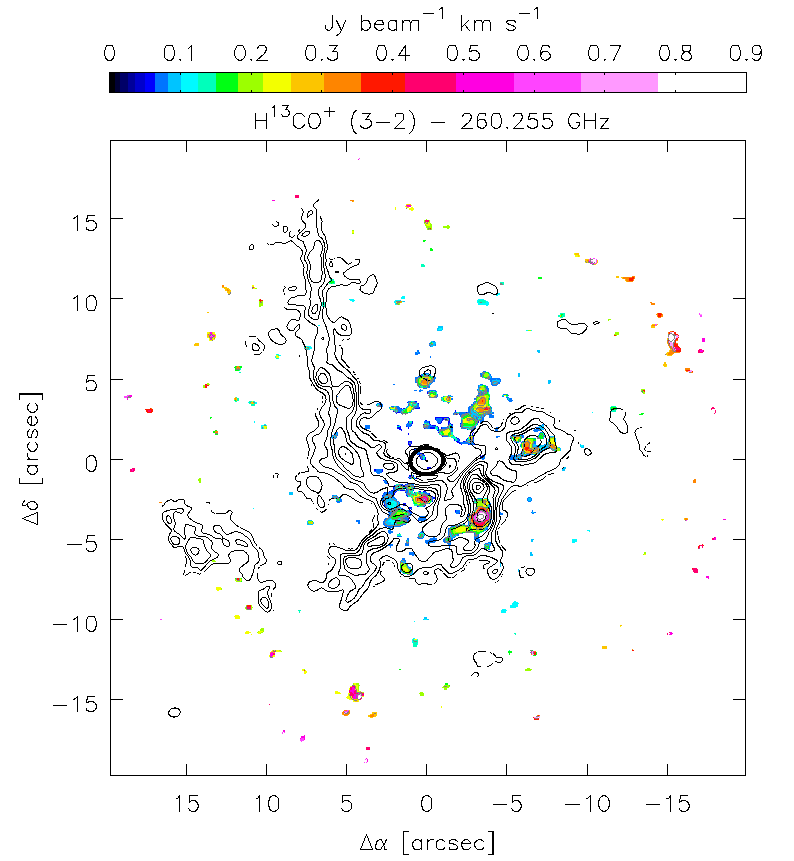}&
	\includegraphics[trim = 5mm 0mm 7mm 0mm, clip, width=0.23\textwidth]{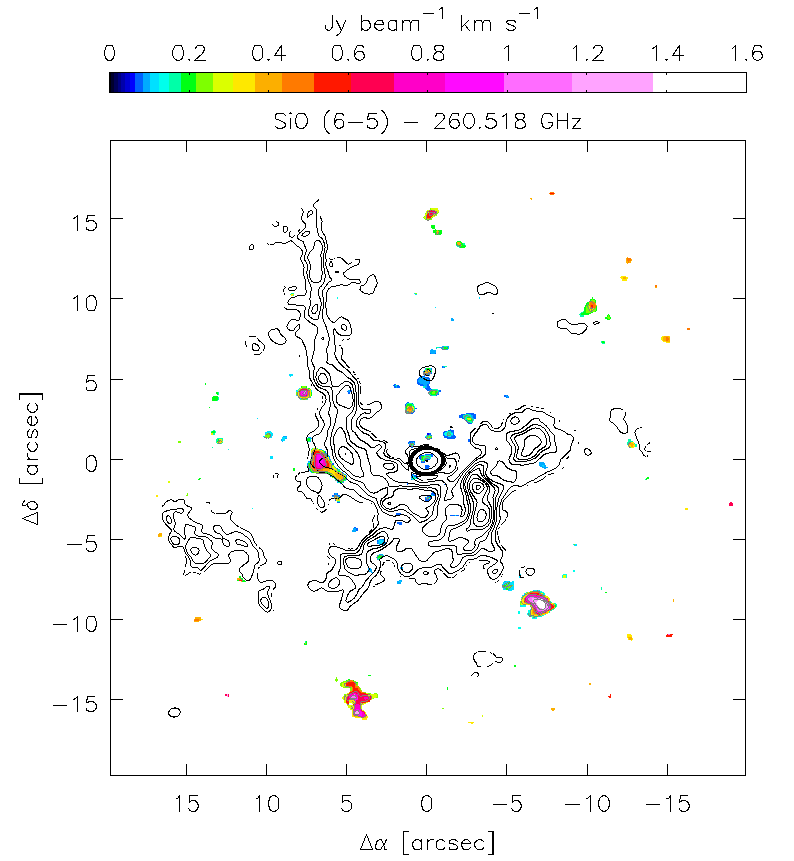}& 
	\includegraphics[trim = 5mm 0mm 7mm 0mm, clip, width=0.23\textwidth]{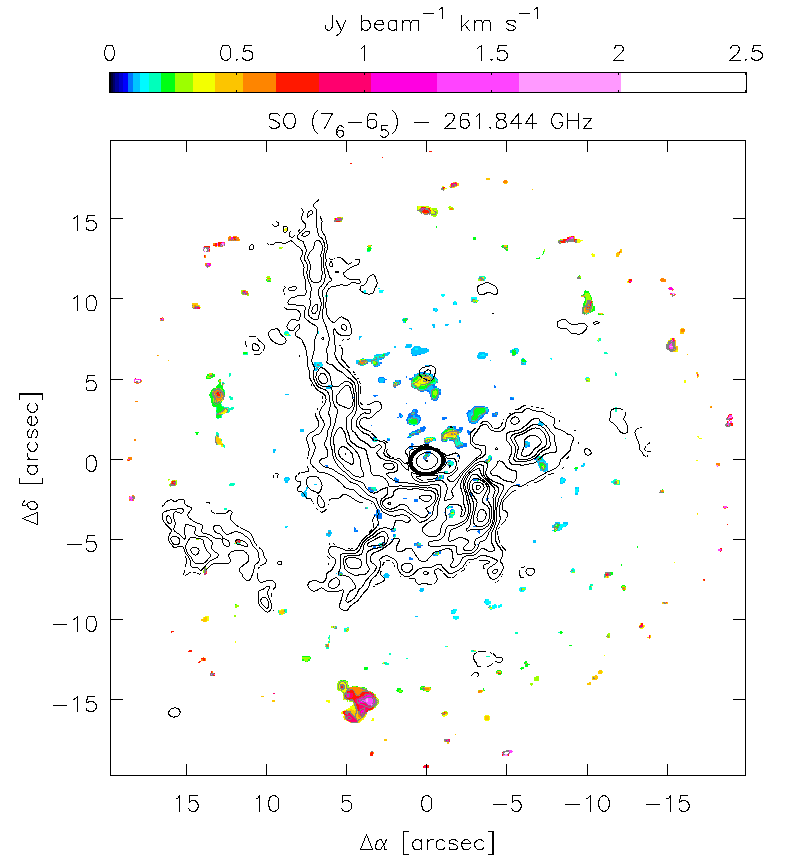}&
	\includegraphics[trim = 5mm 0mm 7mm 0mm, clip, width=0.23\textwidth]{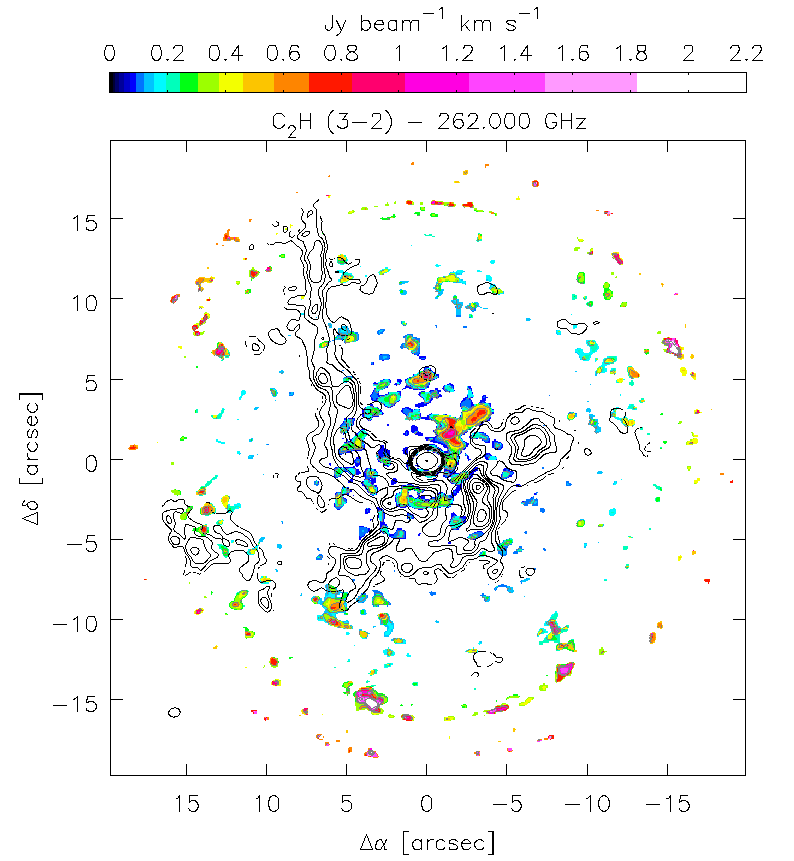}\\
	\end{array} $                                                            \caption{Molecular line emission images of the inner 40$''$  (1.6 pc) compared to the 100 GHz (top) and 250 GHz (bottom) continuum emission (as in Fig. \ref{cont-full}): 
		From left to right: H$^{13}$CO$^+$(3--2), SiO(6--5), SO(7--6), and C$_2$H(3--2) as in Fig. \ref{mols-full1}. 
	}
	\label{mols-100-250}
\end{figure*}

\begin{figure*}[htbp]
	\centering $
	\begin{array}{cccc}
	\includegraphics[trim = 5mm 0mm 7mm 0mm, clip, width=0.23\textwidth]{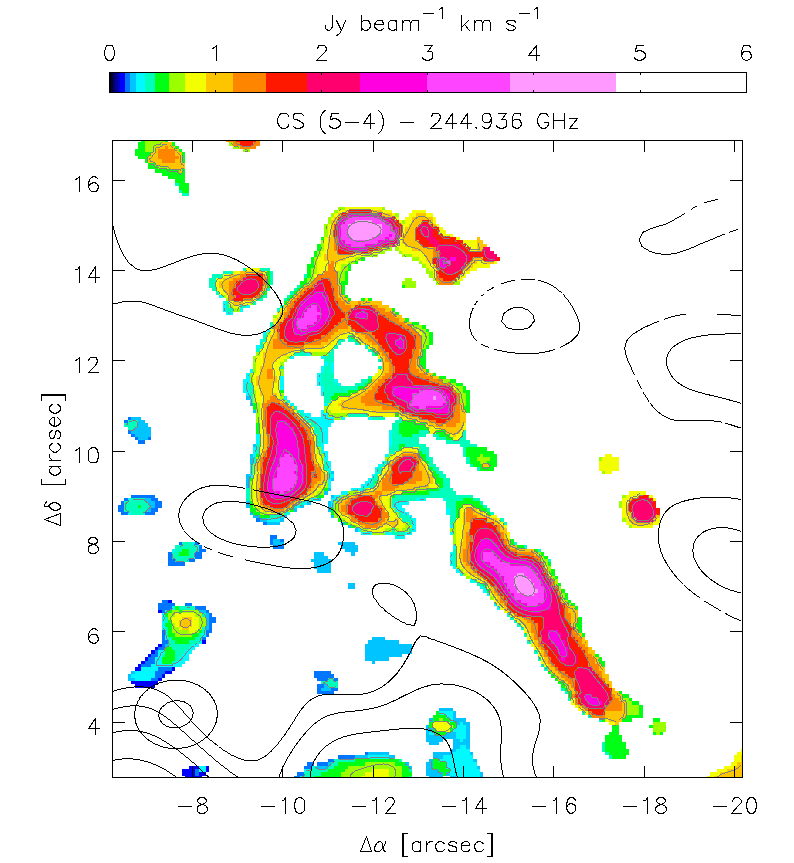}& 
	\includegraphics[trim = 5mm 0mm 7mm 0mm, clip, width=0.23\textwidth]{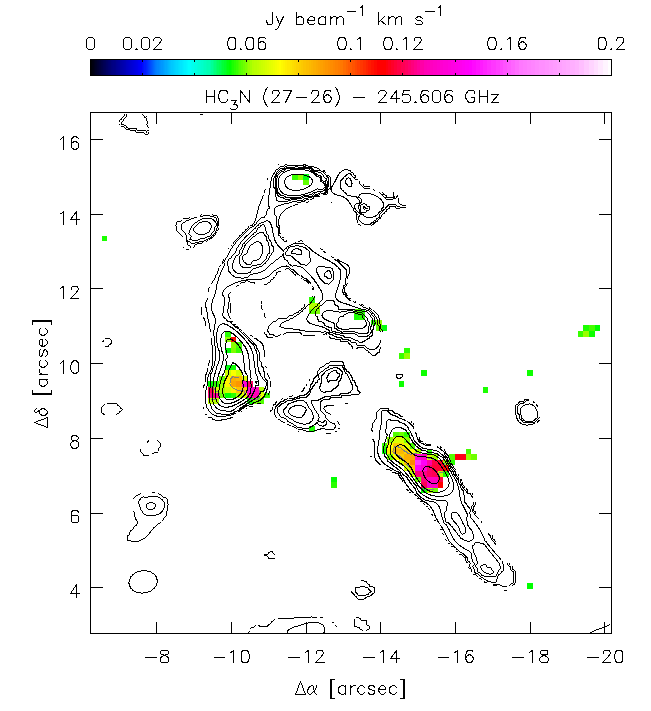}&  
	\includegraphics[trim = 5mm 0mm 7mm 0mm, clip, width=0.23\textwidth]{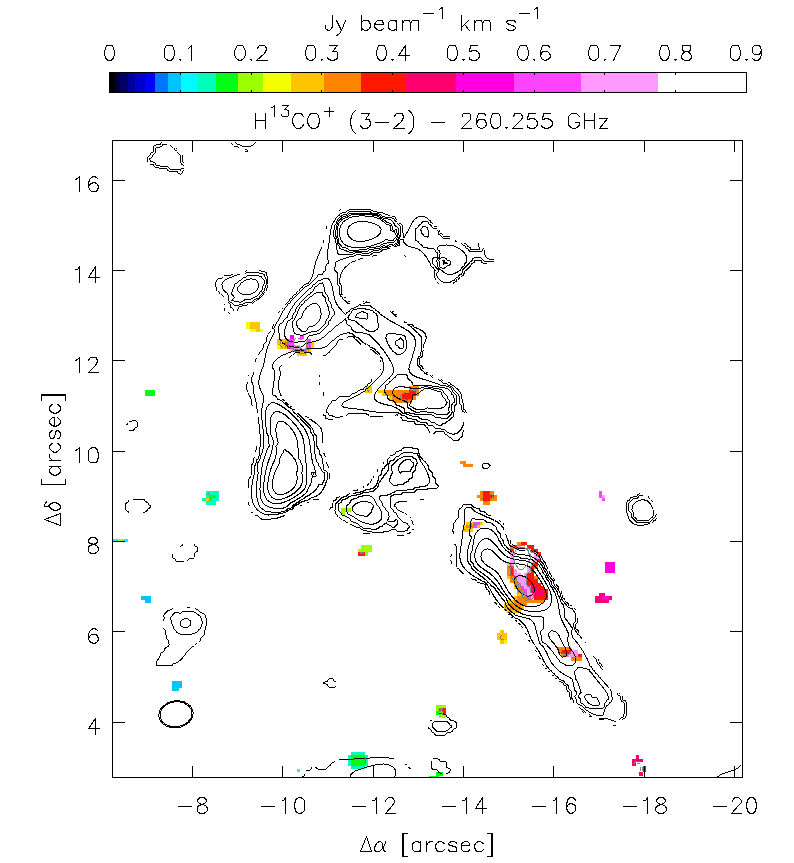}&  
	\includegraphics[trim = 5mm 0mm 7mm 0mm, clip, width=0.23\textwidth]{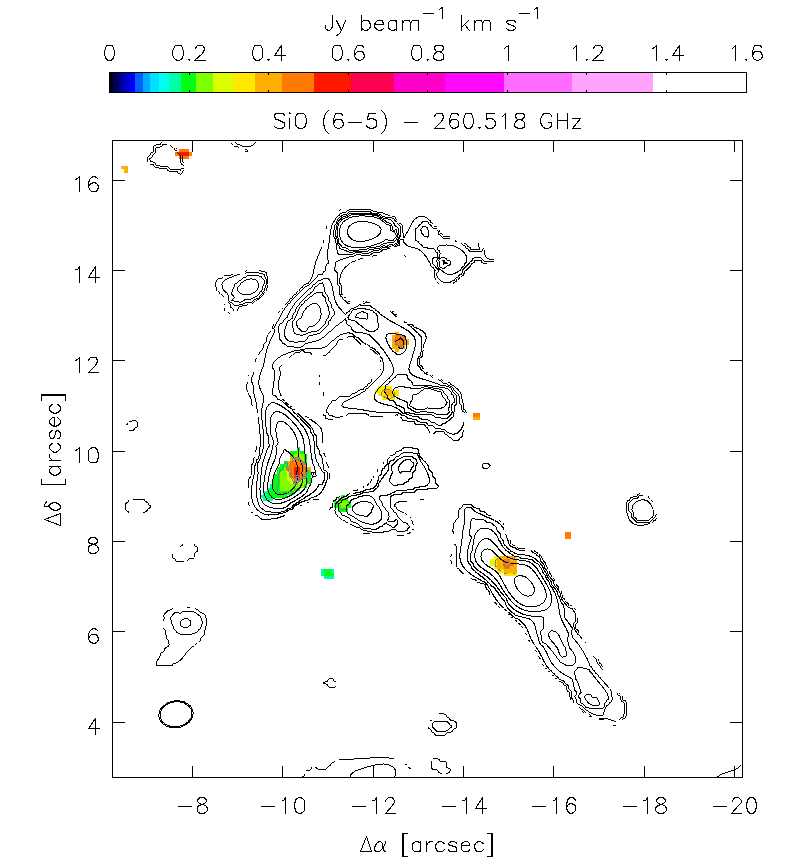}\\
	\includegraphics[trim = 5mm 0mm 7mm 0mm, clip, width=0.23\textwidth]{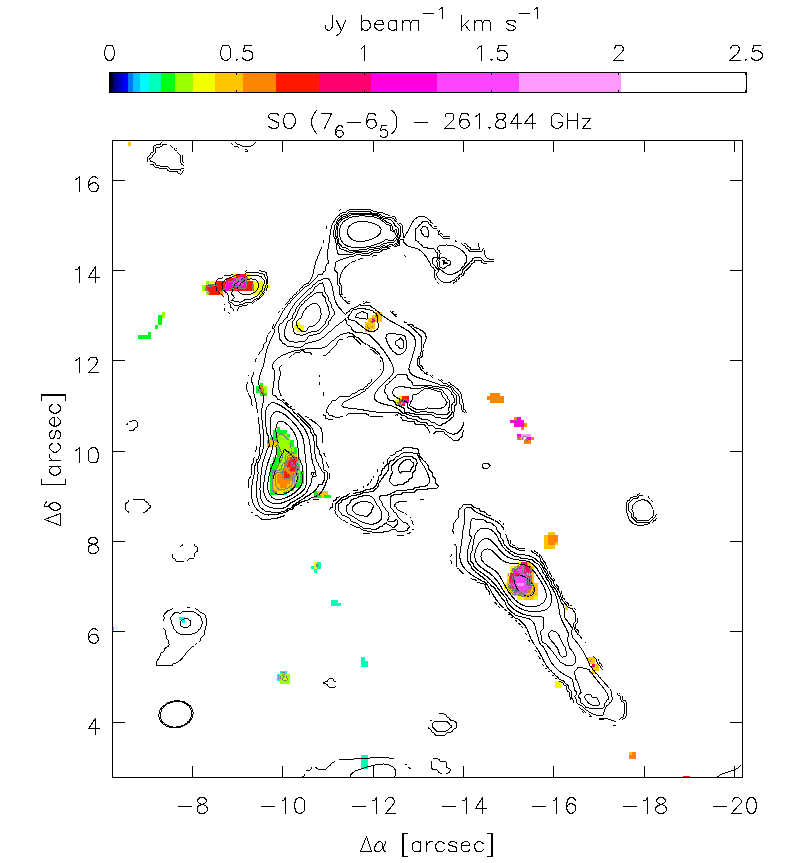}&
	\includegraphics[trim = 5mm 0mm 7mm 0mm, clip, width=0.23\textwidth]{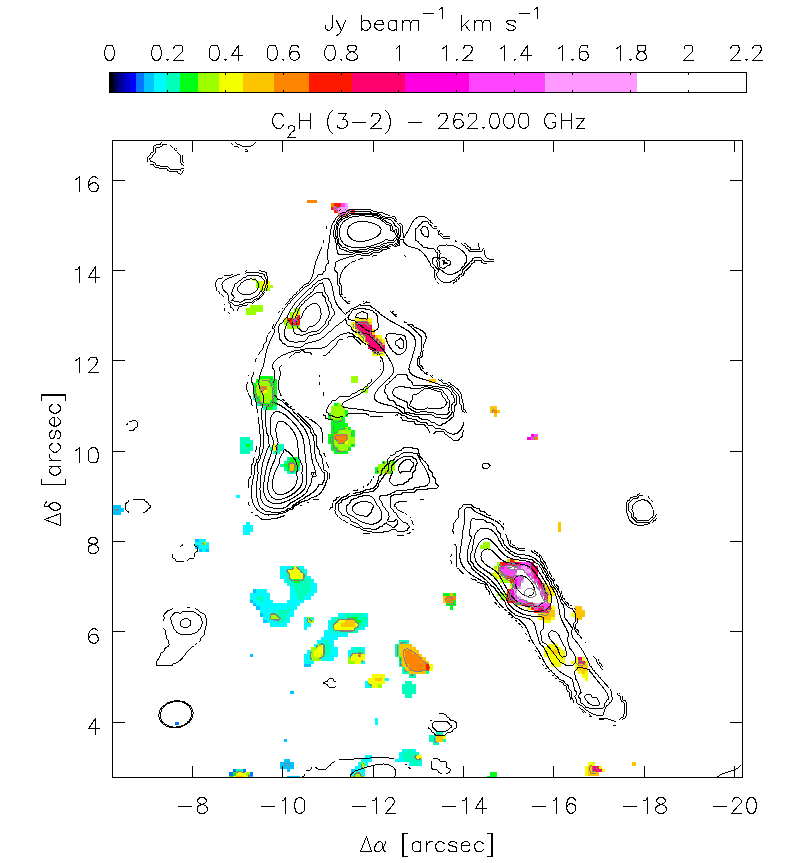}&
	\includegraphics[trim = 5mm 0mm 7mm 0mm, clip, width=0.23\textwidth]{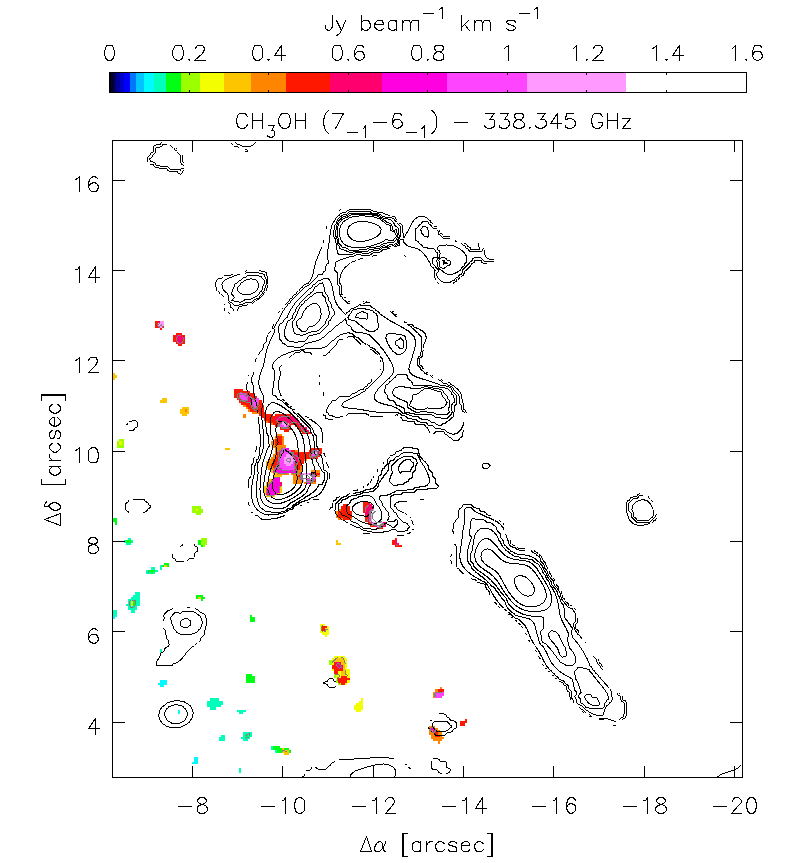}&
	\includegraphics[trim = 5mm 0mm 7mm 0mm, clip, width=0.23\textwidth]{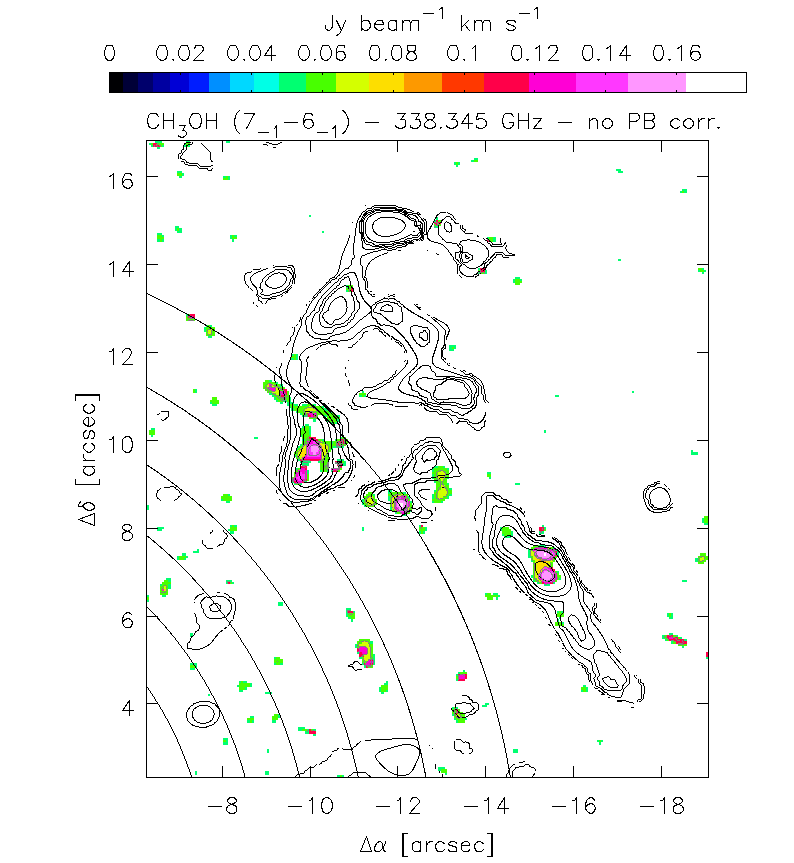}\\
	\includegraphics[trim = 5mm 0mm 7mm 0mm, clip, width=0.23\textwidth]{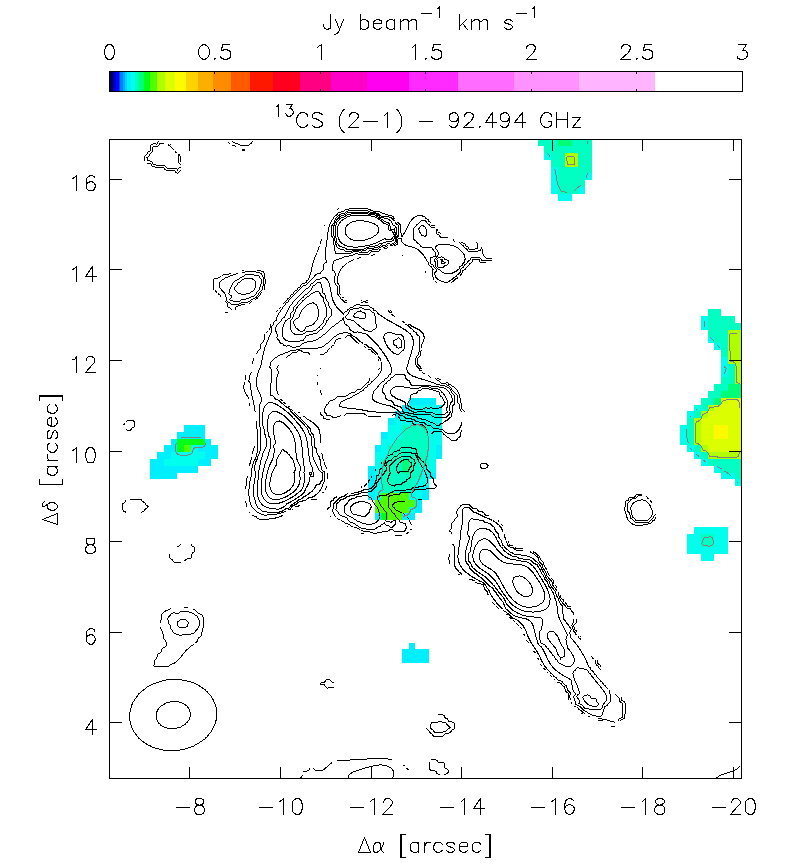}& 
	\includegraphics[trim = 5mm 0mm 7mm 0mm, clip, width=0.23\textwidth]{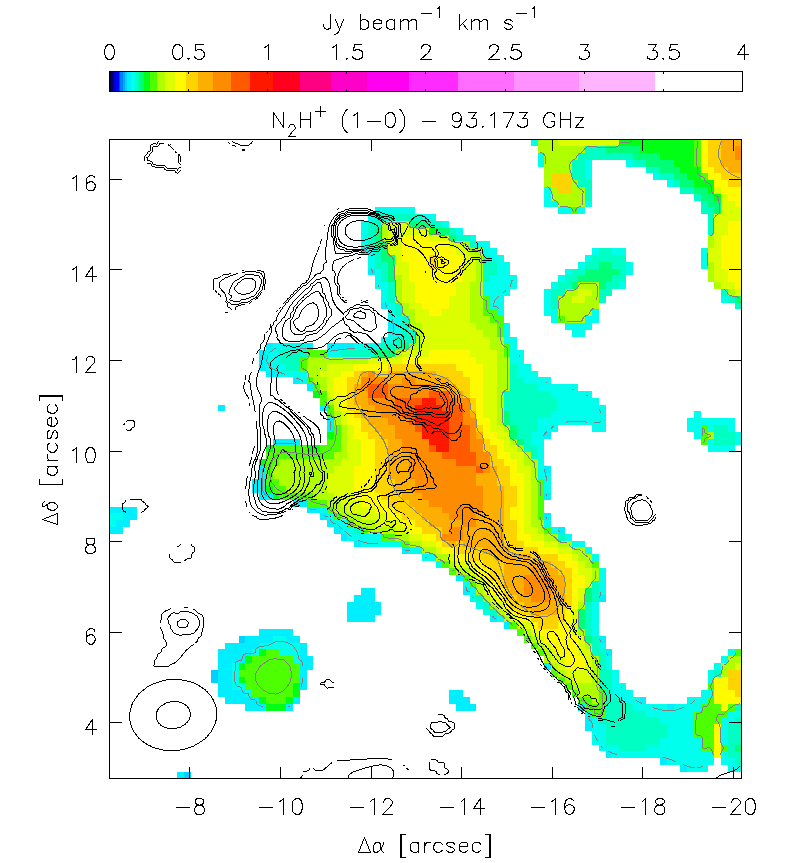}&
	\includegraphics[trim = 5mm 0mm 7mm 0mm, clip, width=0.23\textwidth]{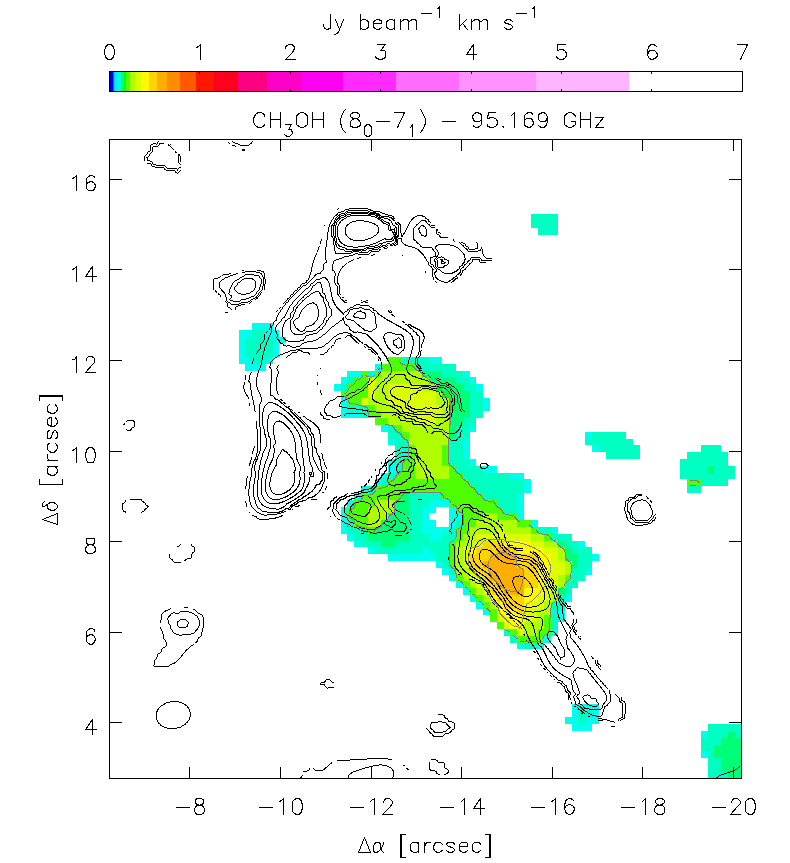}&   
	\includegraphics[trim = 5mm 0mm 7mm 0mm, clip, width=0.23\textwidth]{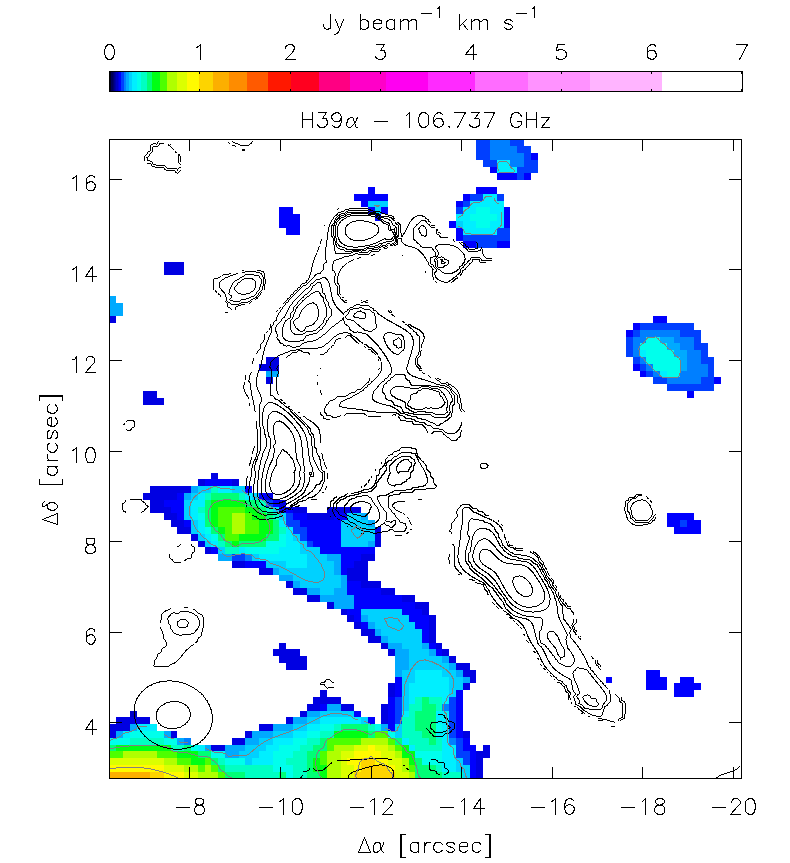}\\ 
	\end{array} $                                                            \caption{\textit{Triop} in the light of molecular line, RRL and continuum emission: 
		Top row (from left to right): CS(5--4), HC$_3$N(27--26), H$^{13}$CO$^+$(3--2), and SiO(6--5) (all as in Fig. \ref{mols-full1}).
		Middle row (from left to right): SO(7--6), C$_2$H(3--2), and CH$_3$OH(7--6) (all as in Fig. \ref{mols-full1}), 
		and CH$_3$OH(7--6) when not corrected for primary beam (PB) with PB contours of 10\% (outmost ring segment) to 60\% (lower left corner), to show the full extend of the emission within the \textit{triop}. 
		Bottom (from left to right): $^{13}$CS(2--1), N$_2$H$^+$(1--0), and CH$_3$OH(8--7) (all as in Fig. \ref{outermol}), 
		and H39$\alpha$ (all as in Fig. \ref{recomb}). 
		Contours show the CS(5--4) emission (as in Fig. \ref{mols-full1}), except from the top outer left, 
		which shows contours of the 100 GHz continuum emission as in Fig. \ref{cont-full}.
	}
	\label{triop}
\end{figure*}

\begin{figure*}[htbp]
	\centering $
	\begin{array}{cccc}
	\includegraphics[trim = 5mm 0mm 7mm 0mm, clip, width=0.23\textwidth]{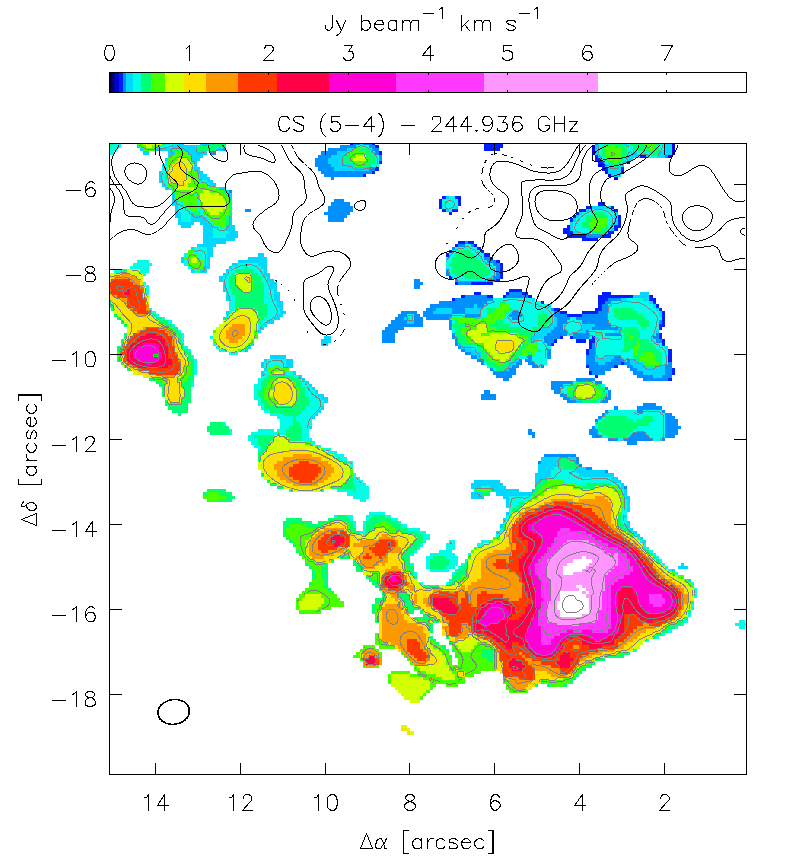}& 
	\includegraphics[trim = 5mm 0mm 7mm 0mm, clip, width=0.23\textwidth]{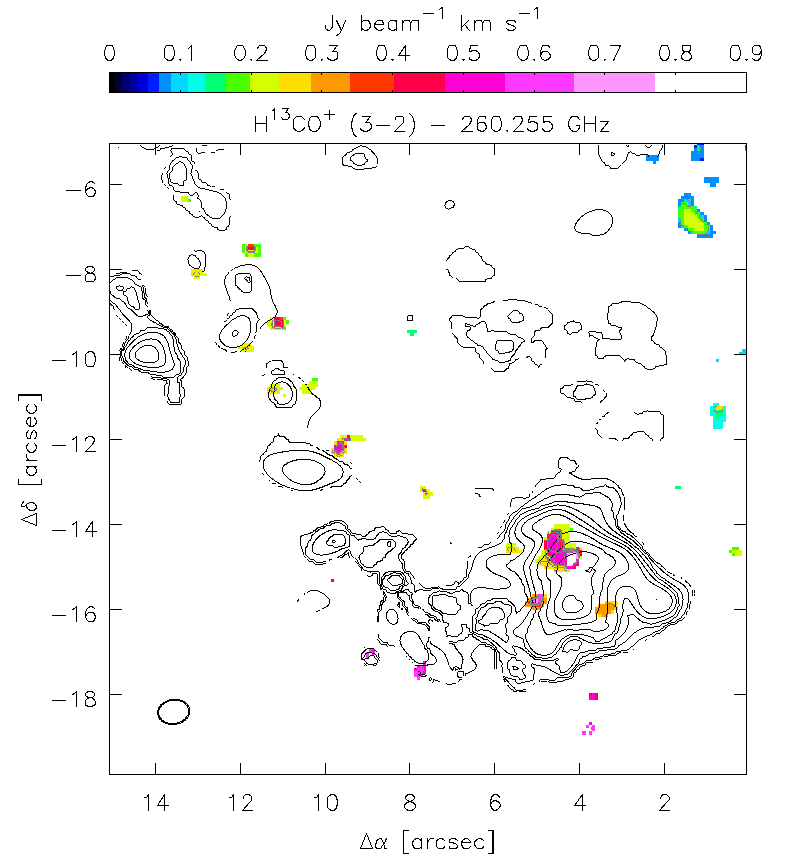}&  
	\includegraphics[trim = 5mm 0mm 7mm 0mm, clip, width=0.23\textwidth]{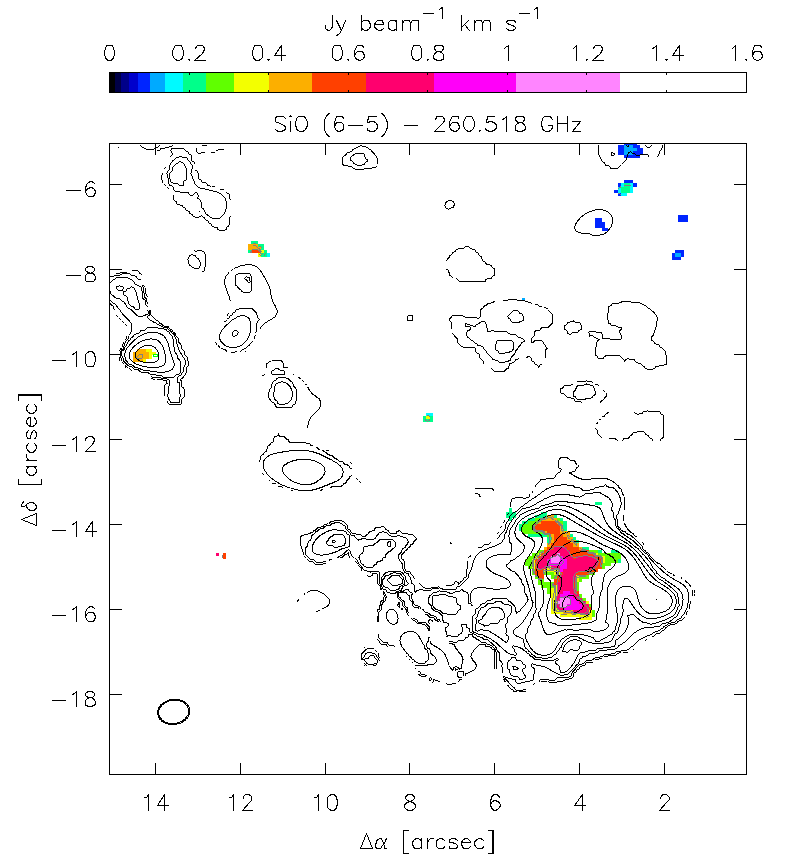}& 
	\includegraphics[trim = 5mm 0mm 7mm 0mm, clip, width=0.23\textwidth]{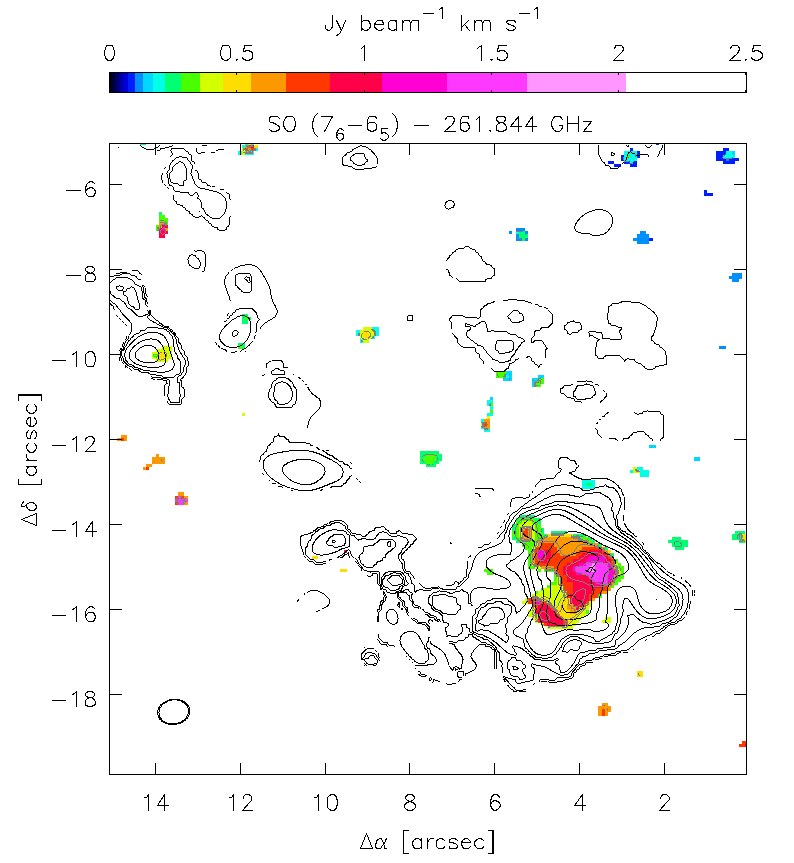}\\
	\includegraphics[trim = 5mm 0mm 7mm 0mm, clip, width=0.23\textwidth]{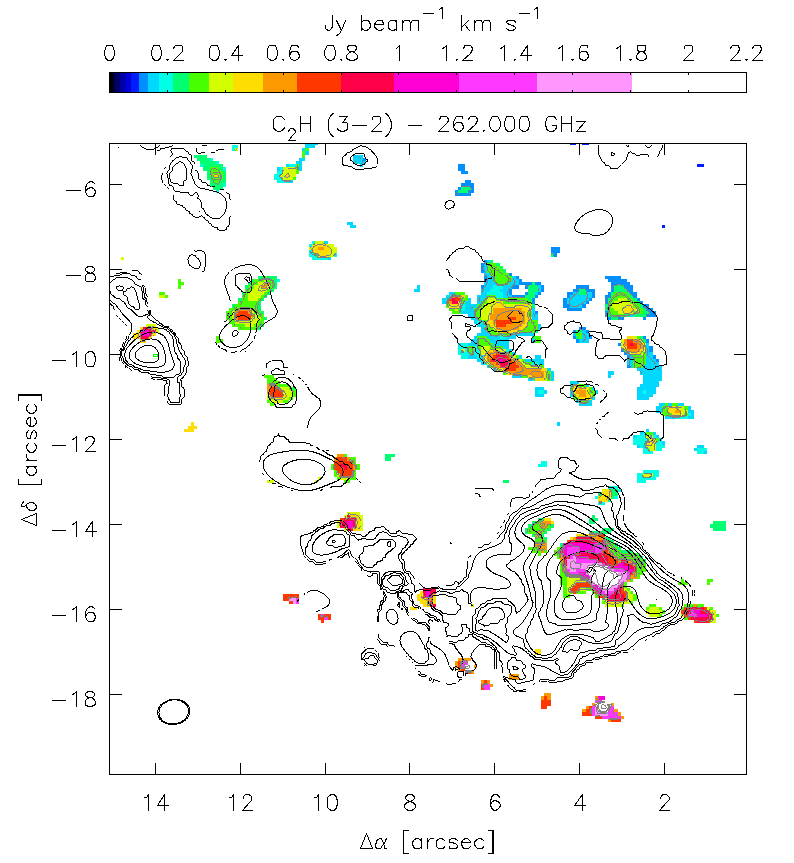}&
	\includegraphics[trim = 5mm 0mm 7mm 0mm, clip, width=0.23\textwidth]{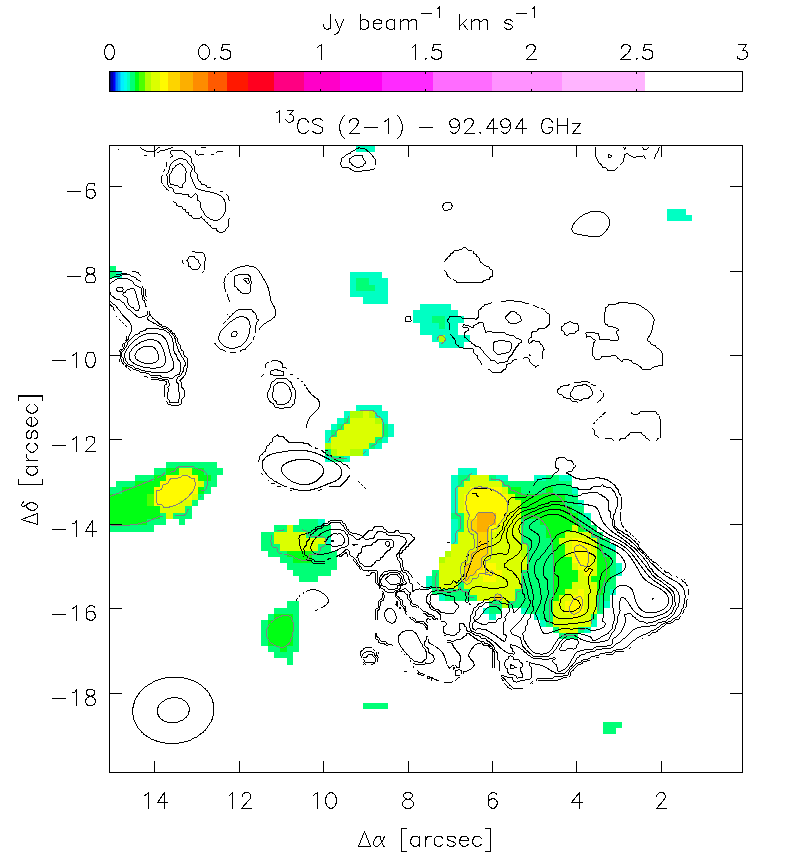}&
	\includegraphics[trim = 5mm 0mm 7mm 0mm, clip, width=0.23\textwidth]{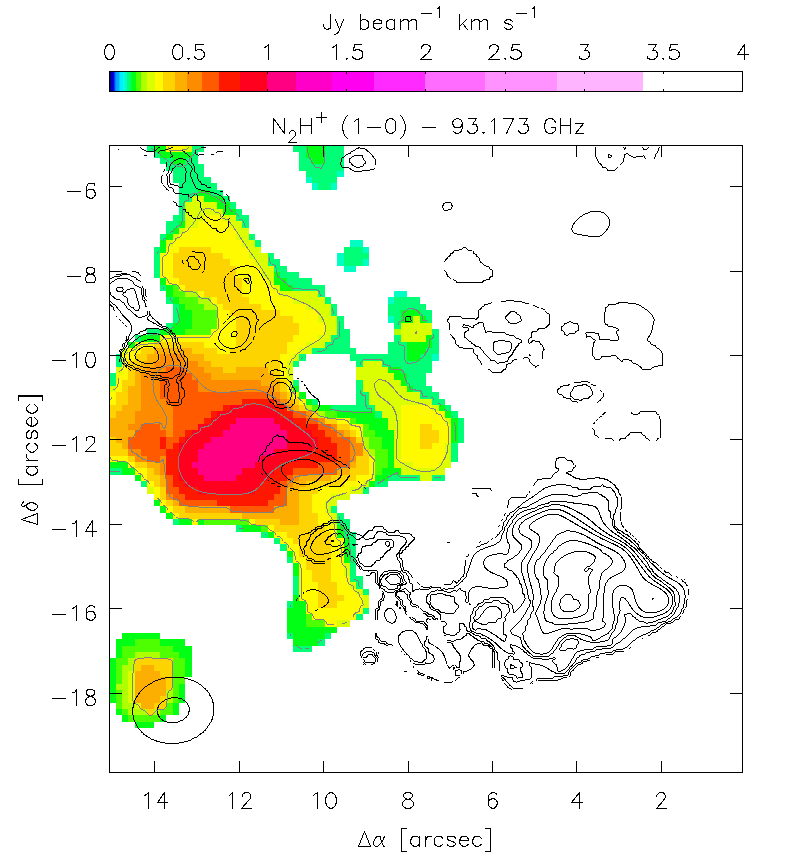}&
	\includegraphics[trim = 5mm 0mm 7mm 0mm, clip, width=0.23\textwidth]{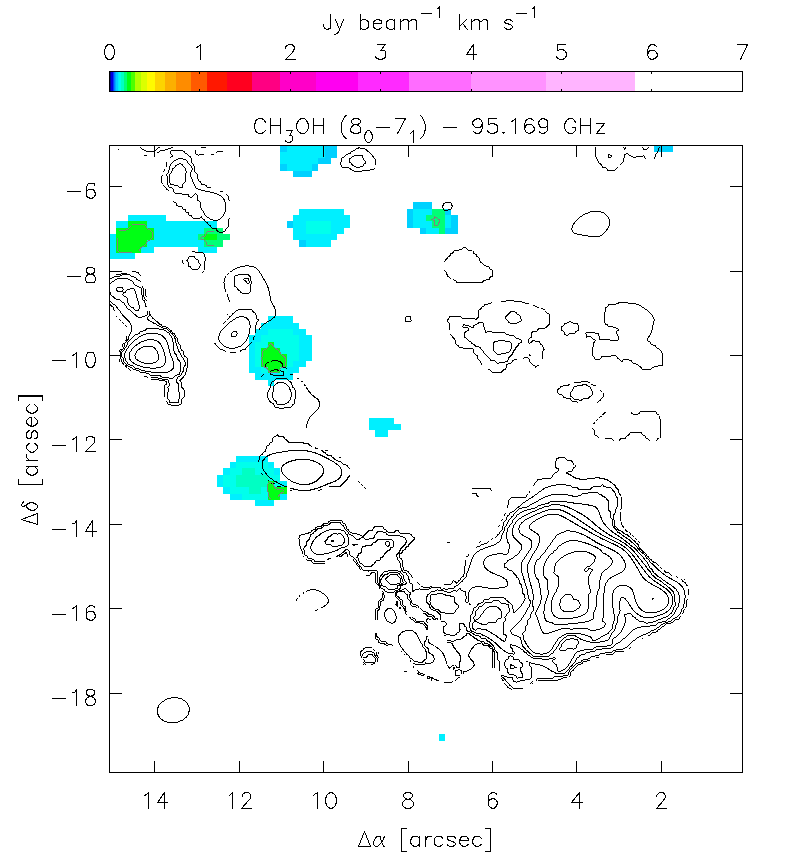}\\ 
	\end{array} $                                                            \caption{SEW cloud in the light of molecular line and continuum emission: 
		Top row (from left to right): CS(5--4), H$^{13}$CO$^+$(3--2), SiO(6--5), and SO(7--6) (all as in Fig. \ref{mols-full1}).
		Bottom - from left to right: C$_2$H(3--2) (as in Fig. \ref{mols-full1}), and $^{13}$CS(2--1), N$_2$H$^+$(1--0), and CH$_3$OH(8--7) (all three as in Fig. \ref{outermol}). 
		Contours show the CS(5--4) emission (as in Fig. \ref{mols-full1}), except from the top outer left, 
		which shows contours of the 250 GHz continuum emission as in Fig. \ref{cont-full}.
	}
	\label{SEW}
\end{figure*}

\begin{figure*}[htbp]
        \centering $
        \begin{array}{cccc}
        \includegraphics[trim = 5mm 0mm 7mm 0mm, clip, width=0.23\textwidth]{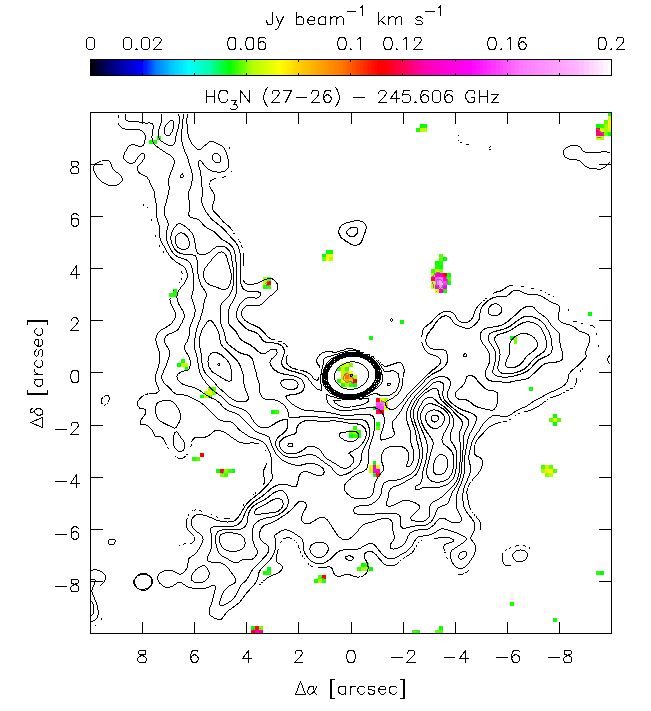}&  
        \includegraphics[trim = 5mm 0mm 7mm 0mm, clip, width=0.23\textwidth]{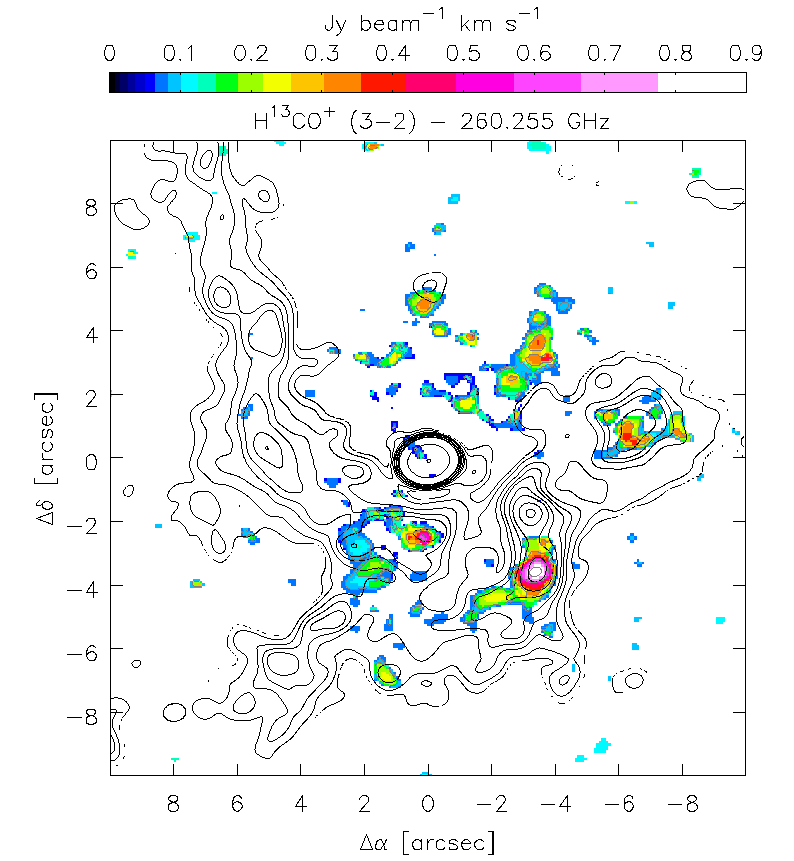}&  
        \includegraphics[trim = 5mm 0mm 7mm 0mm, clip, width=0.23\textwidth]{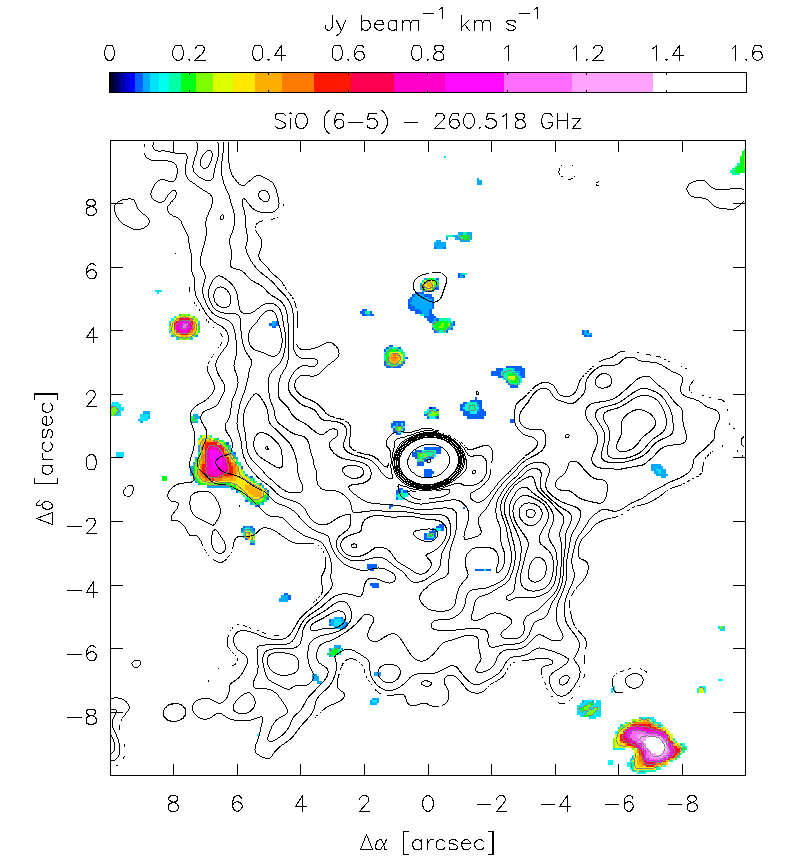}&
        \includegraphics[trim = 5mm 0mm 7mm 0mm, clip, width=0.23\textwidth]{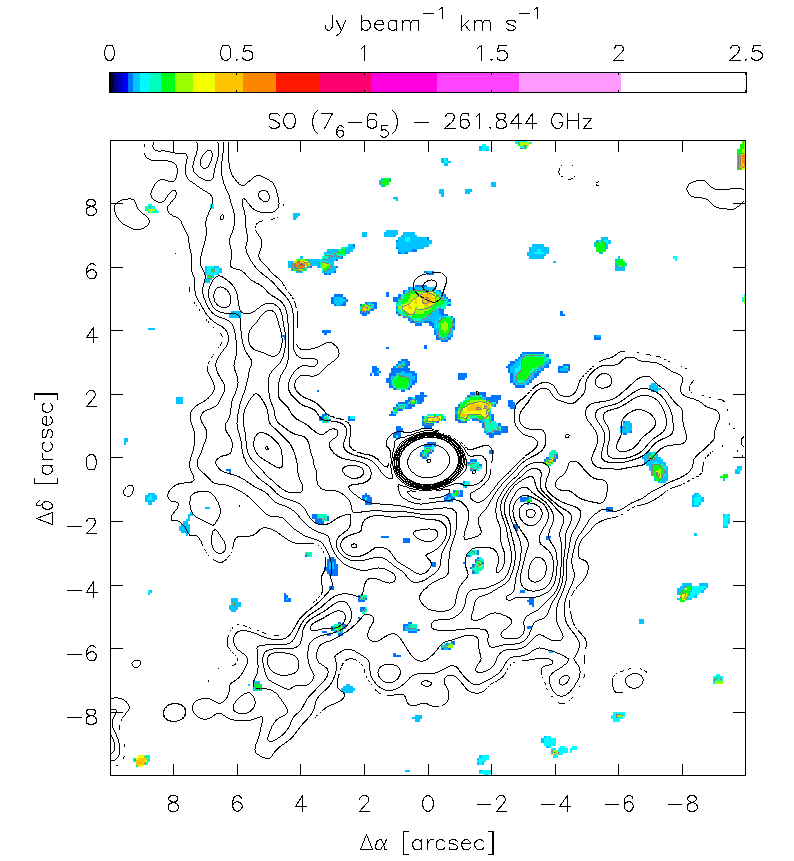}\\ 
        \includegraphics[trim = 5mm 0mm 7mm 0mm, clip, width=0.23\textwidth]{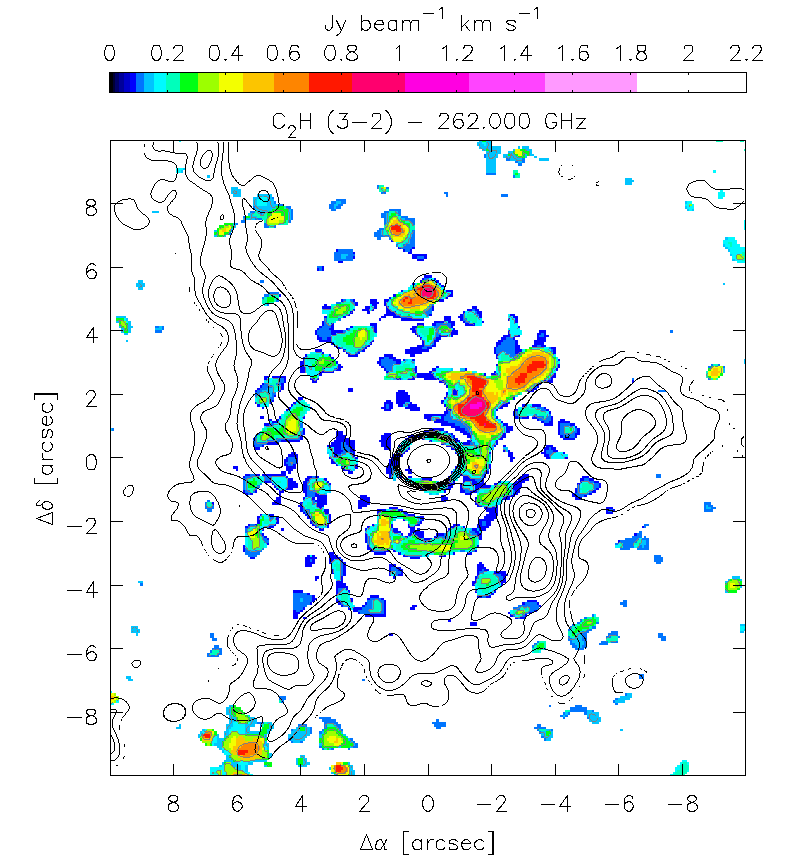}&
        \includegraphics[trim = 5mm 0mm 7mm 0mm, clip, width=0.23\textwidth]{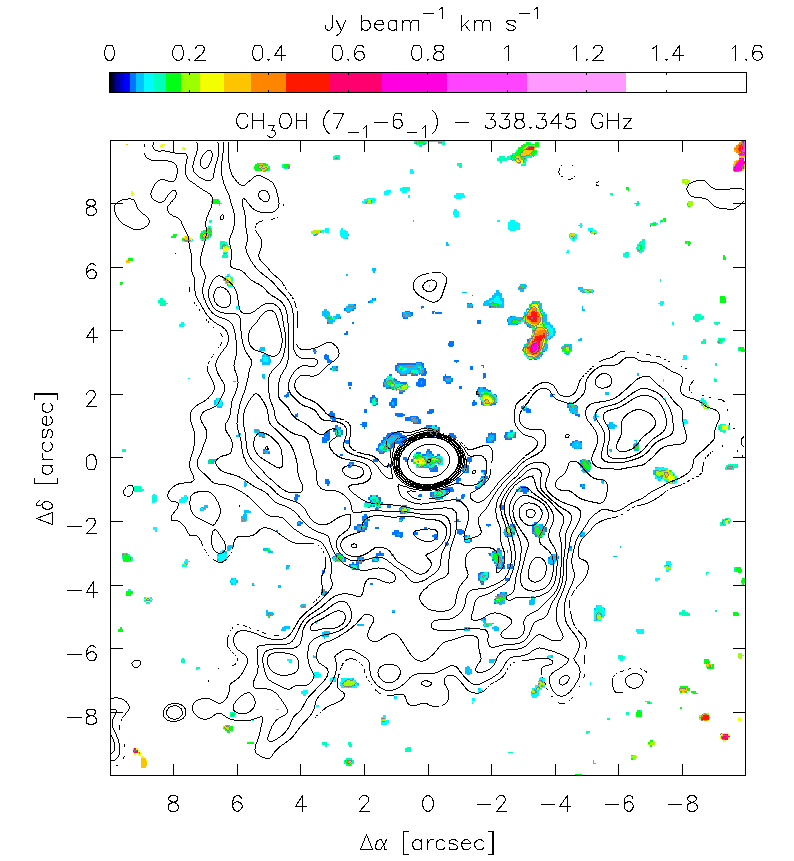}&   
        \includegraphics[trim = 5mm 0mm 7mm 0mm, clip, width=0.23\textwidth]{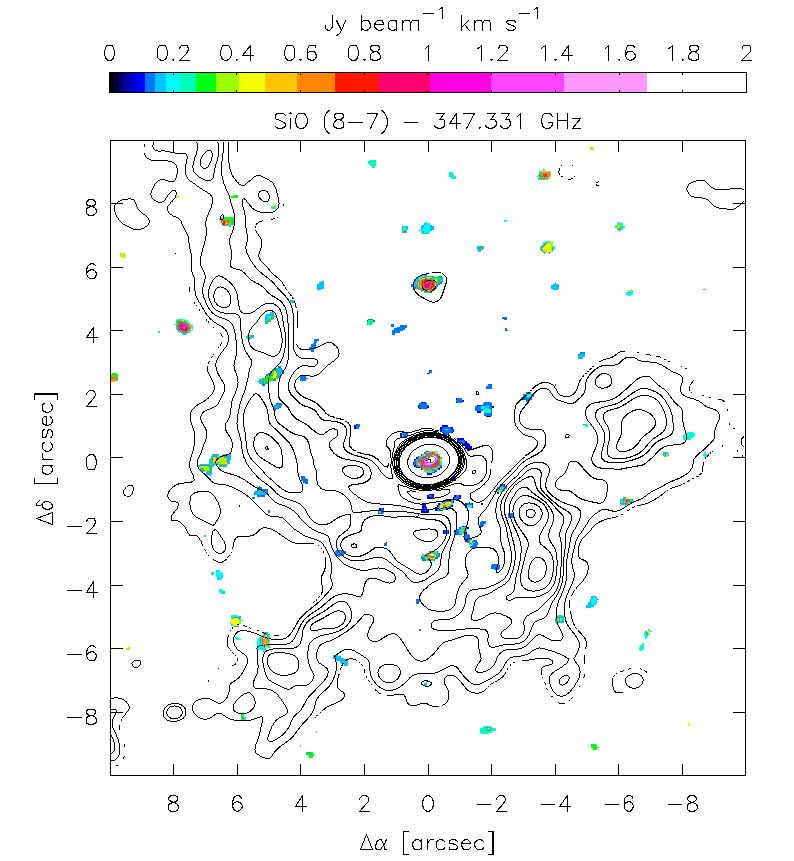}& 
        \includegraphics[trim = 5mm 0mm 7mm 0mm, clip, width=0.23\textwidth]{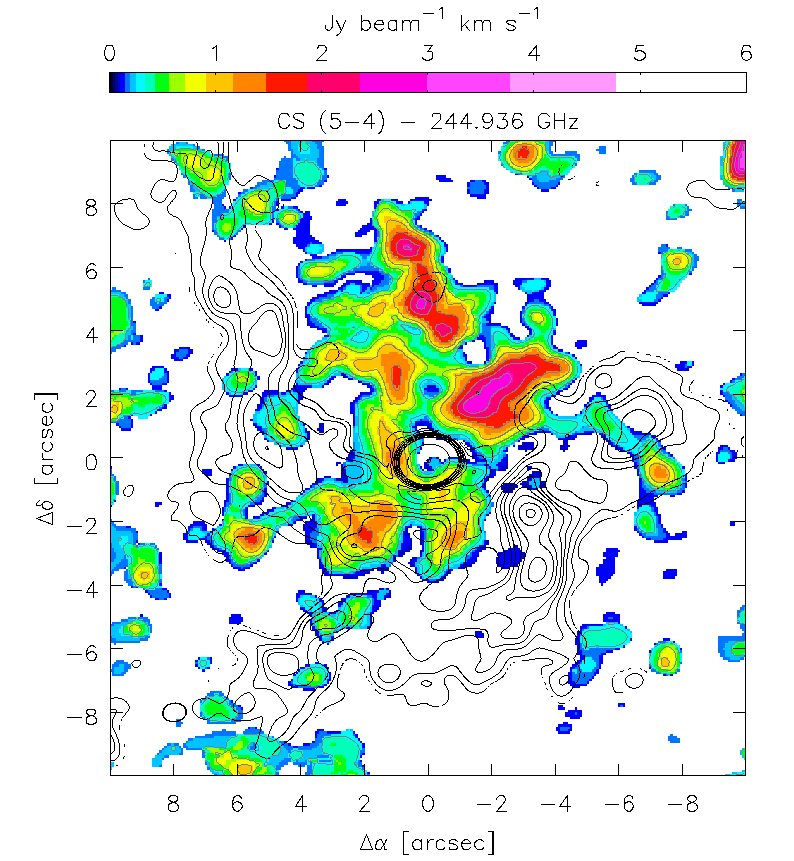}\\                                                                       
        \end{array} $                                                            \caption{Molecular line emission images of the inner 20$''$ (as in Fig. \ref{mols-full1}) compared to the 250 GHz continuum emission in contours (as in Fig. \ref{cont-full}).
                Top row (from left to right): HC$_3$N(27--26), H$^{13}$CO$^+$(3--2), SiO(6--5), and SO(7--6). 
                Bottom row (from left to right): C$_2$H(3--2), CH$_3$OH(7--6), SiO(8--7), and CS(5--4). 
        }
        \label{mols-250-20}
\end{figure*}

\begin{figure*}[htbp]
        \centering $
        \begin{array}{cccc}
        \includegraphics[trim = 5mm 0mm 7mm 0mm, clip, width=0.23\textwidth]{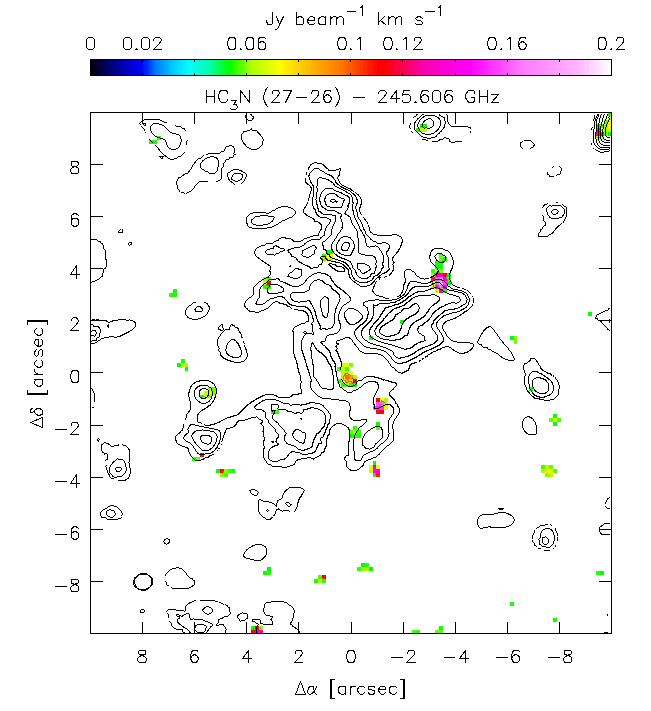}&
        \includegraphics[trim = 5mm 0mm 7mm 0mm, clip, width=0.23\textwidth]{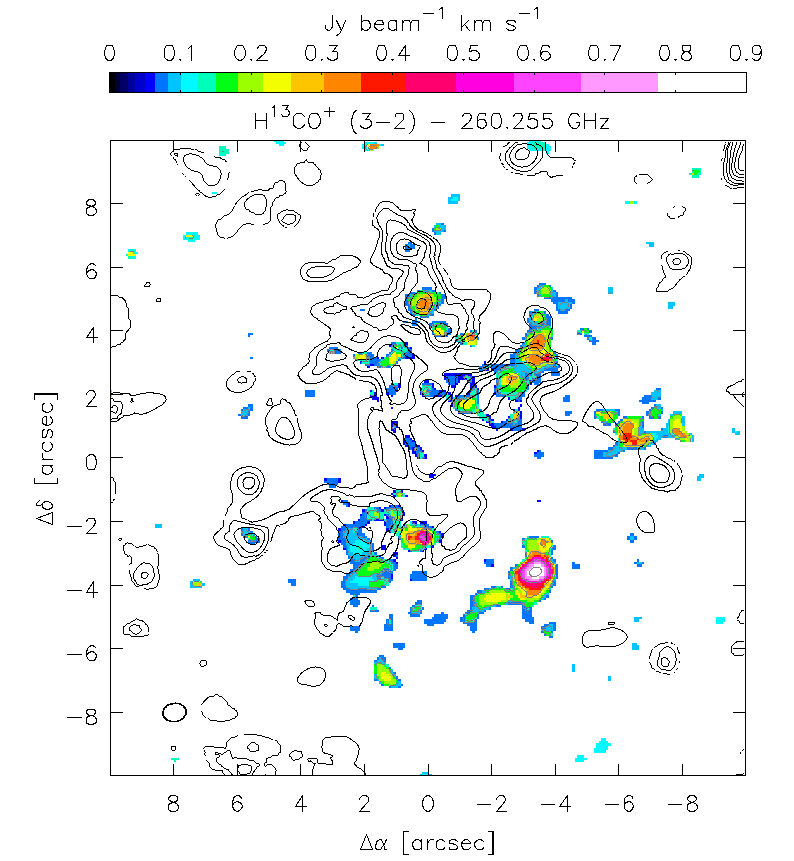}&
        \includegraphics[trim = 5mm 0mm 7mm 0mm, clip, width=0.23\textwidth]{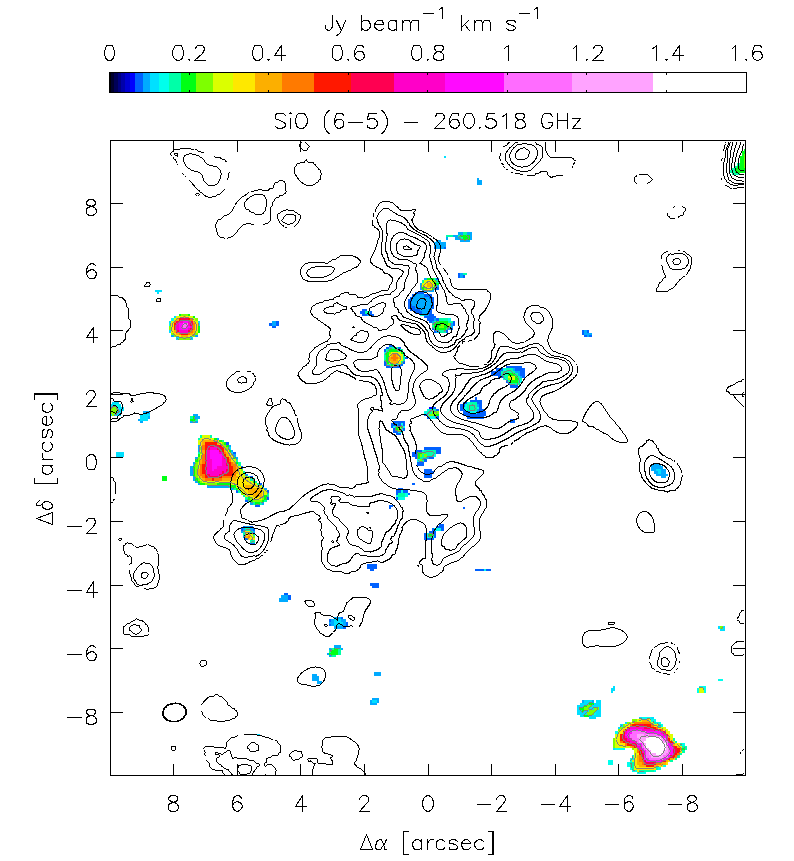}&
        \includegraphics[trim = 5mm 0mm 7mm 0mm, clip, width=0.23\textwidth]{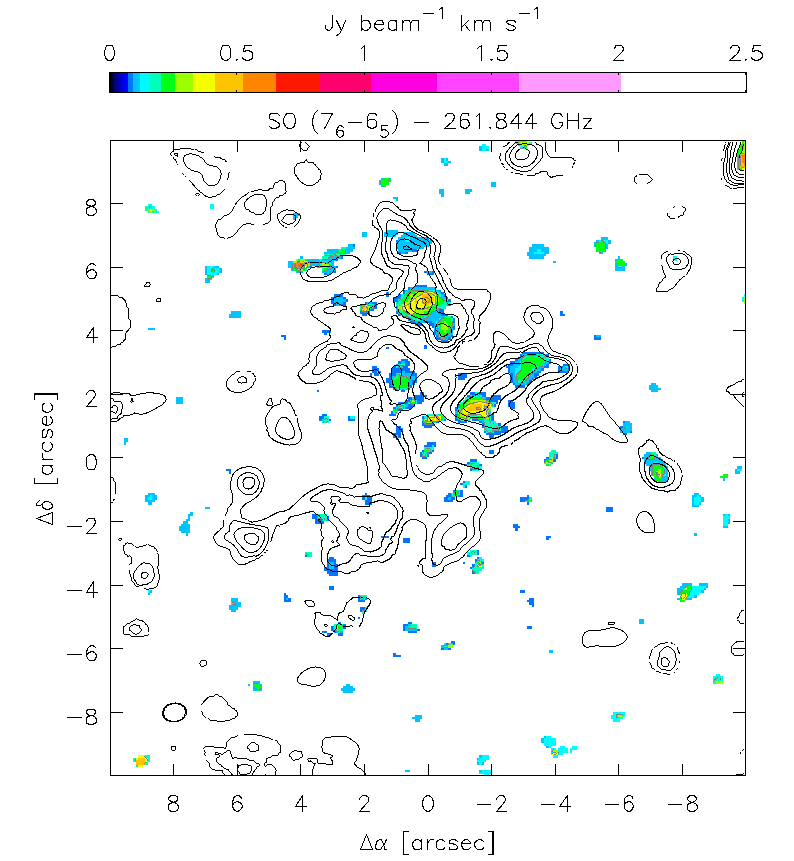}\\
        \includegraphics[trim = 5mm 0mm 7mm 0mm, clip, width=0.23\textwidth]{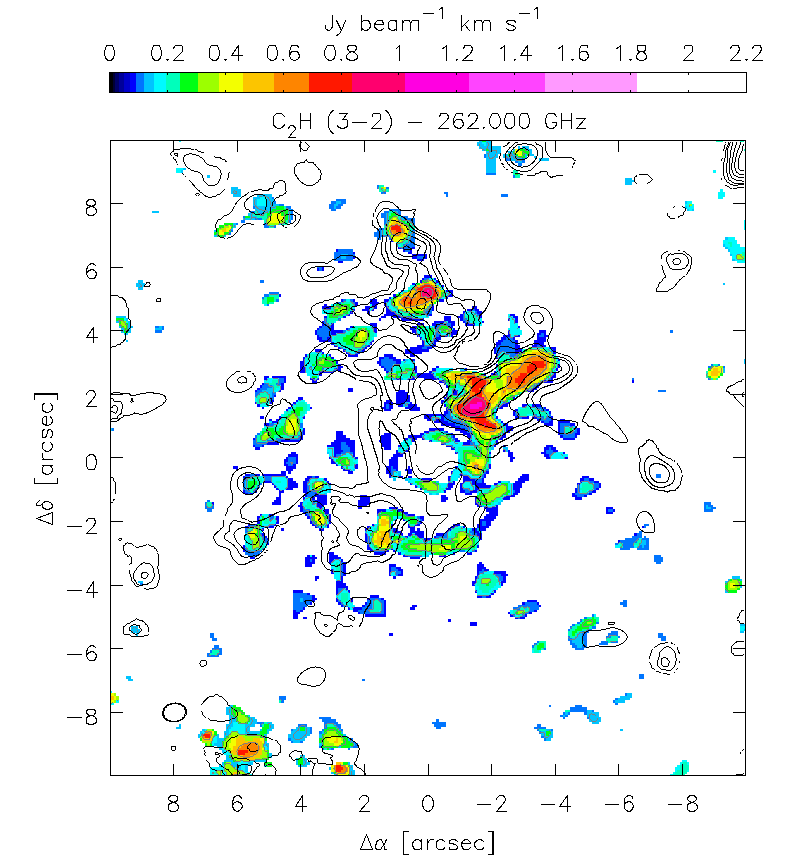}& 
        \includegraphics[trim = 5mm 0mm 7mm 0mm, clip, width=0.23\textwidth]{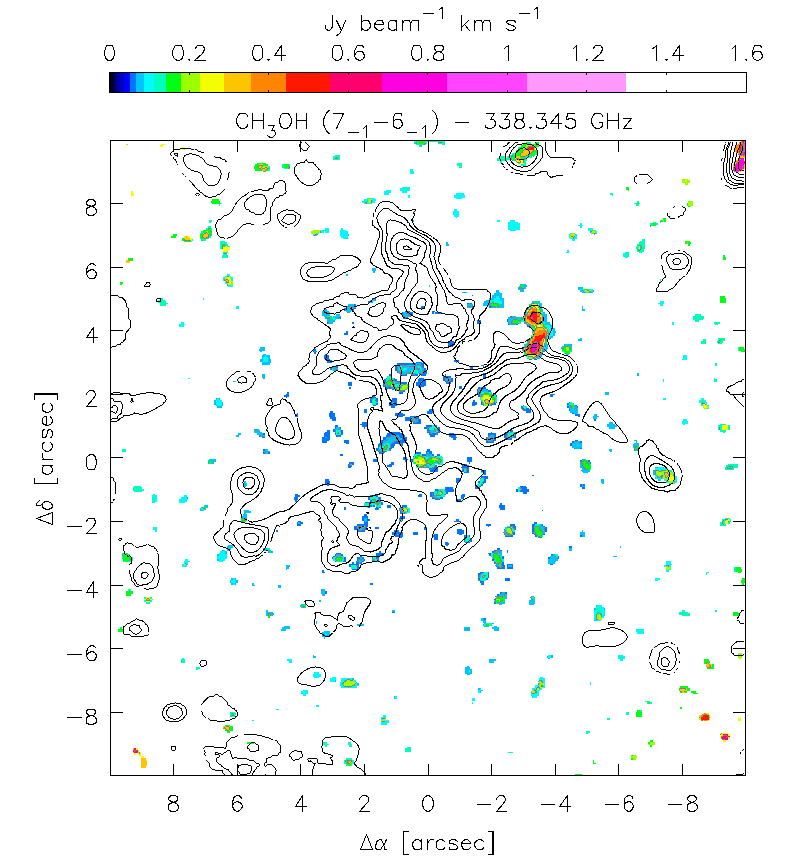}&
        \includegraphics[trim = 5mm 0mm 7mm 0mm, clip, width=0.23\textwidth]{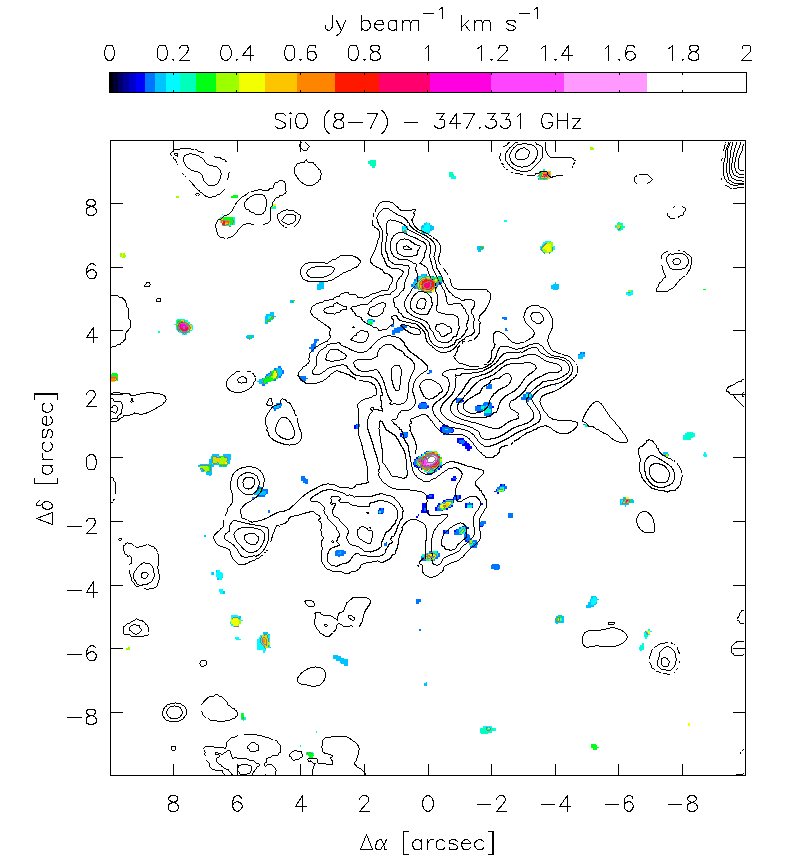}&
        \includegraphics[trim = 5mm 0mm 7mm 0mm, clip, width=0.23\textwidth]{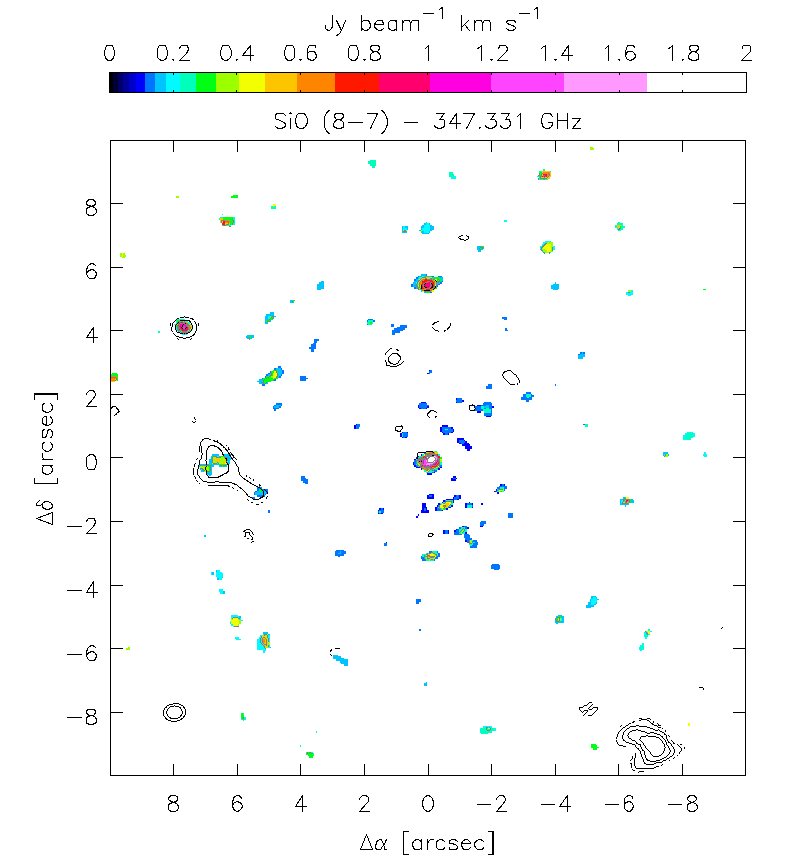}\\
        \end{array} $                                                            \caption{Molecular line emission images of the inner 20$''$ (as in Fig. \ref{mols-full1}) compared to the CS(5--4) emission in contours (as in Fig. \ref{mols-full1}). 
                Top row\ (from left to right): HC$_3$N(27--26), H$^{13}$CO$^+$(3--2), SiO(6--5), and SO(7--6). 
                Bottom row (from left to right): C$_2$H(3--2), CH$_3$OH(7--6), and two times SiO(8--7). 
                The image at the bottom right corner shows SiO(6--5) contours at [2, 4, 8, 12, 16]$\times \sigma$ (= 0.08 Jy~beam$^{-1}$~km~s$^{-1}$) for comparison.
        }
        \label{mols-CS-20}
\end{figure*}

\begin{figure*}[htbp]
        \centering $
        \begin{array}{ccc}
        \includegraphics[trim = 3mm 1mm 6mm 0mm, clip, width=0.31\textwidth]{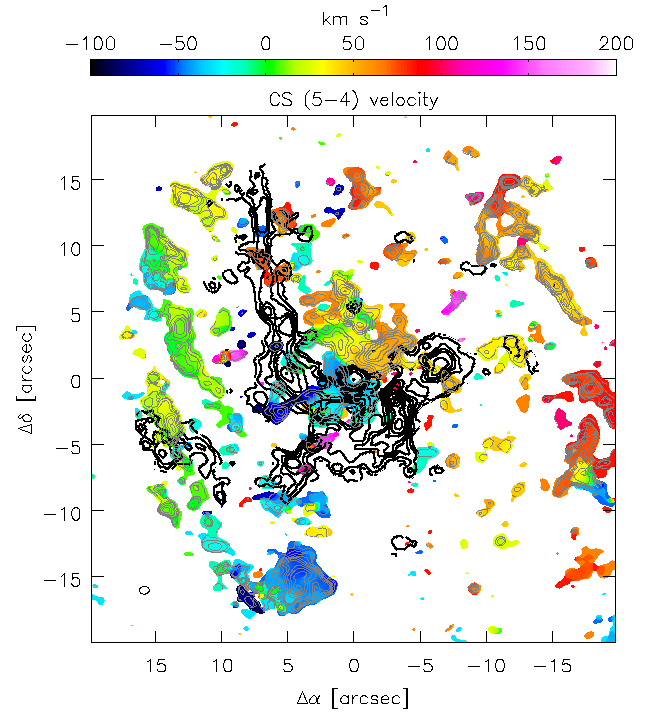}&
        \includegraphics[trim = 3mm 1mm 6mm 0mm, clip, width=0.31\textwidth]{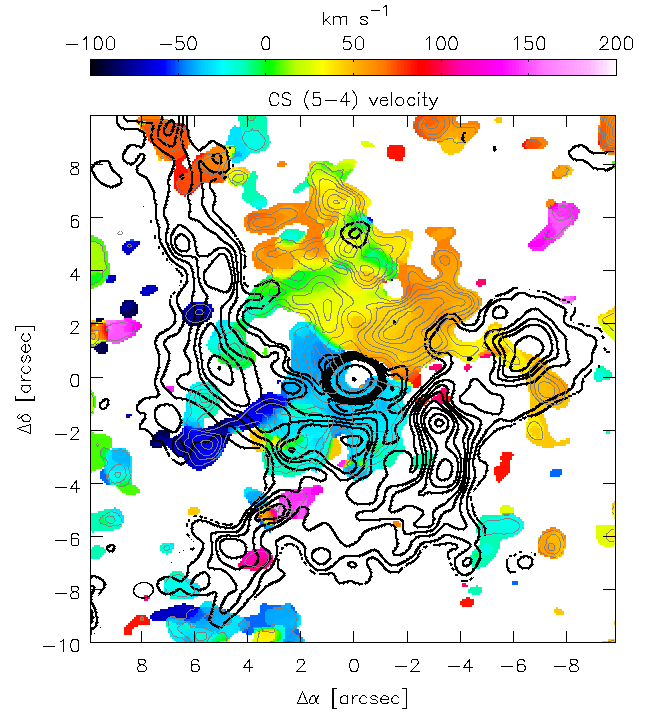}&
        \includegraphics[trim = 0mm 0mm 0mm 0mm, clip, width=0.31\textwidth]{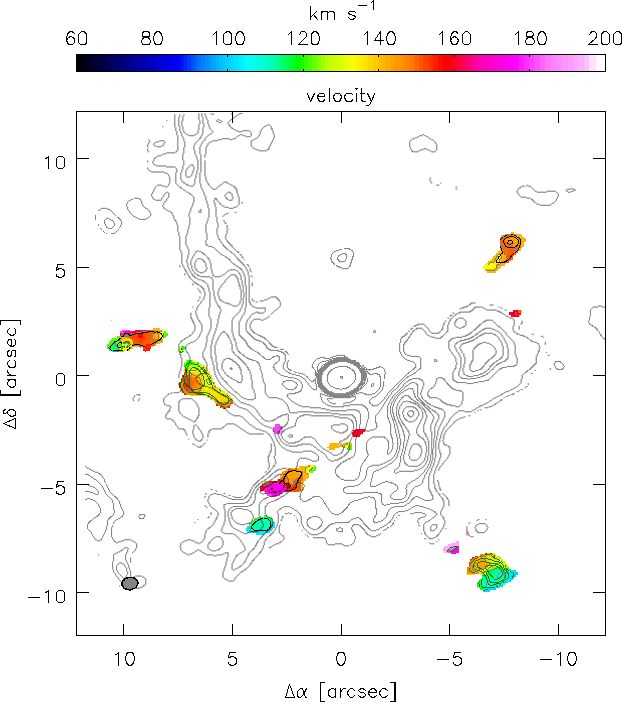}\\
        \end{array} $                                                            \caption{Velocity fields (moment 1) images for CS(5-4) for the inner 40$''$ (left) and 20$''$ (middle) for the velocity range as in Table \ref{ALMA-lines} with gray and black contours showing the CS(5--4) emission as in Fig. \ref{mols-full1} and the 250 GHz emission as in Fig. \ref{cont-full}, respectively.
                Right: Velocity field image of the HVCs overlayed with the emission of CS(5--4) in black contours at the levels of [4, 8, 12]$\times$ $\sigma$ (= 0.08 Jy~beam$^{-1}$~km~s$^{-1}$) integrated over 103 - 198 km~s$^{-1}$ and of SiO(6--5) in green contours at the levels of [2, 4, 8, 12, 16]$\times$ $\sigma$ (= 0.08 Jy~beam$^{-1}$~km~s$^{-1}$) integrated over 66 - 191 km~s$^{-1}$. Gray contours show the 250 GHz emission as in the previous panels.}
        \label{mom1}
\end{figure*}

\begin{figure*}[htbp]
        \centering $
        \begin{array}{ccc}
        \includegraphics[trim = 1mm 0mm 7mm 0mm, clip, width=0.31\textwidth]{images/spx_tap_250-100.png}&
        \includegraphics[trim = 1mm 0mm 7mm 0mm, clip, width=0.31\textwidth]{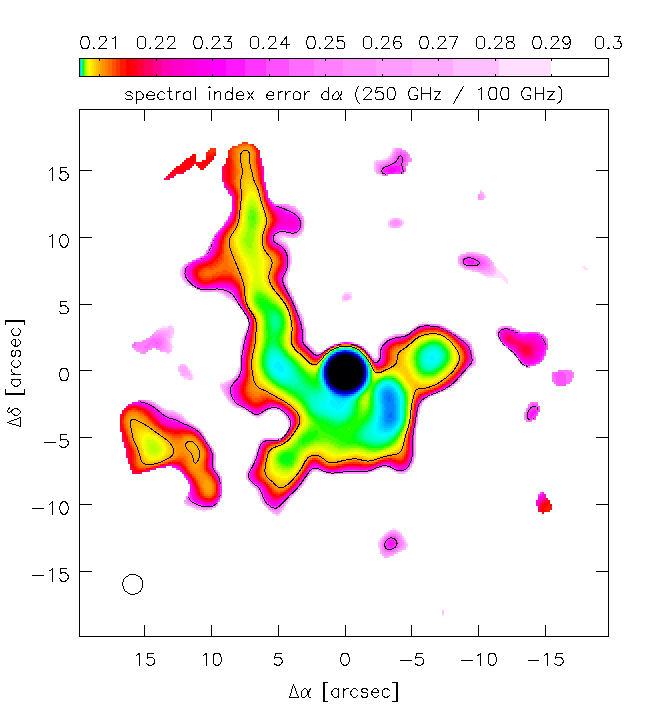}&
        \includegraphics[trim = 1mm 0mm 7mm 0mm, clip, width=0.31\textwidth]{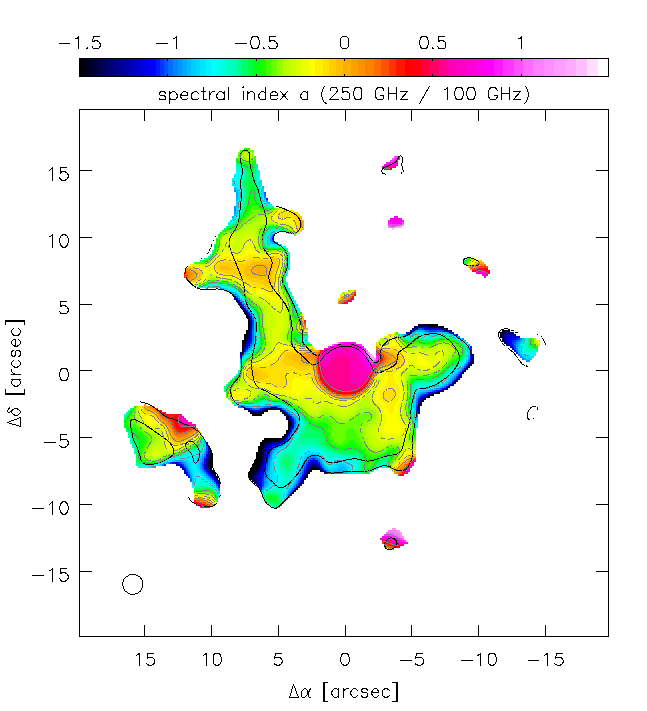}\\
        \includegraphics[trim = 1mm 0mm 4mm 0mm, clip, width=0.31\textwidth]{images/spx_tap_340-250.png}&
        \includegraphics[trim = 1mm 0mm 4mm 0mm, clip, width=0.31\textwidth]{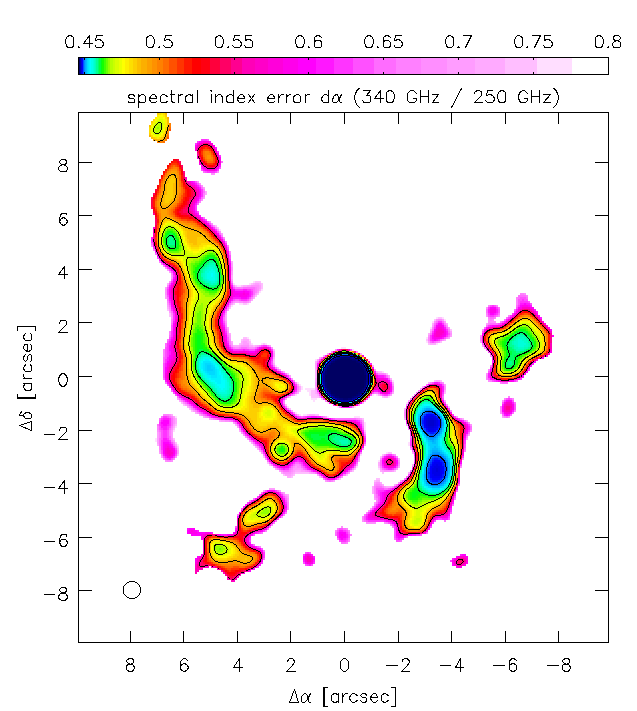}&
        \includegraphics[trim = 1mm 0mm 4mm 0mm, clip, width=0.31\textwidth]{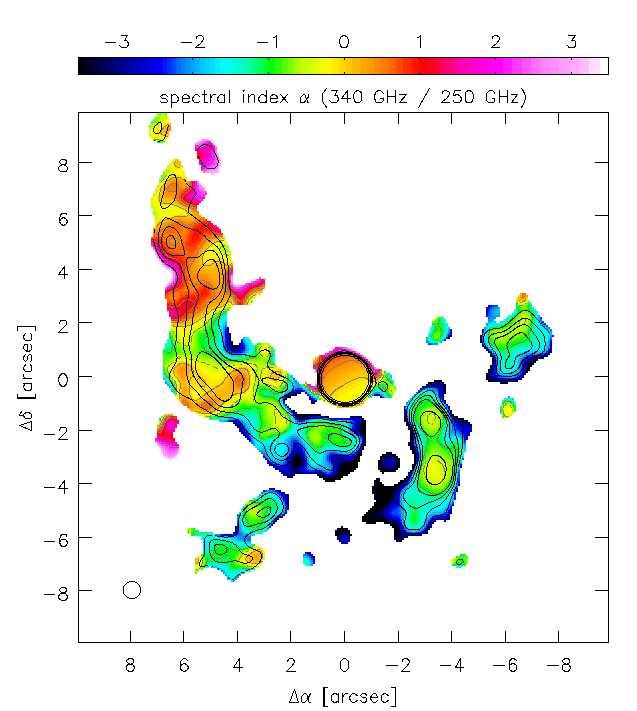}\\
        \end{array} $                                                            \caption{Continuum emission images of the inner $\lesssim $ 3 pc.
                Top: Between 100 and 250 GHz (inner 40$''$) tapered to a resolution of 1.5$''$. 
                From left to right: Spectral index with contours of [-0.75, -0.5, -0.375, -0.25, -0.125, 0, 0.25, 0.5, 1, 1.5], 
                error of the spectral index with contours of [2.068, 2.073, 2.1, 2.4] $\times$ 0.1, and spectral index overlayed with error contours. 
                Bottom: Between 250 and 340 GHz (inner 20$''$) tapered to a resolution of 0.65$''$.
                From left to right: Spectral index with contours of [-2,-1, -0.5, 0, 0.5, 1, 2], 
                error of the spectral index with contours of [4.6, 4.7, 4.9, 5.5] $\times$ 0.1, and spectral index overlayed with error contours
        }
        \label{spex-app}
\end{figure*}

\begin{figure*}[htbp]
        \centering $
        \begin{array}{ccc}
        \includegraphics[trim = 5mm 0mm 7mm 0mm, clip, width=0.31\textwidth]{images/T_e-H39a-100GHz_80as.png}&
        \includegraphics[trim = 5mm 0mm 7mm 0mm, clip, width=0.31\textwidth]{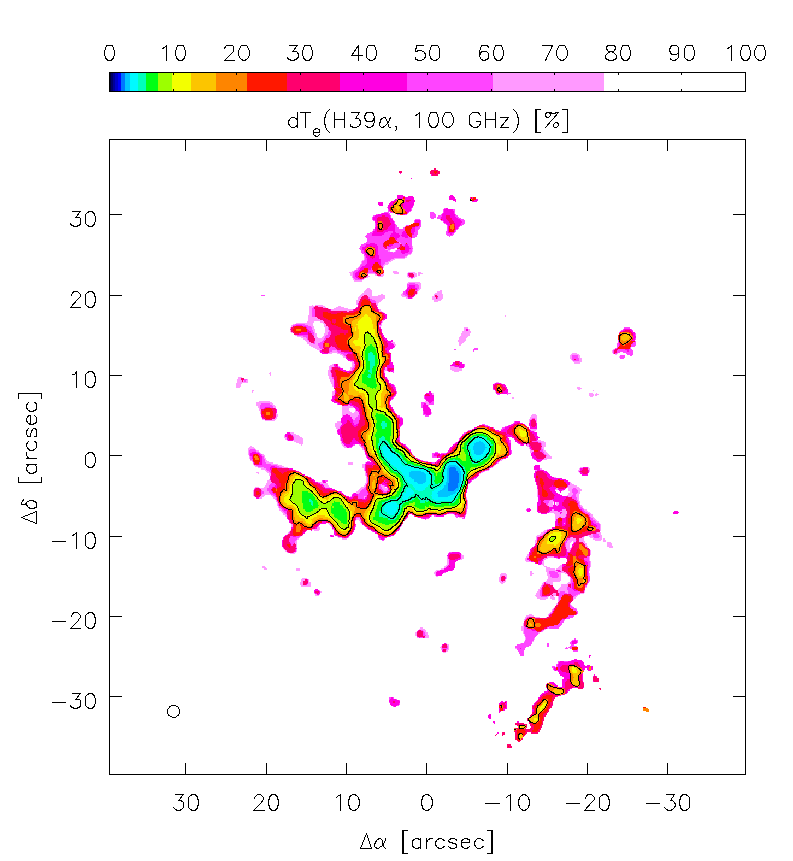}&
        \includegraphics[trim = 5mm 0mm 7mm 0mm, clip, width=0.31\textwidth]{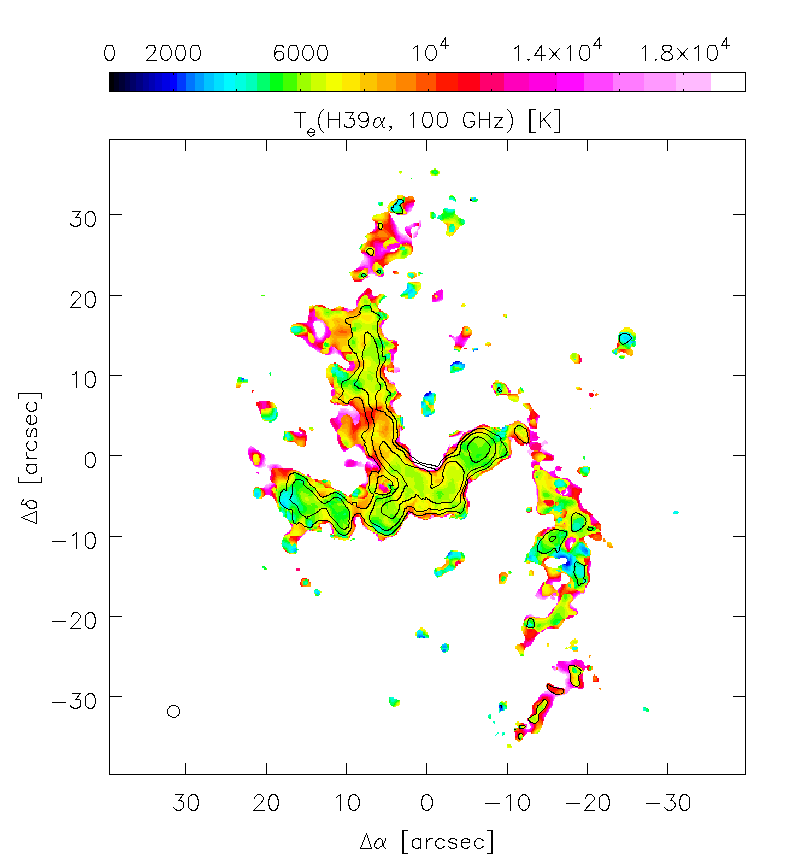}\\
        \includegraphics[trim = 5mm 0mm 7mm 0mm, clip, width=0.31\textwidth]{images/T_e-H51b-100GHz.png}&
        \includegraphics[trim = 5mm 0mm 7mm 0mm, clip, width=0.31\textwidth]{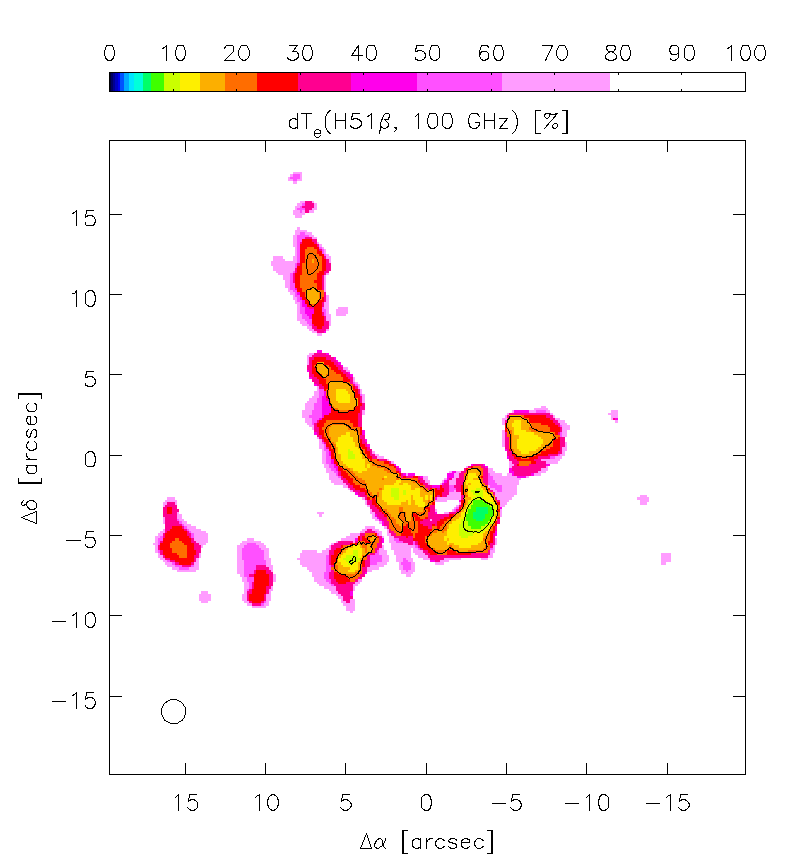}&
        \includegraphics[trim = 5mm 0mm 7mm 0mm, clip, width=0.31\textwidth]{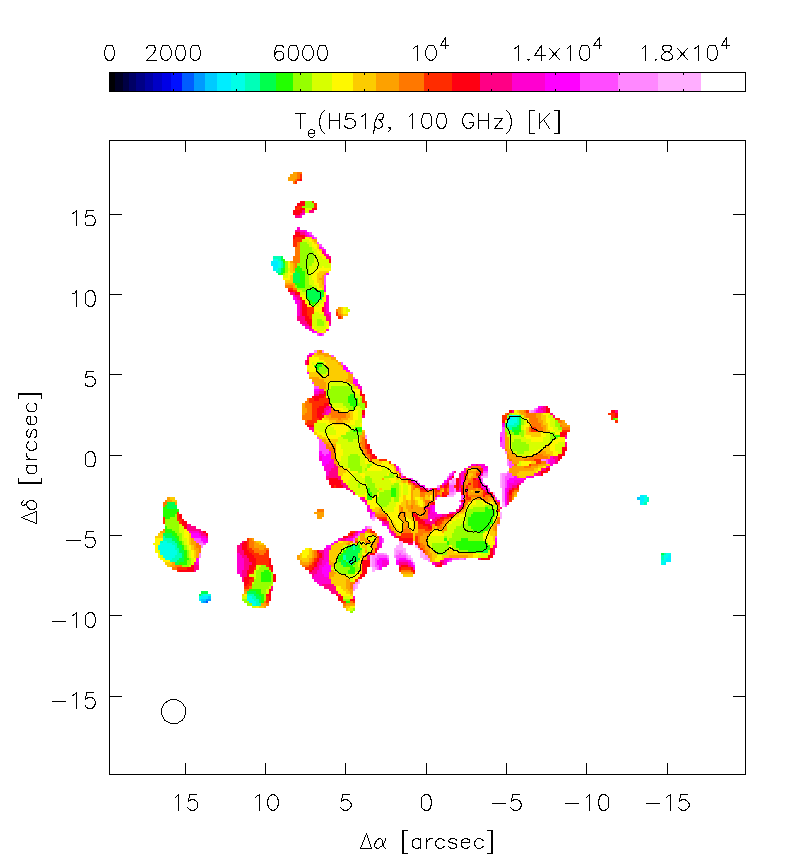}\\
        \end{array} $                                                            \caption{Images of the electron temperature distribution in the inner $\lesssim $ 1.6 pc.
                Top: Based on H39$\alpha$ (inner 40$''$) and tapered to a resolution of 1.5$''$. 
                From left to right: $T_e$ with contours of [4,6,8,10] $\times$ 1000 K, 
                $\Delta T_e$ with contours of 5, 10, and 20\%, and $T_e$ overlayed with $\Delta T_e$ contours. 
                Bottom: Based on H51$\beta$ (inner 20$''$) and tapered to a resolution of 1.5$''$.
                From left to right: $T_e$ with contours of [4,6,8,10] $\times$ 1000 K, 
                $\Delta T_e$ with contours of 5, 10, and 20\%, and $T_e$ overlayed with $\Delta T_e$ contours. 
        }
        \label{Te-app}
\end{figure*}

\begin{figure}[htbp]
        \centering 
        \includegraphics[trim = 5mm 0mm 7mm 0mm, clip, width=0.47\textwidth]{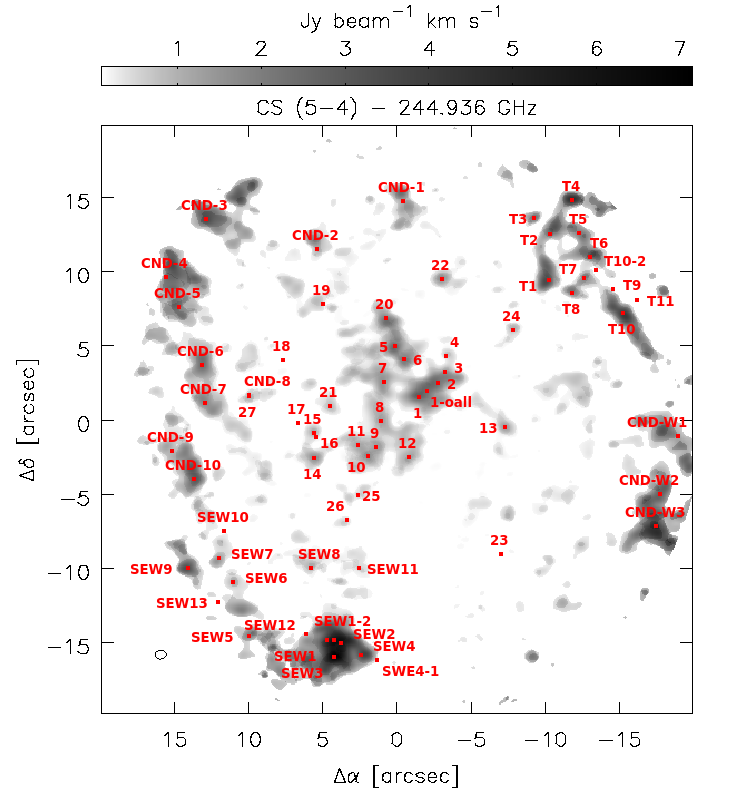}                                                        
        \caption{Finding charts for the clumps in the inner 40$''$ (1.6 pc) based on the CS(5--4) emission.}
        \label{find-app}
\end{figure}

\begin{figure}[htbp]
        \centering 
        \includegraphics[trim = 5mm 0mm 7mm 0mm, clip, width=0.47\textwidth]{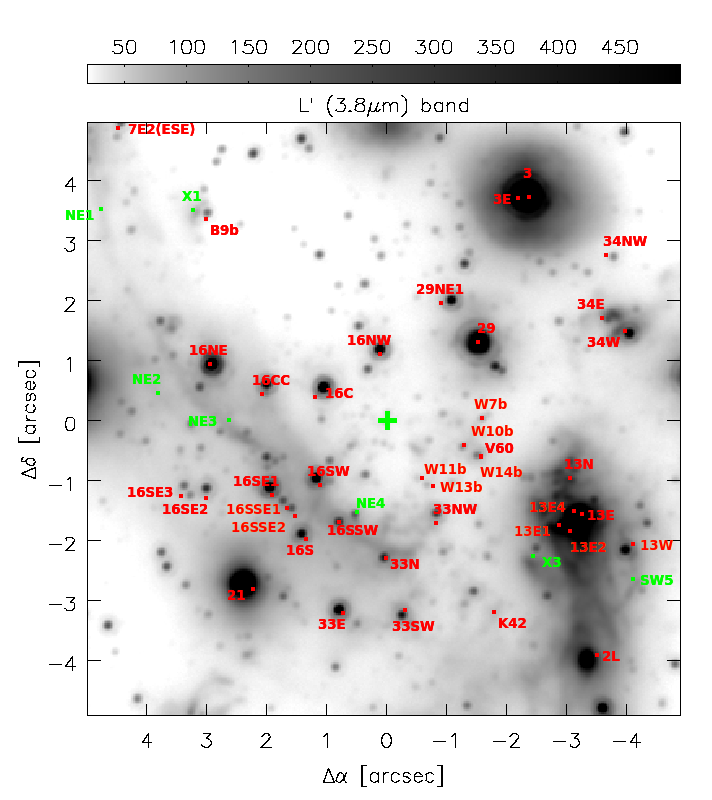}                                                        
        \caption{Finding charts for the stars and filaments \citet{Paumard2006,Viehmann2006,Muzic2007} mentioned in this work, 
                demonstrated for the inner 10$''$ (0.4 pc) on a NIR L' (3.8 $\mu$m) emission image (Sabha, private communication, here: arbitrary units).
        }
        \label{find-IRS-app}
\end{figure}

\clearpage

\section{Source positions and fluxes}


\longtab[1]{					
        
        \setlength{\tabcolsep}{0.25cm} 
        \normalsize				
        
        \begin{landscape}

        \end{landscape}
}					

\section{Molecular source velocities}
\label{sec:mol-src-vel}


\longtab[1]{					
        
        \setlength{\tabcolsep}{0.16cm}  
        \setlength\LTright{-7pt}
        \normalsize					
        
        \begin{landscape}
                                \end{landscape}
                        }						

\section{Velocities of the gas toward the star and minispiral filaments}
\label{sec:mol-IRS-vel}


\longtab[1]{					
        
        \setlength{\tabcolsep}{0.49cm}  
        
        \normalsize					
        
        \begin{landscape}
                \end{landscape}
        }					

\section{Source line ratios}
\label{sec:app-ratio}


\longtab[1]{					
        
        \setlength{\tabcolsep}{0.1cm}  
        
        \normalsize					
        \begin{landscape}
                                                                                                                                                                                                                                                                                                                                                 
                \end{landscape}                                                                                                                                                                                                                                                                 
                
        }                   					

\end{appendix}

\end{document}